%% file: got_article.tex
\let\LaTeXcline\cline
\documentclass[sn-mathphys,Numbered]{sn-jnl}
\let\cline\LaTeXcline

\usepackage{graphicx}%
\usepackage[table]{xcolor}%
\usepackage{tabularx}
\usepackage{multirow}%
\usepackage{amsmath,amssymb,amsfonts}%
\usepackage{amsthm}%
\usepackage{mathrsfs}%
\usepackage[title]{appendix}%
\usepackage{xcolor}%
\usepackage{textcomp}%
\usepackage{manyfoot}%
\usepackage{booktabs}%
\usepackage{algorithm}%
\usepackage{algorithmicx}%
\usepackage{algpseudocode}%
\usepackage{listings}%
\usepackage{subcaption}%
\usepackage{footmisc}                   


\theoremstyle{thmstyleone}%
\theoremstyle{thmstyletwo}%

\theoremstyle{thmstylethree}%

\begin{document}


%

\title{Interconnected Kingdoms: Comparing `A Song of Ice and Fire' 
Adaptations Across Media Using Complex Networks}



\author*[1]{\fnm{Arthur} \sur{Amalvy}}\email{arthur.amalvy@univ-avignon.fr}

\author*[3,4]{\fnm{Madeleine} \sur{Janickyj}}\email{janickym@uni.coventry.ac.uk}

\author*[2]{\fnm{Shane} \sur{Mannion}}\email{shane.mannion@ul.ie}


\author*[2]{\fnm{Pádraig} \sur{MacCarron}}\email{padraig.maccarron@ul.ie}

\author*[1]{\fnm{Vincent} \sur{Labatut}}\email{vincent.labatut@univ-avignon.fr}

\affil*[1]{\orgdiv{Laboratoire Informatique d'Avignon -- UPR 4128}, \orgaddress{\city{Avignon}, \country{France}}}


\affil[2]{\orgdiv{MACSI, Department of Mathematics and Statistics}, \orgname{University of Limerick}, \orgaddress{\city{Limerick}, \country{Ireland}}}

\affil[3]{\orgdiv{Centre for Fluid and Complex Systems}, \orgname{Coventry University}, \orgaddress{\city{Coventry}, \country{United Kingdom}}}

\affil[4]{\orgname{$\mathbb{L}^4$ Collaboration \& Doctoral College for the Statistical Physics of Complex Systems}, \orgaddress{ \city{Leipzig-Lorraine-Lviv-Coventry}, \country{Europe}}}

\abstract{In this article, we propose and apply a method to compare adaptations of the same story across different media. We tackle this task by modelling such adaptations through character networks. We compare them by leveraging two concepts at the core of storytelling: the characters involved, and the dynamics of the story. We propose several methods to match characters between media and compare their position in the networks; and perform narrative matching, i.e. match the sequences of narrative units that constitute the plots. We apply these methods to the novel series \textit{A Song of Ice and Fire}, by G.R.R. Martin, and its comics and TV show adaptations. Our results show that interactions between characters are not sufficient to properly match individual characters between adaptations, but that using some additional information such as character affiliation or gender significantly improves the performance. On the contrary, character interactions convey enough information to perform narrative matching, and allow us to detect the divergence between the original novels and its TV show adaptation.}

\keywords{Character networks, Adaptations across media, Character matching, Narrative matching, Centrality study.}



\maketitle

\section{Introduction}
\label{sec:introduction}
It has been a long-standing tradition that if a story is highly successful in one medium,  it gets  adapted to another. For example, the 13\textsuperscript{th} century folk-tales of \textit{Robin Hood} was turned into stage plays two centuries later, 
and more recently into film and TV shows, 
or even the 10\textsuperscript{th} century \textit{Beowulf} being turned into a 21\textsuperscript{st} century film. Hence, it is no surprise that recent successful stories are adapted into novels, movies, TV shows, computer games, comics, etc. In this study, we are interested in how that adaptation process affects the way the plot and the characters' roles change to conform to the constraints of a different medium. According to Hutcheon's terminology~\citep{Hutcheon2006}, we compare the \textit{adaptation} of a \textit{story} into different \textit{forms}. We consider the \textit{story} itself to be the underlying object (encompassing plot, characters, themes\ldots) that is being transcribed into a \textit{form} (film, book, video game\ldots). For simplicity, we refer to the different versions of the same story as \textit{adaptations}, even when that includes the original version of the story. Meanwhile, the \textit{medium} refers to the physical support (electronic, book\ldots) of each adaptation, while the \textit{plot} is the arrangement of the series of events that is being shown. Following the \textit{Living Handbook of Narratology}\footnote{\url{https://www-archiv.fdm.uni-hamburg.de/lhn/}}, we define a character as a ``media-based figure in a storyworld, usually human or human-like''.

In order to model the adaptations, we leverage character networks, a tool which has frequently been used for this purpose in recent years~\cite{Labatut2019}. Many authors have tried to use such complex networks to compare different adaptations. \citet{Chaturvedi2018} want to detect remakes among a collection of movies. They work with Wikipedia articles describing the movie plots, and extract a number of text- and graph-related features to train a classifier to distinguish between remakes from original films. \citet{Chowdhury2019} tackle the problem of comparing novels and their movie adaptations, through their conversational networks. Their method focuses on four centrality metrics to summarise and compare the networks. \citet{Massey2019} works on the four \textit{Gospels} and the \textit{Acts of the Apostles}. He proposes a method to align their corresponding character networks, in order to compute some similarity scores between these adaptations. These scores are then used to perform a hierarchical clustering which highlights the relationships between the different adaptations. 
Finally, \citet{Zhang2021o} compare two books telling the same ancient Chinese story from the perspective of a historian and from that of a novelist. They work on the English translations to extract both character networks, and compare them in terms of standard topological properties (small-worldness, scale-freeness).

Although the works we are dealing with here are fictional, by having a self-contained dataset we can identify difficulties when working with adaptations, or different tellings of similar events. This is particularly problematic when dealing with historical social networks. A \emph{hagiography} deals with the life of a saint, comparing these to the same events in history texts, different structural properties are observed~\cite{gramsch2017medieval}. This work identified some female characters, central to the events, that were barely mentioned in one source but more integrated in the network in another. Similar issues arise in dealing with medieval sagas, for example some Irish narratives appear in different manuscripts and have large differences~\cite{kinsella2002tain}. These alternative versions have different numbers of characters~\cite{kenna2024}, so it is important to be able to understand how the overall network differs and whether we can determine if this is due to the adaptation process.

Here, we analyse the book series \textit{A Song of Ice and Fire}, the comic of the same name adapted from this, and its better known TV series \emph{Game of Thrones}. When the author, George R. R. Martin -- also a TV show writer -- envisaged the book series as he began writing it, he deemed it too massive for a TV series to capture. However, less than 20 years after its publication, it spawned one of the most successful TV series of all time. As a sprawling epic across a large geographical region, the book series introduces many characters, almost 2,000 named characters by the end of the fifth book (the series is not yet completed). However, the TV adaptation, which is completed, contains over 500 named characters in total. 

There are previous works analysing \textit{A Song of Ice and Fire} through the prism of character networks. \citet{Beveridge2016} use traditional social network analysis tools to study a network they extract from the third novel, \textit{A Storm of Swords}. \citet{Liu2017} use structural balance to study the tensions between the main houses in the TV show. \citet{Beveridge2018} perform analysis through character networks they obtain from the script of the TV show, notably around the concept of \textit{fractal protagonist}. \citet{Garza2020} study the first three seasons of the TV show, and focus on identifying the most important relationships. \citet{Stavanja2019} try to predict the next kill in the TV Show, while \citet{Gessey2020} try to assess the realism of the novels' network. However, to the best of our knowledge, it is the first time that character networks are used to compare three adaptations of the same story, developed for three distinct media. More precisely, we focus on two research questions. First, is it possible to match the characters from one adaptation to the other, based on their interactions (RQ1)? Second, is it possible to use these interactions to align their plots (RQ2)? For instance, to determine whether one TV episode relates to a certain chapter of the novel. Answering these two research questions will allow understanding what tools are applicable when studying adaptations through character networks.

Concretely, our first contribution is to constitute a corpus of networks representing the original novels as well as their comics and TV show adaptations. In order to answer the RQ1, we formulate the task as a \textit{Graph Matching} problem. Our contributions are empirical here, as we apply existing graph matching methods to experiment with various types of information when identifying the best match. We also consider a relaxed version of the problem, and perform a centrality analysis to provide some insight into our matching results. To tackle RQ2, we formulate the task as a many-to-many matching problem, which is called \textit{Narrative Matching} in the literature
~\citep{Pial2023a,Pial2023b}. Our contributions here are more methodological, as we propose three approaches aiming at solving this task: the first is a baseline that relies on textual representations of the stories, the second leverages the structure of the dynamic networks modelling the stories, and the third combines these two types of information. We also propose the notion of block, an intermediary temporal scale to represent the TV show, which leads to improved matching performance.

The rest of this article is organised as follows. We first present the raw data and the methods used to extract the character networks (Section~\ref{sec:Data}). Next, we apply and compare several methods to match characters between networks, and answer RQ1 (Section~\ref{sec:CharComp}). We then turn to RQ2, describing and assessing several approaches to align the dynamics of different adaptations (Section~\ref{sec:Align}). Finally, we discuss our main findings and perspectives (Section~\ref{sec:Conclusion}).


\section{Data Description}
\label{sec:Data}
With three  different forms of media being explored within this work, it is necessary to use three different methods to gather the raw data and extract the networks. Following the common practice regarding character networks~\cite{Labatut2019}, each vertex in our networks represents a character, whereas each edge models an interaction between them. In the following, we first describe how we extract the networks from the adaptations (Section~\ref{sec:DataRel}), and how we represent the dynamics of the adaptations (Section~\ref{sec:DataTime}). We then explain how we supplement the basic networks with vertex and edge attributes (Section~\ref{sec:DataAttr}). Next, we compute and discuss the main topological properties of the networks (Section~\ref{sec:DataDescr}). Finally, we define distinct groups of characters to conduct our experiments (Section~\ref{sec:DataChar}). 
Our data\footnote{\url{https://doi.org/10.5281/zenodo.13893061}} and source code\footnote{\url{https://github.com/CompNet/Sachan}} are available online.

\subsection{Character Interactions}
\label{sec:DataRel}
The network extraction process is specific to each type of medium considered in this work. In order to produce comparable graphs, we select data sets and apply extraction methods that allow us to obtain edges with roughly the same semantics: two connected vertices model two characters that interact during the considered narrative unit.

The first medium corresponds to the five novels constituting George R. R. Martin's original material, as of February 2024. We do not work directly with the raw text, but rather with the manual annotations provided by Gessey-Jones \textit{et al}.~\cite{Gessey2020}. They applied a close-reading approach, and recorded each character appearing in a chapter, and every interaction between two characters in each chapter. Each edge in our network model such an interaction. Note that this does include memories, as frequently important plot information is revealed through a character thinking of past events. As a result, the characters appearing in a chapter are not necessarily present at that particular time point.

The Comics series\footnote{\url{https://georgerrmartin.com/cover-art-gallery/?id=4512}} is constituted of two volumes, which are directly adapted from the first two novels. These volumes contain 24 and 32 issues respectively. Each issue is, in turn, split into two to four chapters matching those of the novels. These chapters were not published in the exact same order as in the novels, though, probably for editorial reasons related to the number of pages in an issue. Although their title is the same as the TV Show, the comics are a direct and  close adaptation of the novels~\cite{AWoIaF2021}. Their publication started with the first season of the TV show, presumably for commercial reasons. We manually annotated the comics using the method from~\cite{Labatut2022}, which uses the scene as the narrative unit. Here, a scene is defined as a sequence of consecutive panels involving the exact same group of characters, without interruption. In our network, we connect all characters interacting in the same scene.

The TV show\footnote{\url{https://www.hbo.com/game-of-thrones}}, entitled \textit{Game of Thrones}, comprises eight seasons that cover the five novels, but also go beyond the original plot. Moreover, it is a much looser adaptation of the novels~\cite{AWoIaF2022} than the comics. The plot is relatively close to Martin's books at first, but gradually diverges, especially after the fifth season. The later seasons were developed from information made known by Martin to the show runners, and under their own direction. Jeffrey Lancaster provides a very rich dataset concerning the TV show\footnote{\url{https://github.com/jeffreylancaster/game-of-thrones}}. We leverage his annotations describing scene co-occurrences. As before, we assume that two characters occurring in the same scene interact, and connect them in our network.

\subsection{Representation of Time}
\label{sec:DataTime}
Based on the annotated interactions, we extract several types of networks. First, we consider \textit{static} networks by integrating all the interactions over certain periods of time, in order to constitute large snapshots of the adaptations. This integration connects two characters if there is at least one interaction between them during the considered period. Moreover, as in many character network-related works from the literature~\cite{Labatut2019}, we compute edge weights corresponding to the number of such interactions over the period of interest. In our experiments, we consider three periods: the first two novels, comic volumes, and TV show seasons, noted \texttt{U2}; the first five novels and seasons, noted \texttt{U5}; and all 8 seasons of the TV show, noted \texttt{U8}. We select these periods so that the concerned adaptations cover the same part of the plot, in order to allow a fair comparison. Table~\ref{tab:NarrStats} in the Supplementary Material provides a summary of the structure of the adaptations in terms of their constitutive narrative units, whereas Figure~\ref{fig:NarrOverlap} shows how much their timelines overlap.

In addition to the static networks, we extract different forms of \textit{dynamic} networks. All of them are based on the same notion of temporal integration as before, but this time using a \textit{sliding window}. As a result, one gets a sequence of graphs called \textit{time slices}, each one representing a position of the sliding window in the adaptation. The size of this window can be expressed in terms of different narrative units, depending on the media. For the novels, the raw data only allows us to use the \textit{chapter}. For comics, from the smallest to the largest unit, we use the \textit{scene}, the \textit{chapter}, and the \textit{issue}. For the TV show, in the same order, we have the \textit{scene} and the \textit{episode}. However, there is a scale problem when comparing the TV show and novels, because a scene is much shorter than a chapter in terms of plot, and an episode is much larger. For this reason, we introduce the concept of the \textit{block} to segment the TV show. Blocks are lists of contiguous scenes from the same episode, that ideally correspond to a \textit{point of view} sequence, i.e. a part of the story narrated from the perspective of a given character. We make the assumption that a new block in the TV show starts when there is a change in geographical location, as it is likely that this indicates that the focus switches to a different character group. Since the annotations for the TV show include the geographical location of each scene, we automatically extract blocks by considering that two contiguous scenes are from the same block if they occur in the same location.

We extract a dynamic network for all available time scales (scene, block, chapter, issue, episode) and period (\texttt{U2}, \texttt{U5}, \texttt{U8}). Moreover, we adopt two representations of time, resulting in two distinct types of dynamic networks. In \textit{cumulative} networks, each time slice integrates all the interactions since the \textit{beginning} of the period. On the contrary, in \textit{instant} networks, each time slice integrates only the interactions occurring in the corresponding \textit{time window}. The former tend to be more stable, whereas the latter better reflect sudden changes in the dynamics, which makes them suitable for different uses.

\subsection{Vertex \& Edge Attributes}
\label{sec:DataAttr}
As mentioned before, edge weights are computed when performing the temporal integration of character interactions, simply by counting the number of such interactions occurring during some time period. Due to the difference between adaptations and time scales, these values cannot be compared directly. For this reason, we max-normalise them, i.e. we divide each weight by the largest weight in its network, which produces values in the interval $[0{;}1]$.

The two data sources that we use to extract the novels and TV show networks also provide some information regarding individual characters. Based on this information, we extract two vertex attributes representing the biological sex and the main affiliation of the characters. We solve any disagreement between the sources thanks to two Wiki Website populated by fans of the novels (\textit{A Wiki of Ice and Fire}\footnote{\url{https://awoiaf.westeros.org/index.php/Main_Page} \label{ftn:awoiaf}}) and of the TV show (\textit{Wiki of Westeros}\footnote{\url{https://gameofthrones.fandom.com/wiki/Wiki_of_Westeros} \label{ftn:wow}}). We also leverage the online resources to complement our own annotations of the comics and define the same vertex attributes for this medium. The \textit{Sex} attribute corresponds to the biological sex as inferred from the adptations. It can take the values \textit{Male} and \textit{Female}, but also \textit{Unknown} in some certain cases, and \textit{Mixed} when dealing with vertices representing groups of characters. The \textit{Affiliation} attribute identifies the main organisation to which the character belongs. The term organisation must be taken in a very broad sense, since they can be institutions such as the \textit{Gold Cloaks} (city watch of \textit{King's Landing}) or the \textit{Faith of the Sevens} (clergy), and noble houses such as \textit{House Stark} or \textit{House Lannister}, but also more informal groups such as the \textit{Brave Companions} (group of sellswords). Table~\ref{tab:CharacterList} in the Supplementary Material shows the main characters with their \textit{Sex} and \textit{Affiliation} attributes.

Finally, a critical vertex attribute is the name of the characters. There is some variability in these names from one adaptation to the other, and even within the same adaptation. For instance, some women are referred to by both their maiden and married names (eg. \textit{Talisa Maegyr} vs. \textit{Talisa Stark}); some characters have a nickname (ex. \textit{Fat Walda} is \textit{Walda Frey}); there are many homonyms that must be distinguished (ex. five persons are named \textit{Walder Frey}). This heterogeneity prevents a direct comparison of the networks based on vertex names. For this reason, we leverage the previously mentioned online Wikis to curate a list of names used as a reference, and normalise the network names so that they all take the exact same form. We also create a \textit{Named} Boolean attribute, which indicates whether a character has a proper noun. Thus, it is false for characters that are mentioned in the novels but referred to with some descriptive expressions only (e.g. \textit{Stable boy}), as well as characters that are shown in the comics or TV show, but never explicitly introduced. We use this attribute to filter out minor characters, as explained later.

As mentioned previously, the TV show is a much looser adaptation of the Novels. Several pages of the online Wikis describe in detail how both adaptations differ, and three of these differences are particularly relevant to the present work. First, the name of certain characters has been completely changed when adapting the novels to TV. For instance, Dothraki warrior \textit{Jhogo} becomes \textit{Kovarro}, while the Yunkai nobleman \textit{Grazdan mo Eraz} is renamed \textit{Razdal mo Eraz}. Unlike the name differences described above, this is not due to natural language variability, but rather to a voluntarily choice of the showrunners. In general, this is meant to avoid confusion between characters that have the similar names, or identical first names~\cite{WoW2023}. We consider them as distinct characters in our networks, as these differences are not just cosmetic, but may also involve behavioural aspects. The second relevant difference in the TV adaptation is the merging of certain minor characters into a single character~\cite{WoW2023a}. For instance, red priestess \textit{Kinvara} from the TV show corresponds to both red priests \textit{Benerro} and \textit{Moqorro} from the novels. In our networks, we consider all of these characters as different. Third, some characters were outright created for the TV Show. We counted 85 of them in our data. By comparison, the names used in the comics are exactly the same as in the novels: the adaptation only affects the number of characters explicitly shown, which is constrained by the medium.

\subsection{Descriptive Analysis}
\label{sec:DataDescr}
\begin{figure}[htb!]
    \centering
    \includegraphics[height=4.8cm]{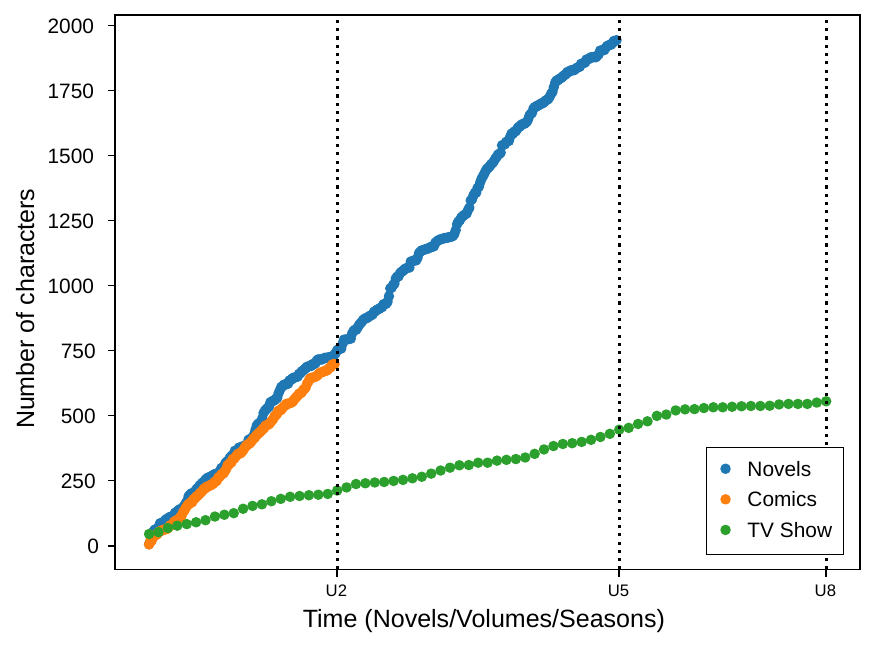}
    \hfill
    \includegraphics[height=4.8cm]{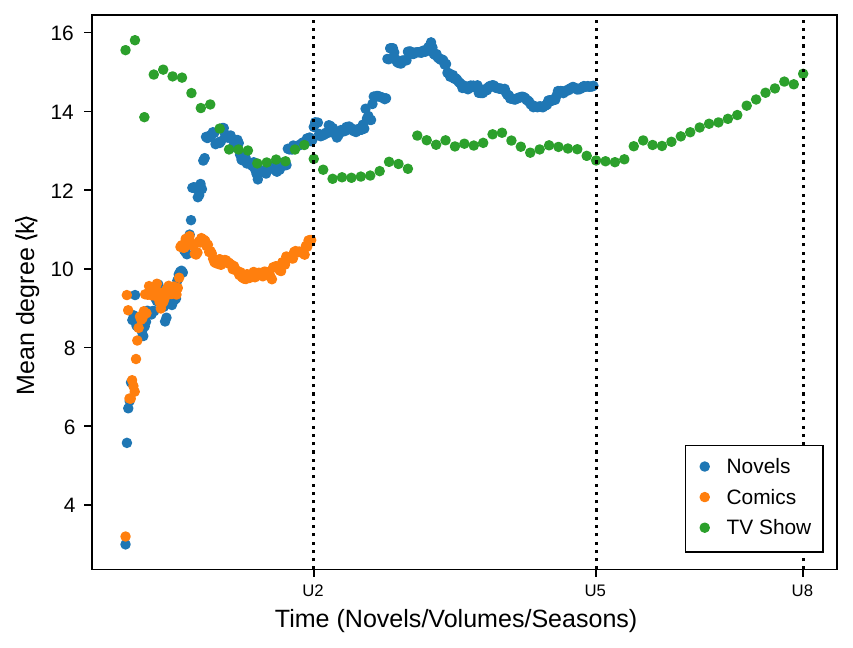}
    \caption{The left panel shows the number of characters in plot time for each of the three media. On the right, the average degree $\langle k \rangle$ is displayed in time}
    \label{fig:topology}
\end{figure}

We initially compare the evolution of the total number of characters in each medium. The left panel in Figure~\ref{fig:topology} displays this for the three time periods. In each medium, the number of characters grows almost linearly, with the only obvious plateau being in the final time-period for the show. The growth of the number of characters is much slower in the show compared to the other two media. 

We create a static network of each time period from the interactions. Table~\ref{tab:networks} displays some of the standard network properties for each of these. While they have different numbers of vertices, complex networks properties tend to not be related to system size (see, for example, \citet{watts1998collective}). Here we observe that mean degree $\langle k \rangle$, average shortest path length $\langle \ell \rangle$, clustering coefficient $C$, and assortativity $r$ do not change much in each time period.

\begin{table}[htb!]
    \caption{Network statistics for each of the time-periods giving the number of vertices $n$ and edges $L$, the density $\delta$, the mean degree $\langle k \rangle$, the average shortest path length $\langle \ell \rangle$, the average clustering coefficient $\langle C \rangle$, the degree assortativity $r$, and the modularity $Q$}
    \label{tab:networks}%
    \begin{tabular}{@{\extracolsep{\fill} }l l r r r r r r r r r r}
        \toprule
        Period & Adaptation & $n$ & $L$ & $\delta$ & $\langle k \rangle $& $ \langle \ell \rangle $& $\langle C \rangle$ & $r$ & $Q$\\
        \midrule
        \texttt{U2} & \textit{Novels} & 777 & 5,216 & 0.017 & 13.43 & 3.01 & 0.60 & $-0.01$ & 0.53 \\
        & \textit{Comics} & 721 & 3,745 & 0.014 & 9.61 & 3.16 & 0.72 & $-0.17$ & 0.68 \\
        & \textit{TV Show} & 199 & 1,308 & 0.066 & 13.15 & 2.88 & 0.75 & $0.07$ & 0.57 \\
        \midrule
         \texttt{U5} & \textit{Novels} & 1,943 & 14,232 & 0.008 & 14.65 & 3.33 & 0.58 & $-0.02$ & 0.58 \\
         & \textit{TV Show} & 430 & 2,767 & 0.030 & 12.87 & 2.90 & 0.78 & $-0.05$ & 0.61 \\
        \midrule
         \texttt{U8} & \textit{TV Show} & 555 & 4,150 & 0.027 & 14.95 & 2.70 & 0.78 & $-0.08$ & 0.44 \\
        \botrule
    \end{tabular}
\end{table}

Comparing the different media, the clustering coefficient is higher in the TV show and comics compared to the novels. It is likely an artefact of how interactions are defined in the annotations. In the comics and TV show, an edge is made between all characters participating in the same scene. In the novels, if there is a large group of characters, edges are only made when it is explicit that two of those characters interact. The former methods will likely yield more closed triads. However, other properties are more similar for the TV show and novels. 

The comics have a lower average degree and lower assortativity than the other two media. It is worth observing that a lot of the action is not shown in the comics, but described in long text boxes. This is quite unusual, as comics' authors tend to apply the \textit{Show, don't tell} principle. This narrative text mainly concerns important characters, and the interactions described in the text do not appear in the comics annotations. The comic networks consequently miss some of these interactions between high degree vertices, which could explain the lower assortativity. The right panel of Figure~\ref{fig:topology} displays the mean degree chapter by chapter. Here we see the average degree stabilise mid-way through the first time period, though the novels have a big increase after the third book. This jump is not present in the TV show, but in the final time period we see this value increase as plots and characters converge.

On a global level, however, the network properties are not too different between each of the media. We might have expected the TV show and novel networks to be more different by the end of period \texttt{U5} due to the stories diverging more, however this is not the case. Despite fewer characters in the TV show, the topological properties are similar.

\subsection{Character Sets}
\label{sec:DataChar}
In the annotated data that we use to extract the networks, the numbers of characters are quite different from one adaptation to the other, as shown in Figure~\ref{fig:topology} and Table~\ref{tab:CharNbrs} (rows noted \textit{All characters}). For period \texttt{U2}, there are 777 characters in the novels, but only 199 in the TV show. The comics contain 721 characters, a number comparable to the novels'. However, only 324 (45\%) of them are named, and therefore likely to be of importance, whereas there are 731 (94\%) named characters in the novels, and 167 (84\%) in the TV show annotations. On the one hand, these variations reflect changes due to the adaptation process. For instance, it is not possible to show too many characters in a Comics panel while keeping them recognisable, and the TV Show needs to show physical persons and is therefore more expensive than the other two mediums. But on the other hand, these variations also show differences in the way the three adaptations are annotated. In particular, (with a handful of exceptions) unnamed characters are not listed in the novels annotations, and only a few of them are explicitly mentioned in the TV show annotations.

\begin{table}[htb!]
    \caption{Numbers of characters in the different character sets and periods, for each adaptation. The rows noted \textit{All characters} correspond to the unfiltered character set. Some periods concern only certain adaptations, hence the empty cells. The \texttt{common} character set depends on the compared networks. There are 123 named characters common to all three \texttt{U2} networks, and 20 in \texttt{top-20}}
    \label{tab:CharNbrs}%
    \begin{tabularx}{\textwidth}{l l X r r r}
        \toprule
        Period & \multicolumn{2}{l}{Character set} & \textit{Novels} & \textit{Comics} & \textit{TV Show} \\
        \midrule
        \texttt{U2} & \multicolumn{2}{l}{All characters} & 777 & 721 & 199 \\
         \cmidrule{2-6}
         & \texttt{named} &  & 731 & 324 & 167 \\
         \cmidrule{2-6}
         & \texttt{common} & Adaptation vs. \textit{Novels} & -- & 297 & 139 \\
         &  & Adaptation vs. \textit{Comics} & 297 & -- & 134 \\
         &  & Adaptation vs. \textit{TV Show} & 139 & 134 & -- \\
        \midrule
        \texttt{U5} & \multicolumn{2}{l}{All characters} & 1,943 & -- & 430 \\
         \cmidrule{2-6}
         & \texttt{named} &  & 1,877 & -- & 285 \\
         \cmidrule{2-6}
         & \texttt{common} & Adaptation vs. \textit{Novels} & -- & -- & 216 \\
         &  & Adaptation vs. \textit{TV Show} & 216 & -- & -- \\
        \midrule
        \texttt{U8} & \multicolumn{2}{l}{All characters} & -- & -- & 555 \\
         \cmidrule{2-6}
         & \texttt{named} &  & -- & -- & 348 \\
        \botrule
    \end{tabularx}
\end{table}

In order to deal with this issue, and also to ease the analysis and comparison of the character networks, we consider three increasingly constrained types of character subsets. The first, which we note \texttt{named}, includes all the characters that are referred to by a proper noun. The second type includes all characters from \texttt{named} that appear in all three adaptations, or in both compared adaptations, depending on the context, and it is noted \texttt{common}. The sizes of these sets are shown in Table~\ref{tab:CharNbrs} (rows \texttt{named} and \texttt{common}, respectively).

The third type of character set contains the 20 most important characters from \texttt{common}, and is noted \texttt{top-20}. In order to assess character importance, we consider the number of occurrences of each character in each adaptation, using the smallest available narrative unit: chapters for the novels, and scenes for the comics and TV show. We max-normalise these values separately for each adaptation, in order to get scores in $[0{;}1]$, where 1 corresponds to the most frequent character in the adaptation. We average the score of each character over the three adaptations to get a mean score representing their overall importance, and use it to rank the characters. Table~\ref{tab:CharacterList} from the Supplementary Material shows the \texttt{top-20} frequent characters for \texttt{U2} and \texttt{U5}.

\section{Character Comparison}
\label{sec:CharComp}
In this section, we want to answer our first research question: \textit{Is it possible to match characters between adaptations, based on the character network structure?} (RQ1). For this purpose, we focus on period \texttt{U2}, i.e. the first two books, volumes and seasons, in order to cover comparable stories. First, we formulate the task as a \textit{Graph Matching} (GM) problem, and leverage GM methods to assess its feasibility (Section~\ref{sec:CharCompMatch}). In order to provide some insight to these results, we next conduct a centrality analysis aiming at studying the structural differences between the characters (Section~\ref{sec:CharCompCentr}). 

\subsection{Character Matching}
\label{sec:CharCompMatch}
Le us consider two graphs $G_1$ and $G_2$ that are similar, but not perfectly identical. They contain the same vertices, connected in roughly the same way. The GM problem consists in identifying a permutation of vertices allowing us to turn one graph into the other while minimising edge disagreement~\cite{Qiao2021}. Put differently, one wants to match each vertex of one graph to one vertex of the other graph, in such a way that an edge (or absence of edge) between two vertices of one graph matches an edge (or absence of edge) between their counterparts in the other graph. The problem can be generalised straightforwardly to weighted edges, and to graphs with different vertex sets. 

In order to solve this problem and answer RQ1, we first leverage existing GM methods (Section~\ref{sec:CharCompMatchSota}), before relaxing the problem in order to apply a simpler approach based on neighbourhood similarity (Section~\ref{sec:CharCompMatchSim}).

\subsubsection{Graph Matching Methods}
\label{sec:CharCompMatchSota}
We apply five state-of-the-art methods implemented in the \texttt{iGraphMatch} R library\footnote{\url{https://cran.r-project.org/package=iGraphMatch}}, covering three families of approaches. The principle underlying the first family is to relax the objective function, which can be done in different ways. Here, we select the methods based on \textit{convex}, \textit{indefinite}~\cite{Lyzinski2016}, and \textit{concave}~\cite{Zaslavskiy2009} relaxations. The \textit{Percolation} method~\cite{Kazemi2015} belongs to the second family, which gathers iterative methods. Those start with a few seeds (vertices matched \textit{a priori}) and propagate to the rest of the graph, adding the new best match at each iteration. Finally, \textit{Umeyama}'s algorithm~\cite{Umeyama1988} belongs to the third family, which relies on the spectral properties of the adjacency matrices.

Two preprocessing steps may affect the resolution of the GM problem. First, it is possible that the compared graphs do not contain exactly the same vertices. It is then necessary to perform an operation called \textit{padding}~\cite{Qiao2021}, which consists of adding to each graph, as \textit{isolates}, the vertices that are only present in the other graph. This makes the GM problem harder, as any isolate of one graph could be incorrectly matched to any isolate of the other graph. 
The second preprocessing step is \textit{centring}~\cite{Sussman2019}, which consists in transforming the adjacency matrices in order to give more importance to mismatches concerning the absence (vs. presence) of edges. In certain cases, this is known to improve the results~\cite{Qiao2021}.

We apply the five methods to all pairs of networks, with and without centring, and considering three versions of increasingly limited sets of characters, as defined in Section~\ref{sec:DataChar}: 1) all named characters present in at least one of the two compared graphs (noted \texttt{named}); 2) only the characters present in both compared networks (\texttt{common}); and 3) only the 20 most important characters present in both networks (\texttt{top-20}). The \texttt{named} character set requires padding, whereas \texttt{common} and \texttt{top-20} do not. Table~\ref{tab:MatchingResults} shows the best results obtained over all considered methods and parameters, and expressed as proportions of correctly matched characters. For the sake of completeness, a comprehensive presentation of the results is provided in the Supplementary Material (Section~\ref{sec:ApdxCharCompMatch}).

The top part of Table~\ref{tab:MatchingResults} (row 1--2) focuses on the matching performed with and without matrix centring. There is no clear improvement, and some additional experiments even show a decrease in performance when centring our networks (cf. Section~\ref{sec:ApdxCharCompMatch}, Supplementary Material), so in the rest of this section we do not use centring anymore. Another interesting result is the performance increase when the considered character set gets narrower: the scores are always very low for all named characters, gets better when focusing on characters common to both considered networks, and even better when using only the 20 most important characters. In addition, the performance scores are similar when comparing the \textit{Novels} vs. \textit{Comics} or \textit{TV Show} networks, but get better when comparing \textit{Comics} vs. \textit{TV Show}, which suggest the latter are more similar. Overall, the scores are very low: in the best case, we only match half of the top 20 characters correctly.  

\begin{table}[htb!]
    \caption{Best results obtained for the graph matching problem. The left columns indicate the method parameters: considered vertex attributes (\textit{Attr.}), centring of the adjacency matrices (\textit{Centr}), nature or number of the seeds (\textit{Seeds}). The other columns show the proportions of characters correctly matched for each pair of networks, considering three character sets (cf. main text)}
    \label{tab:MatchingResults}%
    \begin{tabular}{l@{~}l@{~}l r@{~}r@{~}r r@{~}r@{~}r r@{~}r@{~}r}
        \toprule
        \multicolumn{3}{c}{Method Parameters} & \multicolumn{3}{c}{\textit{Novels} vs. \textit{Comics}} & \multicolumn{3}{c}{\textit{Novels} vs. \textit{TV Show}} & \multicolumn{3}{c}{\textit{Comics} vs. \textit{TV Show}} \\
        Attr. & Centr. & Seeds & \texttt{named} & \texttt{common} & \texttt{top-20} & \texttt{named} & \texttt{common} & \texttt{top-20} & \texttt{named} & \texttt{common} & \texttt{top-20} \\
        \midrule
        None & Yes &  0 &  0.13\% &  3.42\% & 20.00\% & 0.13\% & 10.07\% & 20.00\% &  0.57\% & 21.97\% &  50.00\% \\
        --   & No  &  0 &  0.53\% &  3.77\% & 20.00\% & 0.26\% &  3.60\% & 25.00\% &  5.10\% & 21.21\% &  35.00\% \\
        \midrule
        None & No  & AH &  0.00\% & 25.34\% & 15.00\% & 0.00\% & 11.51\% &  5.00\% &  0.00\% & 34.85\% &  50.00\% \\
        --   & --  & AS &  0.00\% &  4.79\% & 15.00\% & 0.00\% & 12.23\% & 25.00\% &  0.00\% & 30.30\% &  15.00\% \\
        \midrule
        None & No  &  5 &  4.26\% & 14.63\% & 60.00\% & 2.39\% & 20.90\% & 40.00\% & 12.36\% & 38.58\% &  86.67\% \\
        --   & --  & 10 &  6.43\% & 23.40\% & 80.00\% & 2.54\% & 27.91\% & 60.00\% & 11.66\% & 42.62\% & 100.00\% \\
        --   & --  & 15 &  7.01\% & 26.35\% & 100.00\% & 3.76\% & 33.87\% & 60.00\% & 10.95\% & 39.32\% & 100.00\% \\
        \midrule
        Sex  & No  &  0 &  0.40\% &  3.77\% & 15.00\% & 0.66\% &  7.91\% & 15.00\% &  0.85\% & 16.67\% &  80.00\% \\
        Aff. & --  &  0 & 14.93\% & 65.75\% & 60.00\% & 6.32\% & 64.75\% & 40.00\% & 15.58\% & 81.06\% &  80.00\% \\
        Both & --  &  0 & 16.12\% & 73.97\% & 65.00\% & 6.46\% & 64.75\% & 55.00\% & 15.86\% & 83.33\% & 100.00\% \\
        \botrule
    \end{tabular}
\end{table}

One way to improve the matching performance is to use the so-called \textit{adaptive} seeds method~\cite{Qiao2021}. A \textit{seed} is a match assumed to be correct, which can be used as an input of a graph matching method, making its work easier. In the general case, a seed is a trusted match because it comes from some ground truth. The adaptive seeds method proceeds differently, though. It consists in applying a matching algorithm to get a first tentative vertex map, as well as a score that estimates the reliability of the corresponding matches. The most reliable matches are then used as seeds when applying the same method again. It is possible to consider the seed as \textit{hard}, i.e. it is assumed to be a perfectly exact match, which should be absolutely respected, or as \textit{soft}, i.e. the match is just viewed as very probable, and does not have to be respected. The second part of Table~\ref{tab:MatchingResults} (rows 3--4) shows the results obtained using adaptive hard (noted \textit{AH}) and soft (\textit{SH}) seeds. The performance of both approaches is poor on \texttt{named} characters; close to zero matches. On the contrary, adaptive seeds improve the results for all pairs of networks when considering \texttt{common} characters. There is not much effect when considering \texttt{top-20} characters, probably because of the size of this vertex set. These results confirm that the task of matching characters based only on the graph structure is a difficult, as the best case performance stays at 50\%.

It is possible that incorporating extra information in addition to the graph structure could help to improve results. To test this assumption, we first leverage \textit{ground truth} hard seeds, i.e. we provide the methods with a few exact matches as inputs. We experiment with 5, 10, and 15 seeds for \texttt{named} and \texttt{common} characters, which represent between 1\% and 11\% of the vertices, depending on the considered networks. The seeds are selected among the most important characters. In any case, the performance is assessed only on the remaining non-seed vertices. The third part of Table~\ref{tab:MatchingResults} (rows 5--7) shows the matching performance; it improves in all cases compared to our previous methods. It remains unsatisfying, though, with scores under 50\% for \texttt{named} and \texttt{common} characters. The values are much higher for \texttt{top-20} characters (up to 100\% correct matches), but the seeds proportionally represent a large part of the network.

Another way of leveraging some extra information to improve matching is to use vertex attributes. In our case, we know the sex and affiliation of each character (cf. Section~\ref{sec:ApdxDescrAnal} in the Supplementary Material for some examples). We use them to compute similarity scores between the characters, and fetch them as soft inputs to the matching methods. The last part of Table~\ref{tab:MatchingResults} (rows 8--10) shows the results obtained when using sex and affiliation (noted \textit{Aff.}) separately, and together (noted \textit{Both}). Sex alone does not bring any noticeable improvement compared to no attribute at all, while the affiliation strongly helps to match the characters in all cases, especially when comparing the \textit{Comics} and \textit{TV Show} networks. Interestingly, combining both attributes leads to even better results, on par or better than the scores obtained using ground truth seeds.

\subsubsection{Similarity-Based Matching}
\label{sec:CharCompMatchSim}
Our results from the previous section reveal that it is difficult to match characters based on graph structure alone. To explore this question further, we consider a relaxed version of this problem: we assume that, when comparing two characters, their neighbours are known and can be used to perform this comparison. 

Our proposed method directly leverages this additional information. For each pair of characters $(v_1,v_2)$ belonging to the two compared networks $G_1$ and $G_2$, we assess the inter-character similarity by computing the weighted version of Jaccard's index, also called Ru\v{z}i\v{c}ka's similarity~\cite{Ruzicka1958, DeCceres2013}, of their respective neighbourhoods. Let us note $\mathbf{x}$ and $\mathbf{y}$ the rows (or columns) representing $v_1$ and $v_2$ in the adjacency matrices of their respective graphs. We assume that $G_1$ and $G_2$ contain the same characters, in the same order (which may require some padding and reordering). The similarity between $v_1$ and $v_2$ is defined as
\begin{equation}
    R(\mathbf{x},\mathbf{y}) = \sum_{i} \min(x_i, y_i) \Big/ \sum_{i} \max(x_i, y_i).
\end{equation}
Like the original Jaccard index, this measure ranges from 0 (completely different vectors) to 1 (identical). The most likely match of a character $v_1$ is the most similar character in $G_2$. Of course, the opposite might not be true: there may be another character $v'_1 \neq v_1$ in $G_1$ that is more similar to $v_2$ than $v_1$. For this reason, we consider that we have a correct match only if it works in both directions.

\begin{table}[htb!]
    \caption{Results obtained when matching vertices using their neighborhood. The left column indicates how time is represented in the considered graphs: static, or dynamic (cumulative vs. instant networks). As in Table~\ref{tab:MatchingResults}, the other columns show the proportions of characters correctly matched for each pair of networks, considering three character sets}
    \label{tab:SimResults}%
    \begin{tabular}{l r@{~}r@{~}r r@{~}r@{~}r r@{~}r@{~}r}
        \toprule
        Time & \multicolumn{3}{c}{\textit{Novels} vs. \textit{Comics}} & \multicolumn{3}{c}{\textit{Novels} vs. \textit{TV Show}} & \multicolumn{3}{c}{\textit{Comics} vs. \textit{TV Show}} \\
         & \texttt{named} & \texttt{common} & \texttt{top-20} & \texttt{named} & \texttt{common} & \texttt{top-20} & \texttt{named} & \texttt{common} & \texttt{top-20} \\
        \midrule
        Static     & 10.57\% & 42.81\% &  70.00\% & 4.87\% & 43.88\% & 70.00\% & 16.71\% & 54.55\% & 95.00\% \\
        \midrule
        Cumulative & 13.64\% & 47.95\% &  70.00\% &     -- &      -- &      -- &      -- &      -- &      -- \\
        Instant    & 10.28\% & 64.56\% & 100.00\% &     -- &      -- &      -- &      -- &      -- &      -- \\
        \botrule
    \end{tabular}
\end{table}

\begin{figure}[htb!]
    \centering
    \includegraphics[width=0.32\textwidth]{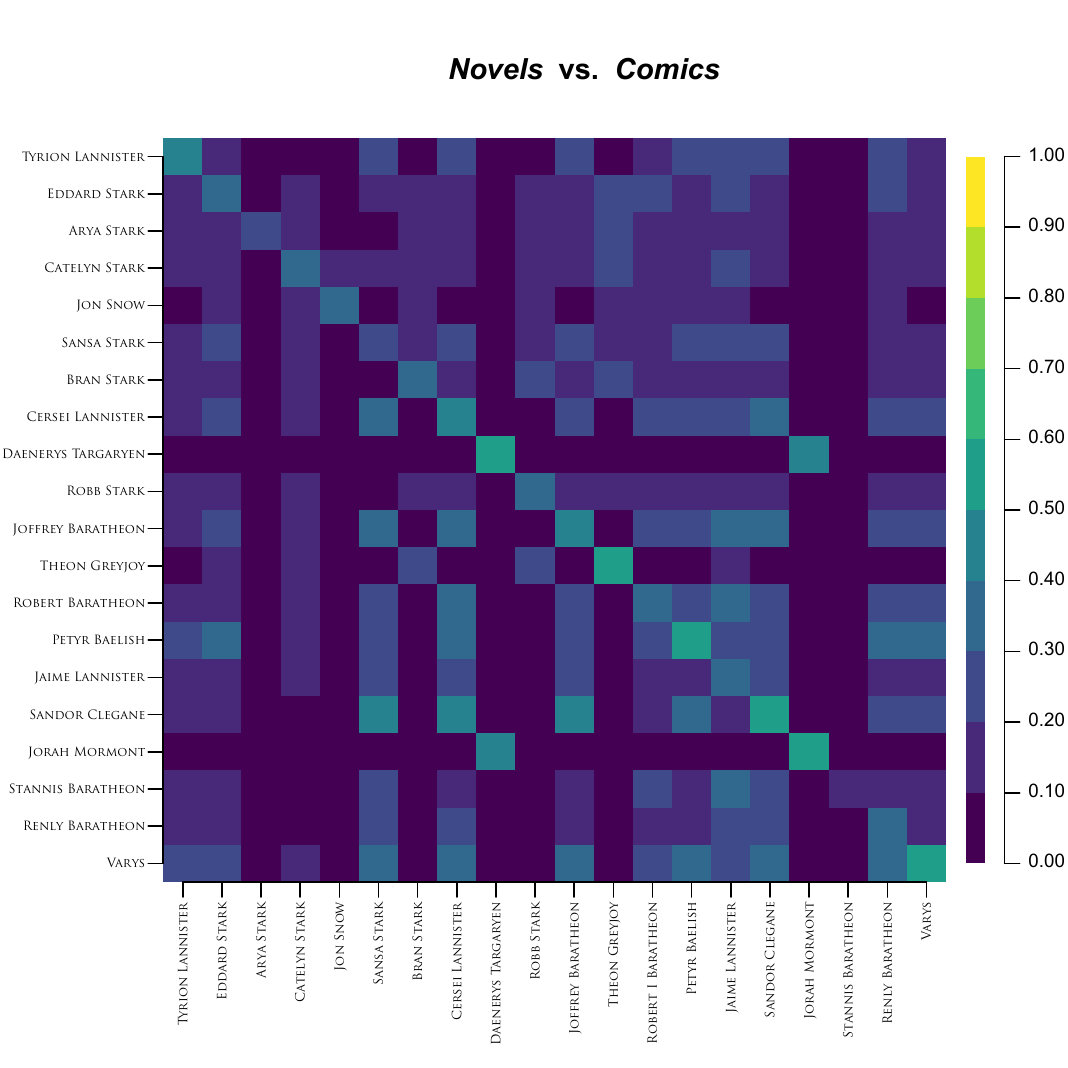}
    \hfill
    \includegraphics[width=0.32\textwidth]{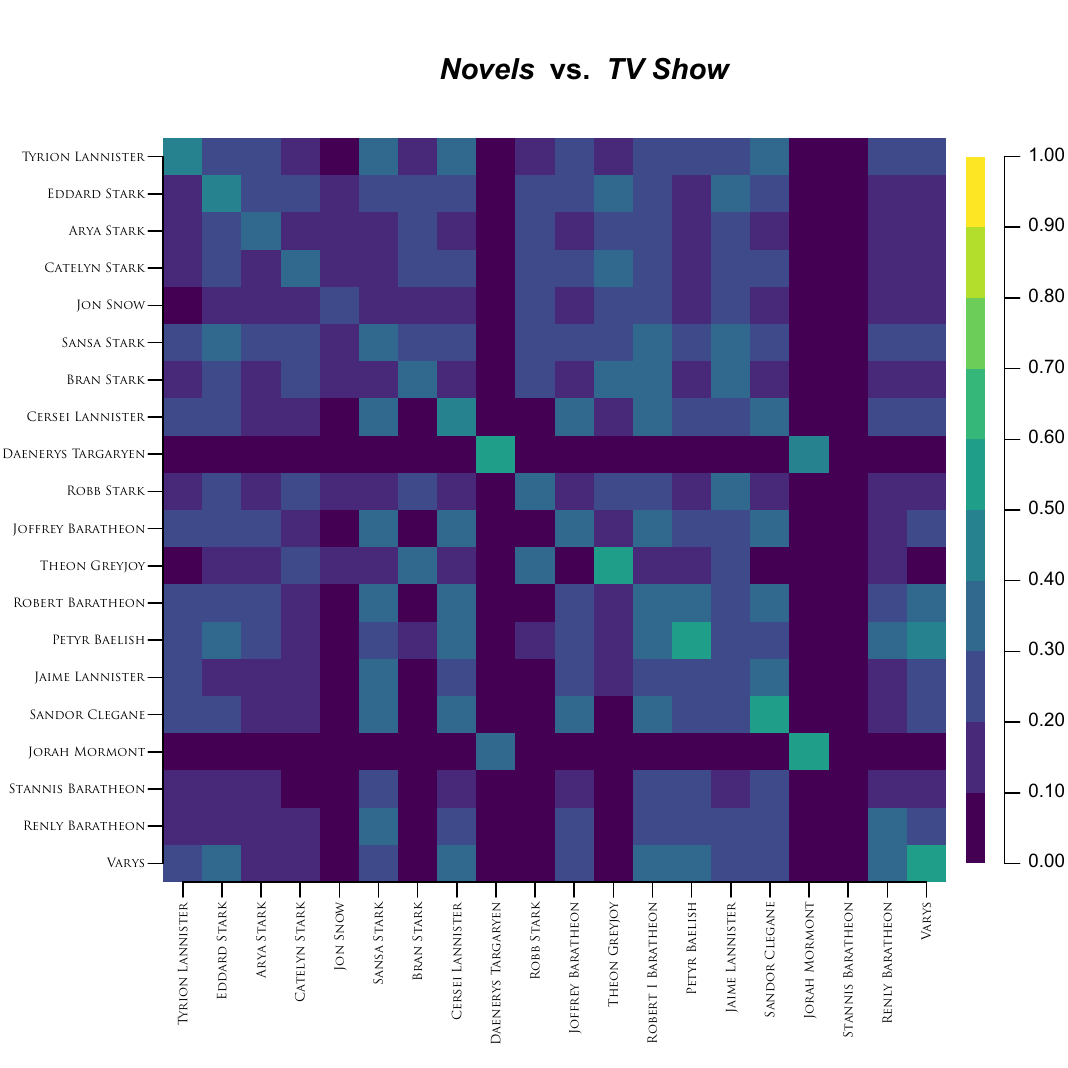}
    \hfill
    \includegraphics[width=0.32\textwidth]{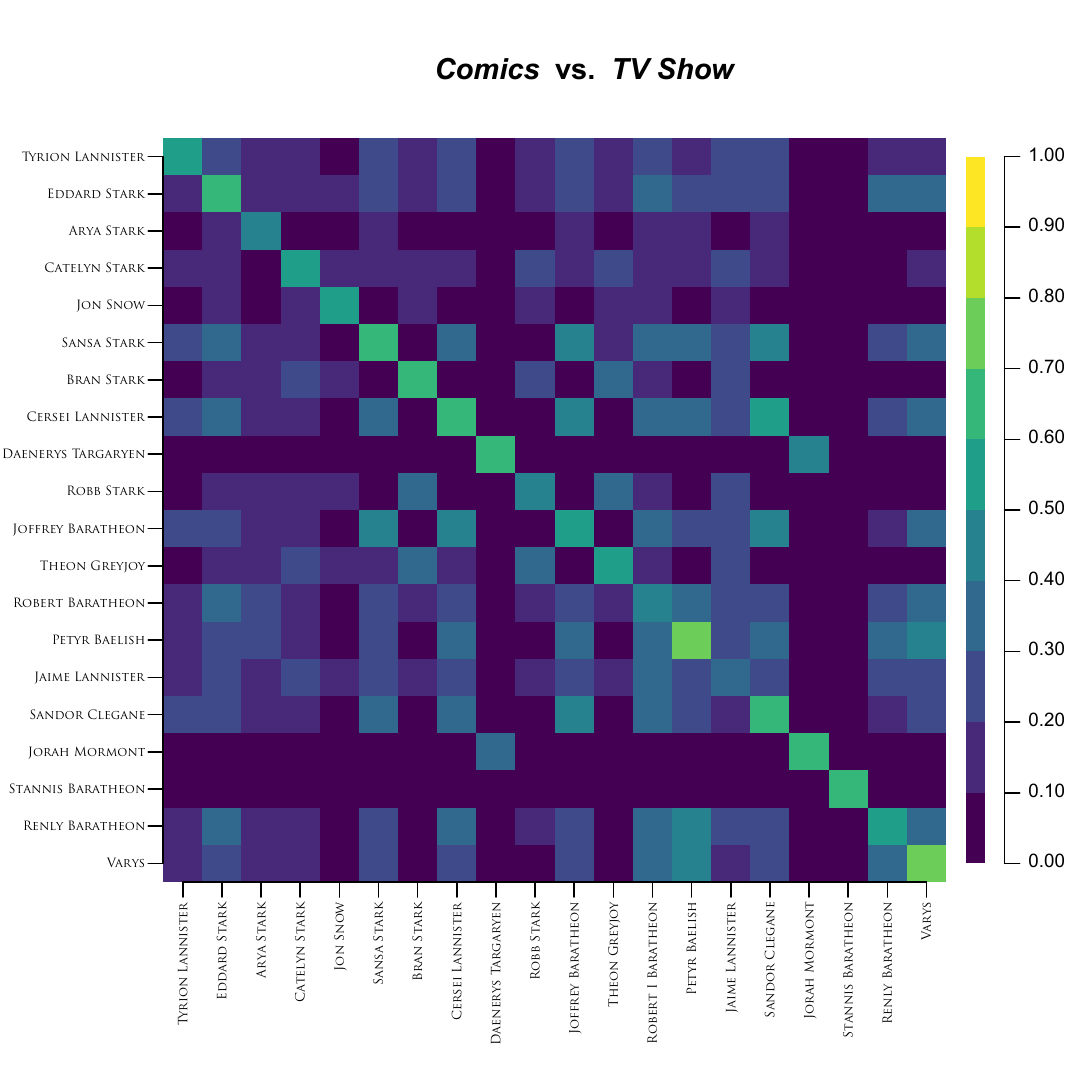}
    \caption{Similarity matrices obtained with Ru\v{z}i\v{c}ka's similarity, for the 20 most important characters of each pair of adaptations}
    \label{fig:MatchingResultsSimTop}
\end{figure}

The first row of Table~\ref{tab:SimResults} shows the matching results obtained with this method. We observe the same differences as before: 1) the narrower the character set, the higher the performance; and 2) the best match seems to be obtained when comparing the \textit{Comics} and \textit{TV Show} networks. The former point is clear when considering the similarity matrices shown in Figure~\ref{fig:MatchingResultsSimTop}, as the highest values are generally located on their diagonals. By comparison, the full matrices exhibit many off-diagonal high similarity values (cf. Figure~\ref{fig:MatchingResultsSimAll} in the SM). Interestingly, even though the amount of information provided in addition to the structure could be considered as larger than when we leverage ground truth seeds or vertex attributes to perform graph matching in Section~\ref{sec:CharCompMatch}, the results are not clearly better here. Interestingly, the method works well to identify characters undergoing major transformations during the adaptation process, e.g. splitting or merging as discussed in Section~\ref{sec:DataAttr}: see the Supplementary Material for more detail (Section~\ref{sec:ApdxCharCompMatchChars}).

We propose a sequential extension of this similarity-based method, in order to take advantage of the dynamic networks. At each time step, we match the characters using Ru\v{z}i\v{c}ka's similarity, as before. For a given character, we thus obtain a series of (estimated) matches over the whole timeline. We select the statistical mode of these matches as the overall match, and randomly break possible ties. This approach requires two dynamic networks with the exact same numbers of time slices, so we can only apply it to the chapter-based \textit{Novels} and \textit{Comics} networks. The bottom part of Table~\ref{tab:SimResults} shows the obtained results. The performance obtained with the cumulative networks is similar to that of the static network, but the instant networks lead to much improvement, on par with what we got earlier when leveraging vertex attributes.

\subsubsection{Takeaways}
To summarise the main results obtained in this section, it appears that matching characters based only on the graph structure cannot be performed reliably with current state-of-the-art methods. The \textit{Novels} network, in particular, generally leads to fewer matches when compared to the other adaptations. On the one hand, this is a bit surprising, as the comics are a more faithful adaptation of the novels than the TV show. On the other hand, both the comics and TV show are visual media and therefore share a number of adaptation constraints. In any case, the difficulty to match the characters suggests that their positions and roles vary from one adaptation to the other.

However, the performance gets better when focusing on characters common to both compared adaptations, and even more so with only the most important characters. This could reveal that the differences between the networks mainly concern minor characters, but it could also be due to a size effect. Finally, it is worth stressing that leveraging additional information (seeds, vertex attributes, neighbourhood) in addition to structure greatly improves the matching performance.

\subsection{Centrality Analysis}
\label{sec:CharCompCentr}
In order to better understand the differences between the networks, we perform a descriptive analysis of the characters. We compute a selection of standard metrics to assess their centrality: degree, betweenness, closeness, and Eigenvector centrality. As our networks are weighted, we consider both unweighted and weighted variants for each metric\footnote{Note that the weighted version of the degree is called the \textit{strength}.}. We use the normalised version of the weights, and standardise ($z$-score) the centrality scores, in order to get values that are comparable over all adaptations and metrics.

We first compare the behavior of the selected centrality metrics over the three adaptations (Section~\ref{sec:CharCompCentrCorr}), before identifying and discussing the characters' centrality profiles (Section~\ref{sec:CharCompCentrProf}).

\subsubsection{Centrality Correlation}
\label{sec:CharCompCentrCorr}
\begin{figure}[htb!]
    \centering
    \includegraphics[width=0.32\textwidth]{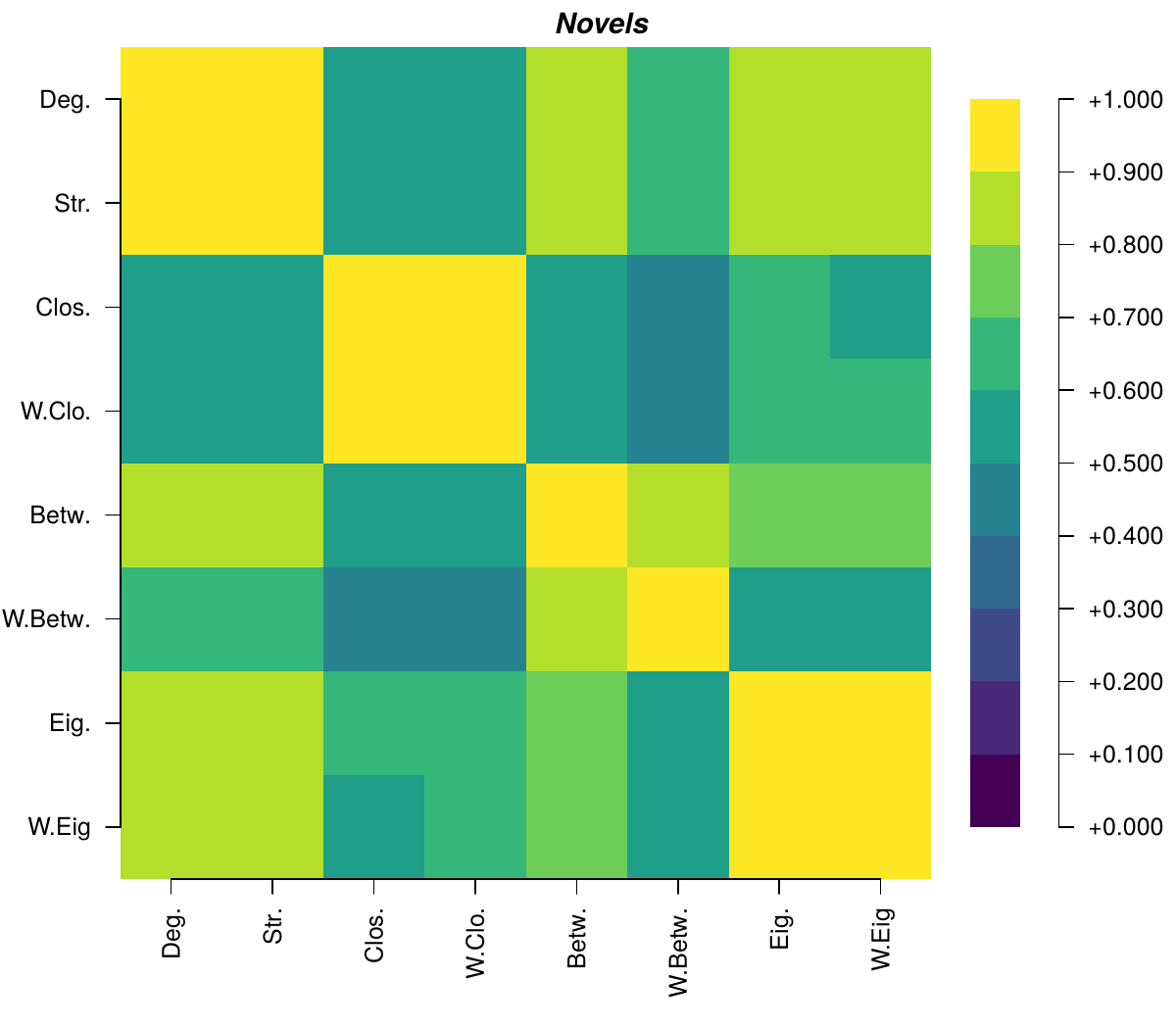}
    \hfill
    \includegraphics[width=0.32\textwidth]{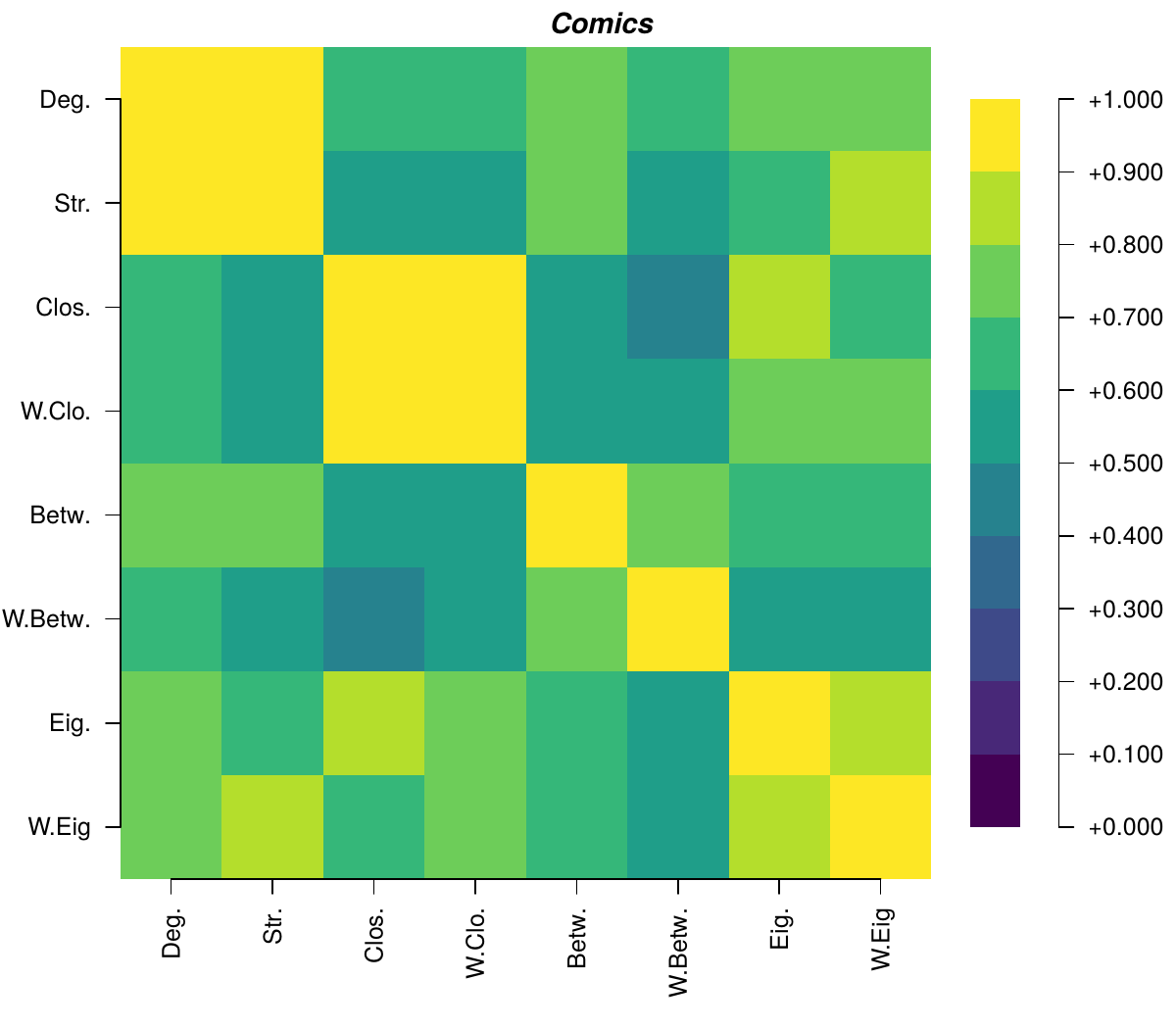}
    \hfill
    \includegraphics[width=0.32\textwidth]{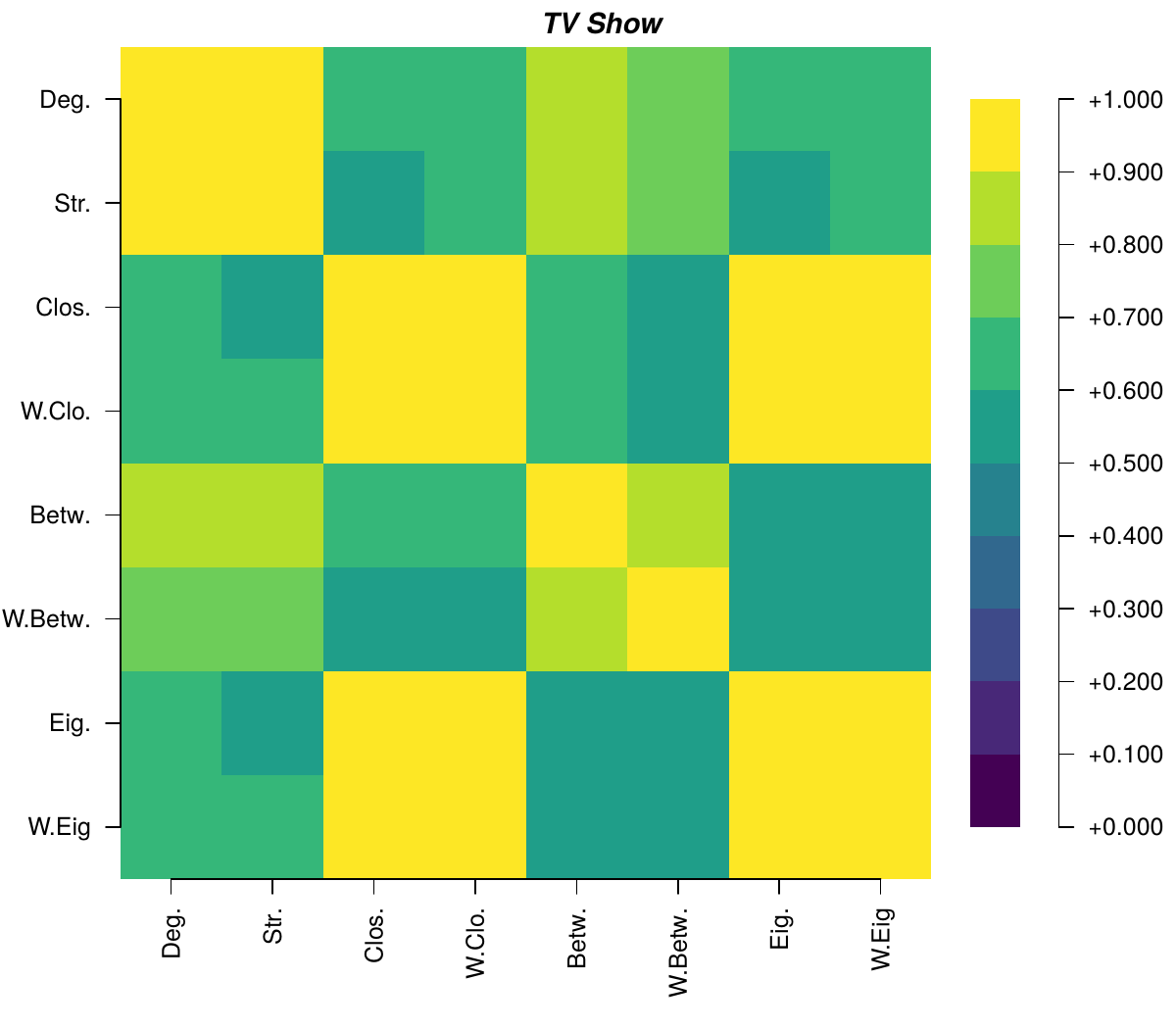}
    \includegraphics[width=0.32\textwidth]{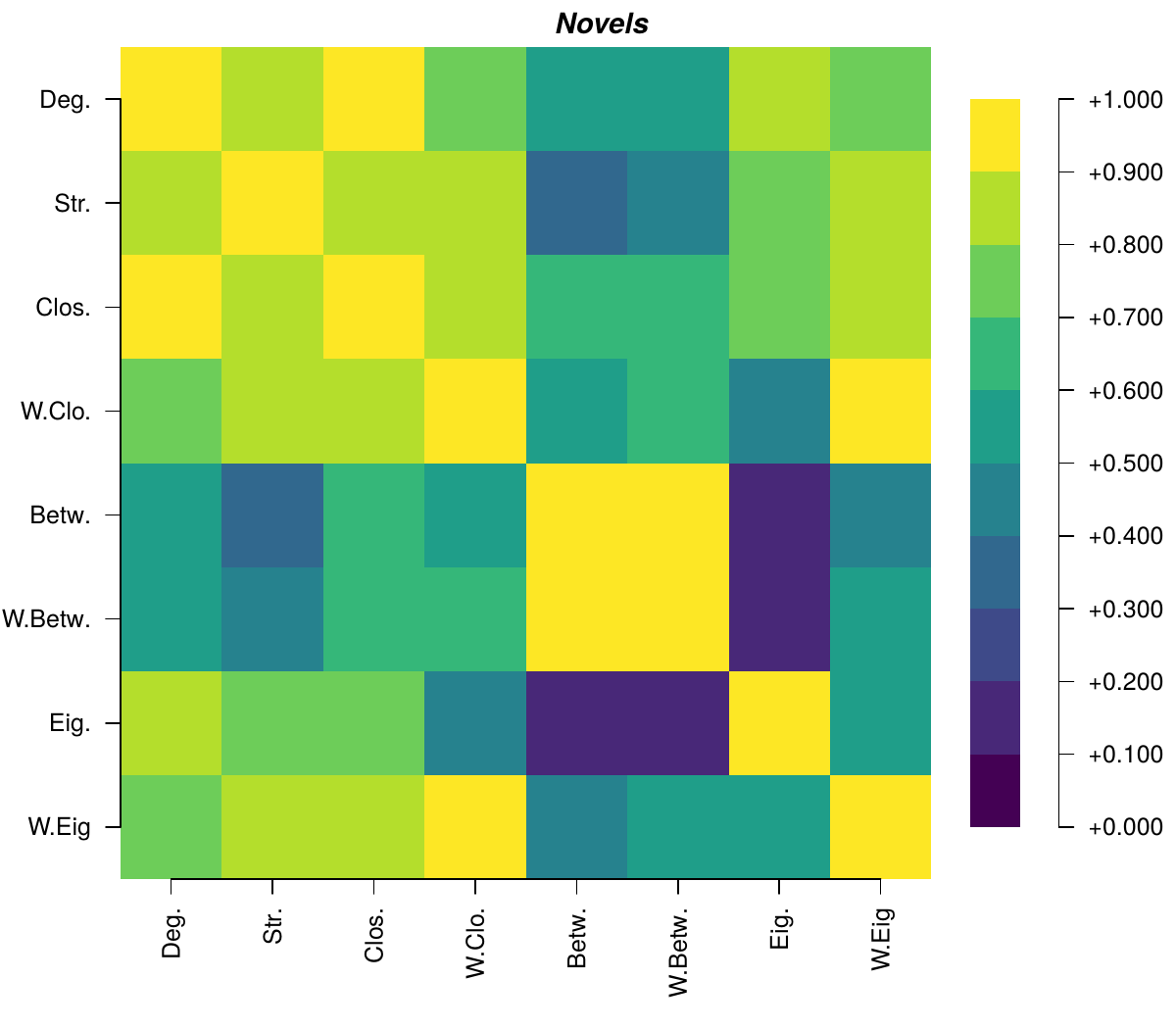}
    \hfill
    \includegraphics[width=0.32\textwidth]{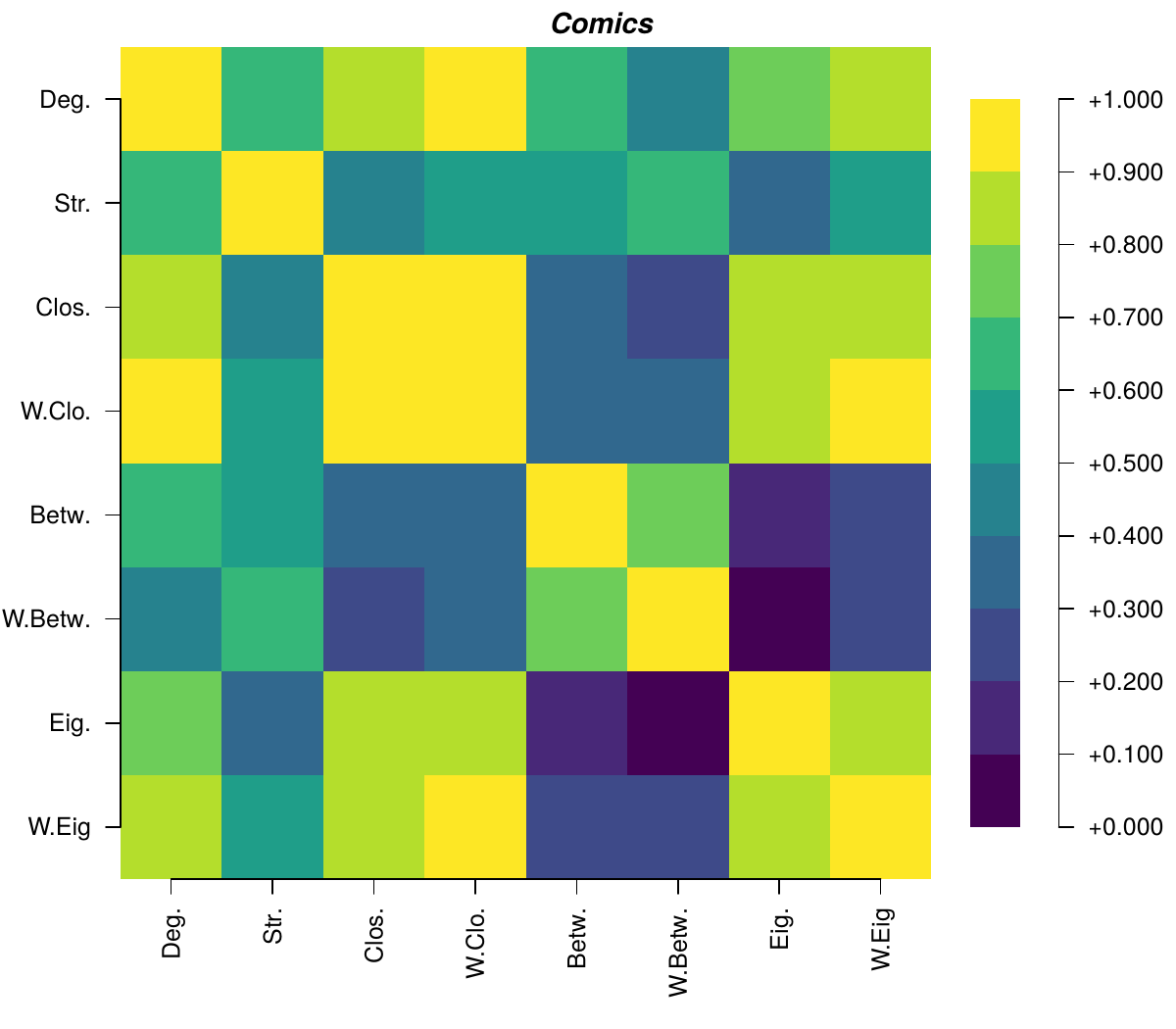}
    \hfill
    \includegraphics[width=0.32\textwidth]{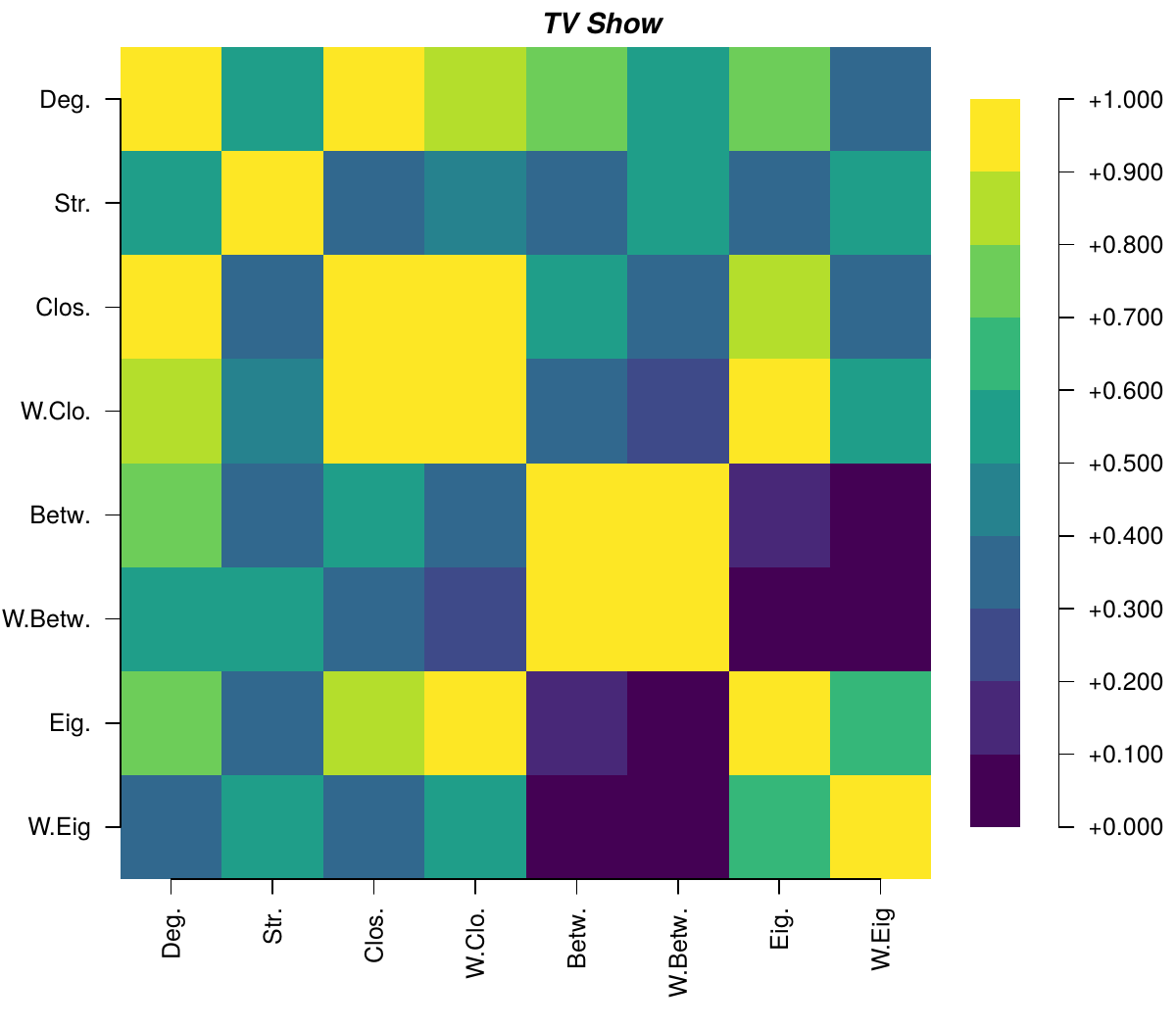}
    \caption{Spearman's correlation between the selected centrality metrics, for each of the three adaptations over period \texttt{U2}, considering the \texttt{named} (top row) and \texttt{top-20} (bottom row) character set}
    \label{fig:VtxMatch_CentrNamedCorr}
\end{figure}

Figure~\ref{fig:VtxMatch_CentrNamedCorr} shows Spearman's correlation between the selected centrality metrics. As in Section~\ref{sec:CharCompMatch}, we focus on period \texttt{U2}, i.e. the first two books, volumes and seasons. Moreover, the figure focuses on \texttt{named} and \texttt{top-20} character sets. However, results concerning the \texttt{common} characters, as well as period \texttt{U5} (the first five books and seasons), are presented in the Supplementary Material (Section~\ref{sec:ApdxCharCompCentrCorr}).

When considering \texttt{named} characters (top row of the figure), all three adaptations exhibit similar matrices, with diagonal $2 \times 2$ blocks. These show that the unweighted and weighted versions of the same metric are very correlated (to a lesser extent, in the case of the betweenness), and therefore possibly redundant. In addition, in the case of the \textit{TV Show}, the closeness and Eigenvector centrality metrics are also highly correlated (off-diagonal block).

Focusing only on the \texttt{top-20} characters (bottom row of the figure) provides a different picture. First, each \texttt{top-20} matrix differs from its \texttt{named} counterpart; the correlation between the unweighted and weighted versions of the same metric is a bit weaker, but more metrics are correlated (e.g. degree and closeness). Most characters are very minor, and are consequently attached to the rest of the network through low weight edges. Consequently, for these characters, the weighted and unweighted versions of the same metric yield very similar scores. This can explain the higher correlation they exhibit when considering \texttt{named} characters. Important characters tend to be considered central according to several metrics at once, which could explain the higher correlation between distinct metrics when focusing only on \texttt{top-20} characters. Our second observation is that matrices from the bottom row exhibit more variability than those of the top row. This hints at differences in the way important characters are interconnected within the three adaptations. As already observed in Section~\ref{sec:CharCompMatch}, the comics and TV show exhibit a higher similarity, compared to the novels.

\subsubsection{Centrality Profiles}
\label{sec:CharCompCentrProf}
\begin{figure}[htb!]
    \centering
    \includegraphics[width=0.7\textwidth]{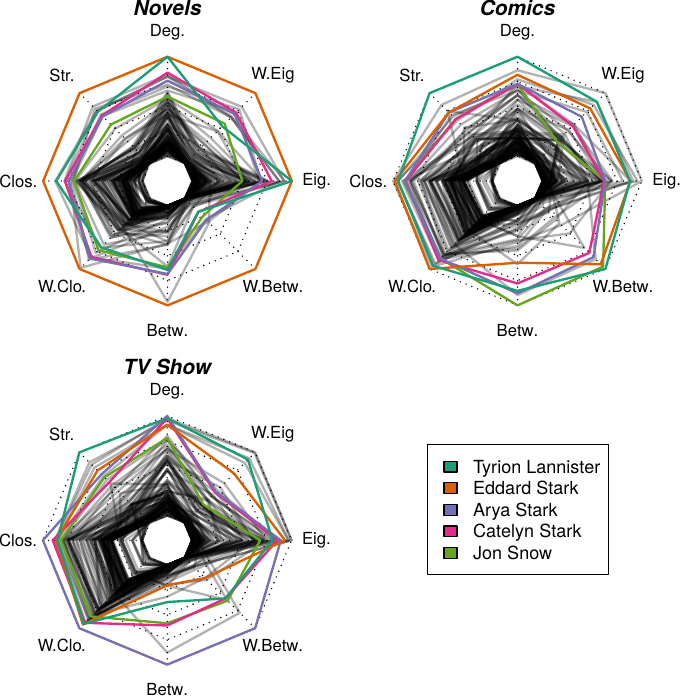}
    \caption{Centrality profiles of the \texttt{common} character set, over period \texttt{U2}, for all three adaptations. The five most important characters are represented in color}
    \label{fig:VtxMatch_CentrCommRadar}
\end{figure}

For each character, we now leverage the selected centrality metrics to constitute a so-called \textit{centrality profile}, which corresponds to the vector of its centrality scores. Put differently, each character is represented as a point in an 8D centrality space. Figure~\ref{fig:VtxMatch_CentrCommRadar} shows these centrality profiles as radar plots, for the \texttt{common} character set, considering period \texttt{U2}. The five most important characters are represented in color, whereas the others are in gray. 

Let us compare the metrics first: it appears that the degree, Eigenvector centrality and Closeness exhibit similar behaviours, with centrality scores that cover the whole range, including low, intermediary and high values. By comparison, only a very few characters have a high betweenness, and it is very low for the others. Regarding the adaptations, one can observe that the separation between the most central characters and the others is much clearer in the novels, and that the comics tend to exhibit higher betweenness scores. In all three adaptations, the most important characters are also the most central. However, they exhibit very different centrality profiles. For instance, in the novels, \textit{Eddard Stark} is the most central character according to all metrics, whereas he is just a central character among others in the comics and TV show. 

\begin{figure}[htb!]
    \centering
    \includegraphics[width=1\textwidth]{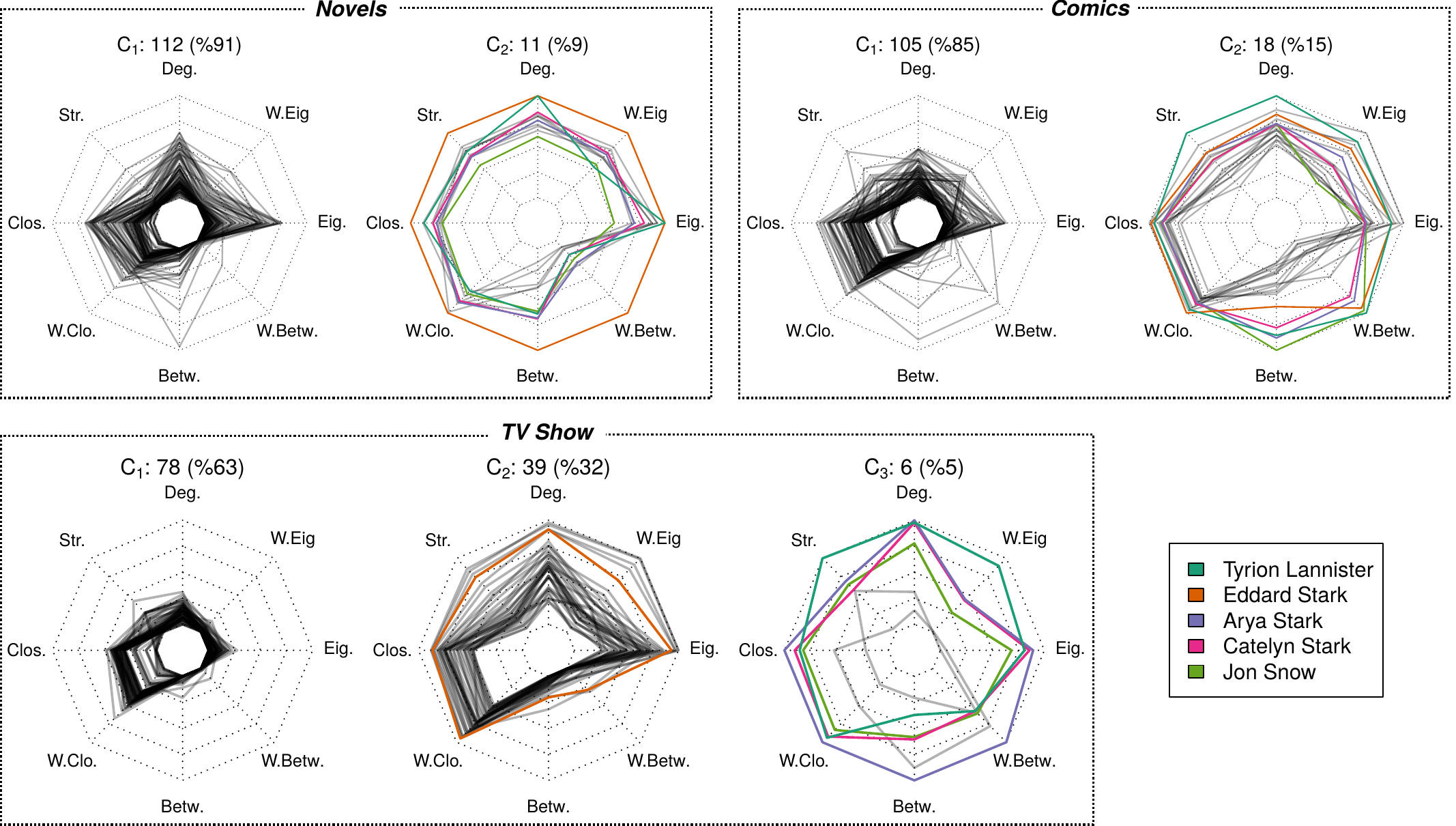}
    \caption{Classes of \texttt{common} characters, detected based on their centrality profiles, over period \texttt{U2}, for all three adaptations. The five most important characters are represented in color}
    \label{fig:VtxMatch_CentrCommClust}
\end{figure}

Using a standard hierarchical method, we perform a cluster analysis in the centrality space. We select the most appropriate cuts in the resulting dendrograms based on the Silhouette measure~\cite{Rousseeuw1987}. This allows us to identify classes of characters holding similar positions in their network. Instead of directly comparing the networks as we do with matching methods in Section~\ref{sec:CharCompMatch}, here we want to do so through these similarity classes. Figure~\ref{fig:VtxMatch_CentrCommClust} shows the clusters obtained for the first two seasons and books, as radar plots. Additional results are provided in the Supplementary Material (Section~\ref{sec:ApdxCharCompCentrProf}). For novels and comics, we observe a separation between minor (clusters $C_1$) and major ($C_2$) characters. In the case of the TV show, major characters are split depending on whether they possess a low ($C_2$) or a high ($C_3$) betweenness. This difference allows us to distinguish between characters that have a local importance, like \textit{Eddard Stark}, from those that connect independent storylines, such as \textit{Arya Stark}. 

The Adjusted Rand Index (ARI)~\cite{Hubert1985} is a standard metric when it comes to comparing partitions. It reaches 1 for a perfect match, and 0 for orthogonal partitions. We use it to compare the adaptations as described by their clusters. Note that this comparison does not rely on some typical centrality scores of the clusters, but rather on which characters they contain. It turns out the pairs of adaptations exhibiting the most similar clusters are, by decreasing similarity: \textit{Novels} vs. \textit{Comics} (0.68), \textit{Comics} vs. \textit{TV Show} (0.31), and \textit{Novels} vs. \textit{TV Show} (0.21). This fits the intuition one could get by visually inspecting Figure~\ref{fig:VtxMatch_CentrCommClust}.

\subsubsection{Takeaways}
In summary, this section confirms some of the observations made in Section~\ref{sec:CharCompMatch} based on character matching. First, the three adaptations are only partially similar in terms of their character network structure. Second, they appear to be more similar when focusing on \texttt{common} and \texttt{top-20} characters. This suggests that the many minor characters are handled differently in both adaptations (comics and TV show) of the novels, but that they tend to respect more the original structure when it comes to the main characters. 

However, a closer look at the characters through their individual centrality profiles reveals that, if the main characters are indeed clearly distinguishable from the rest of the cast in all three adaptations, they also hold different positions depending on the adaptation. This could reflect the fact that the adaptation process does not necessarily put the same emphasis on the same characters as in the original material. Moreover, when looking at the characters in terms of centrality classes, it appears that the novels and comics form the most similar pair of adaptations, contrarily to what was observed in Section~\ref{sec:CharCompMatch} for graph matching.

\section{Narrative Alignment \& Media Divergence}
\label{sec:Align}
As explained in Section~\ref{sec:DataRel}, the comics are a direct adaptation of the first two novels and closely follow them. In contrast to this, the TV show is a looser adaptation, which increasingly diverges from the novels, especially after season five. To determine how much the adaptation of a medium diverges from its original source, one can study whether each narrative unit from the original medium (here, the novel chapters) appears in the target medium (e.g. comics issues and TV episodes), and if the chronology of these units is preserved. In this section, our objective is to tackle this task, which we call \textit{Narrative Alignment}. This will allow us to answer our second research question: \textit{Can we use character interactions to align the plots of adaptations from different media?} (RQ2).

We first provide the general framework for our problem and approach in Section~\ref{sec:AlignAuto}, before proposing and testing two narrative matching methods: one based on text (Section~\ref{sec:AlignText}), and the other on network structure (Section~\ref{sec:AlignStruct}). Finally, we combine both approaches in Section~\ref{sec:AlignCombo}, in order to assess their complementarity. Section~\ref{sec:AlignConclu} summarises our findings.

\subsection{General Framework}
\label{sec:AlignAuto}
We first formulate the narrative alignment problem and the metrics and ground truth that we use for performance assessment (Section~\ref{sec:AlignAutoProb}). We then describe the general approach that we adopt to tackle this problem (Section~\ref{sec:AlignAutoMeth}), on which we elaborate later to build three matching methods.

\subsubsection{Problem Formulation}
\label{sec:AlignAutoProb}
The narrative alignment problem is defined as follows: given two media, let $\mathbf{a} = (a_1, ..., a_n)$ and $\mathbf{b} = (b_1, ..., b_m)$ be the vectors of their constitutive narrative units, with respective lengths $n$ and $m$. For example, $\mathbf{a}$ could represent episodes from the TV show while $\mathbf{b}$ could represent chapters from the novels. We must find a matrix $\mathbf{M} \in \{0, 1\}^{n \times m}$ where $M_{ij} = 1$ if narrative unit $a_i$ corresponds to narrative unit $b_i$, and $M_{ij} = 0$ otherwise. Note that $\mathbf{M}$ describes a many-to-many relationship: a narrative unit from $\mathbf{a}$ may correspond to several narrative units from $\mathbf{b}$, and vice versa. As an example, an episode from the TV show usually adapts several chapters from the novels, while parts of the same chapter may appear in distinct episodes. It is worth stressing that many-to-many matching is much harder than one-to-one matching, as the relation arity is unknown (i.e. one does not know how many narrative units are involved on either sides). 

In order to evaluate our narrative matching methods, we gather gold alignments for all the pairs of media. After manual verification, we adapt the \textit{Novels vs. TV Show} alignment from a fan annotation\footnote{\url{https://joeltronics.github.io/got-book-show/bookshow.html}} who matched chapters and episodes. We create the \textit{Novels vs. Comics} ourselves, at the levels of chapters. We use both of these to automatically produce the \textit{Comics vs. TV Show} alignment. It is worth stressing that manually extracting such alignments necessitates significant effort, so proposing a method to automate this task with satisfying performance would be very useful.

When assessing a matching $\mathbf{M}$ given a gold standard matrix $\mathbf{M}^* \in \{0, 1\}^{n \times m}$, we have four possible outcomes for each pair of narrative units:
\begin{itemize}
    \item A \textit{True Positive} (TP) occurs when $M_{ij} = 1$ and $M^*_{ij} = 1$.
    \item A \textit{False Positive} (FP) occurs when $M_{ij} = 1$ and $M^*_{ij} = 0$.
    \item A \textit{True Negative} (TN) occurs when $M_{ij} = 0$ and $M^*_{ij} = 0$.
    \item A \textit{False Negative} (FN) occurs when $M_{ij} = 0$ and $M^*_{ij} = 1$.
\end{itemize}
These matrices are sparse, and the true negatives are consequently the most frequent outcome, by far. For this reason, we use the \textit{F1-score}, a standard measure from the field of Information Retrieval (IR)~\cite{Manning2008}, to compute our matching performance. It is the harmonic mean of two other well-known IR measures, \textit{Precision} and \textit{Recall}:
\begin{align}
    Pre &= TP/(TP+FP) \\
    Rec &= TP/(TP+FN) \\
    F_1 &= 2 \cdot Pre \cdot Rec / (Pre + Rec).
\end{align}
A higher F1-score indicates better performance.

\subsubsection{Proposed Methods}
\label{sec:AlignAutoMeth}
Our general approach to perform automatic narrative alignment requires first to compute a similarity matrix $\mathbf{S} \in \mathbb{R}^{n \times m}$ between the two considered media. We elaborate further on this point in the next subsections, but in summary, we experiment with two different types of information to obtain this matrix. First, we leverage textual representations of the three adaptations to derive a \textit{textual similarity} matrix. Second, we take advantage of character interactions to compute a \textit{structural similarity} matrix. Finally, we also try to combine textual and structural similarity to compute a \textit{hybrid similarity}, in order to understand if they are complementary.

After computing the similarity matrix $\mathbf{S}$ between two media, we use an alignment algorithm to derive a matching from it. We experiment with two different approaches: basic thresholding vs. the Smith--Waterman algorithm~\citep{Smith1981}. The former is straightforward: to produce a matching $\mathbf{M}$, we set a threshold $t$ and we compute $M_{ij}$ as follows:
\begin{equation}
    M_{ij} =
    \begin{cases}
        1,& \text{if } S_{ij} > t\\
        0,& \text{otherwise.}
    \end{cases}
\end{equation}
For each pair of media, we estimate the optimal threshold $t$ by tuning it on the two other pairs. We consider thresholding as a simple alignment baseline.

The Smith--Waterman alignment algorithm~\citep{Smith1981} is a dynamic programming algorithm that was originally proposed to perform molecular sequence alignment. It identifies the best local alignment between two sequences by first scoring matches and gaps, and then backtracking through a dynamic programming matrix. Recently, \citet{Pial2023b} propose GNAT, a tool that adapts the Smith--Waterman algorithm by enabling many-to-many matching to perform narrative alignment. \citet{Pial2023a} apply this tool to align plots between novels and the scripts of their film adaptations. However, in both articles, the authors limit themselves to textual similarity when extracting their alignments. In order to also experiment with structural similarity, we re-implement their many-to-many adaptation of the Smith--Waterman algorithm. Similarly to the thresholding alignment method, we tune the parameters of the algorithm for each pair of media on the two other media pairs. 

In summary, our experimental setup involves three different ways of computing the similarity matrix between adaptations, and two algorithms to leverage this matrix and estimate the match, which makes a total of 6 methods. 
We apply them to the three possible pairs of media: \textit{Novels vs Comics}, \textit{Novels vs. TV Show} and \textit{Comics vs. TV Show}. As stated in Section~\ref{sec:DataTime}, each medium covers a different time period, so we work on the longest common period for each pair: \texttt{U2} for the \textit{Novels vs. Comics} and \textit{Comics vs. TV Show} pairs, and \texttt{U5} for the \textit{Novels vs. TV Show} pair. Since the TV show diverges more and more from the novels as time progresses, we additionally study the alignment of the \textit{Novels vs TV Show} pair over the \texttt{U2} period in the Supplementary Material (see Section~\ref{sec:ApdxNarrMatchNovelsTVShowU2}).

\subsection{Text-Based Method}
\label{sec:AlignText}
Our first method leverages textual representations of the adaptations to compute the similarity between them. We originally experimented with long textual representations: entire chapters, dialogues from TV Show episodes, and texts extracted using OCR for the comics. However, we found a much better matching performance when using the following short texts instead.
\begin{itemize}
    \item \textbf{Novels Chapters:} We scrape the existing chapter summaries from the previously mentioned fan Wiki~\cite{AWoIaF2023}. With a mean length of $4.56$ sentences, these summaries are usually much shorter than episode summaries, and longer than comics summaries.
    \item \textbf{Comics Issues:} Most (41/56) issues of the comics contain a summary of the next issue that acts as a teaser. We use OCR to extract these texts, and correct them manually. For the 15 remaining issues, we manually write summaries of approximately the same length. The mean length of these summaries is $2.86$ sentences, the shortest amongst all media.
    \item \textbf{TV Show Episodes:} We scrape the episode summaries of the first five seasons available from Wikipedia~\cite{Wikipedia2023, Wikipedia2023a, Wikipedia2023b, Wikipedia2023c, Wikipedia2024}. The mean number of sentences in these summaries is $11$: since episodes encompass different points of views and a number of subplots, each of these is represented by one or more sentences.
\end{itemize}
These textual sources constrain the narrative units used to represent the plots: the alignment is performed at the \textit{chapter} level for the novels, at the \textit{episode} level for the TV show, and at the \textit{issue} level for the comics.

To compute the similarity matrix $\mathbf{S}$ between the narrative units of the pairs of media, we try two different textual similarity functions:
\begin{itemize}
    \item \textit{tfidf}: compute the cosine similarity of the bag-of-words representation of two summaries weighted by standard TF-IDF~\cite{Manning2008}. This scheme weights each word according to its \textit{term frequency} (TF) and \textit{inverse document frequency} (IDF)~\citep{SparckJones1972}, the latter being an indication of how characteristic of a document a term is.
    \item \textit{sbert}: embed both summaries using SentenceBERT~\citep{Reimers2019}, and compute their cosine similarity. SentenceBERT is a variation of the language model BERT~\citep{Devlin2019} specifically fine-tuned to extract semantic sentence embeddings, so that two sentences that are semantically similar should be close in the embedding space.
\end{itemize}

The results of our textual alignment can be found in Table~\ref{tab:textual_matching}. Matching performance varies highly between media pairs, with the \textit{Novels vs. TV Show} pair being the hardest to match (best F1-score: $22.55$) and the \textit{Novels vs. Comics} being the easiest (best F1-score: $55.24$). The Smith--Waterman alignment algorithm largely outperforms the thresholding baseline in almost all configurations. The role of the similarity function is more puzzling: while \textit{tfidf} performs better when using the thresholding baseline, \textit{sbert} outperforms it when using the Smith--Waterman alignment algorithm. It is important to stress, however, that embedding sentences using SentenceBERT is computationally much more expensive than using TF-IDF.

\begin{table}[htb!]
    \caption{Performance obtained when using \textit{text}-based representations to tackle the narrative matching task, expressed in terms of F1-score. Values in bold indicate the best performance for a pair of media}
    \centering
    \input{tables/narr_match/perf_textual}
    \label{tab:textual_matching}
\end{table}

\subsection{Structure-Based Method}
\label{sec:AlignStruct}
Instead of using text as in the previous section, we now leverage character \textit{interactions} to compute a similarity matrix between the narrative units of a media pair. We use the dynamic networks extracted for each adaptation, as described in Section~\ref{sec:DataTime}. Such a network is constituted of a sequence of graphs, each one representing a narrative unit. As explained before, there are two types of dynamic networks: instant vs. cumulative. The best results, which we present here, are obtained with the former. We deal with the latter in the Supplementary Material (see Section~\ref{sec:ApdxNarrMatchCumNets}).

Given a character network per narrative unit for two media, we experiment with several variants of Jaccard's index to assess their similarity. First, we consider the standard index computed over the sets of vertices present in the compared networks, as well as their sets of edges. Second, we use the \textit{weighted} version of the index, mentioned in Section~\ref{sec:CharCompMatchSim}, a.k.a. Ru\v{z}i\v{c}ka's similarity~\cite{Ruzicka1958}. The most straightforward approach to weight each vertex/edge is to use the number of occurrences of the characters/interactions. However, we obtain much better results when weighting according to the \textit{inverse} number of occurrences. We present only these results in the following. We explain this difference by analogy with the IDF part of TF-IDF: using the inverse effectively gives more importance to less frequent characters/interactions. These are generally more typical of a narrative unit, compared to frequent characters, which are more likely to be involved in many narrative units. In the end, we have four ways of measuring the similarity, depending on whether we compare the vertex or edge sets, using the unweighted or weighted version of the index. 

Computing all four of these similarity measures requires an exact mapping between characters from a media to another in order to obtain a meaningful score: we compute these using the normalised \textit{name} vertex attribute that we described in~\ref{sec:DataAttr}. As explained in Section~\ref{sec:DataChar}, our networks contain characters that are unnamed or not necessarily present in all media. Therefore, we carry on matching experiments by progressively filtering our networks using the \texttt{named}, \texttt{common} and \texttt{top-20} character sets. 

In the rest of this section, we present the results of structural matching as follows. First, in Section~\ref{sec:AlignStructOrig}, we apply our methods to the same narrative units as with text in Section~\ref{sec:AlignText}: novel chapters, TV show episodes, and comic issues. The structural approach is not bound to these specific units, though, as they were determined by the sources of the textual content, in the first place. One possible issue with them is that they are not on the same scale, i.e. the chunks of plot they cover are not of comparable sizes. For example, TV show episodes usually adapt several chapters of the novels. To study the effect of the narrative unit scale, we experiment with smaller, commensurate narrative units in Section~\ref{sec:AlignStructAlter}.

\subsubsection{Text-Constrained Narrative Units}
\label{sec:AlignStructOrig}
The best results of structural alignment with the same narrative units as for text, using the thresholding and Smith--Waterman alignment methods, can be found in Table~\ref{tab:perf_structural_best}, while detailed results with all the configurations are provided in the Supplementary Material (cf. Table~\ref{tab:perf_structural}). The performance varies widely, with F1-scores ranging from $1.54$ to $62.94$, depending on the configuration. Overall, by choosing the best configuration, we obtain F1-scores of $62.94$, $32.63$ and $51.40$ for the \textit{Novels vs Comics}, \textit{Novels vs. TV Show} and \textit{Comics vs. TV Show} pairs, respectively. Structural matching leads to better performance than textual matching, for all three pairs of media: $+7.7$, $+10.1$ and $+15.4$ points, respectively. This is further confirmed in results for the \textit{Novels vs. TV Show} pair over the \texttt{U2} period. As shown in Table~\ref{tab:best_matching_tvshow-novels_U2} of the Supplementary Material, structural matching obtains an F1-score of $45.00$, while textual matching gets $36.65$.

\begin{table}[htb!]
    \centering
    \caption{Performance obtained when using \textit{structure}-based representations and the \textit{text-constrained} narrative units to tackle the narrative matching task, expressed in terms of F1-score. Only the best results over all possible configurations are shown}
    \input{tables/narr_match/perf_structural_best}
    \label{tab:perf_structural_best}
\end{table}

Based on these results, and as confirmed by Table~\ref{tab:perf_structural}, it appears that the Smith--Waterman algorithm largely outperforms thresholding for all configurations. Additionally, the \texttt{common} character set obtains the best results, while \texttt{top-20} severely underperforms and is almost always the worst option. It may be because more minor characters are specific to certain narrative arcs, while main characters are often together as groups, blending several narrative units together which renders matching more difficult. We cannot draw any conclusion regarding our Jaccard weighting scheme or whether computing Jaccard similarity on edges or vertices is better, as the behaviour of these parameters is inconsistent across configurations.

\subsubsection{Commensurate Narrative Units}
\label{sec:AlignStructAlter}
The chapter constitutes the natural narrative unit of the original material (the novels), and the sole we have for this medium. Both other text-constrained narrative units, comic issues and TV show episodes, have a larger scale. Put differently, the piece of plot they convey is longer than for a chapter: both issues and episodes adapt several novel chapters (or, in the case of episodes, sometimes parts of chapters). We hypothesise this hinders the matching performance. 

A solution is to adopt smaller narrative units that are closer to chapters. Scenes are available for both media, however, this unit is much smaller compared to novel chapters, which would cause the same scale difference problem as before. Instead, we turn to the intermediary units described in Section~\ref{sec:DataTime}: comic chapters, which are comparable to novel chapters, and TV show blocks, which are sequences of contiguous scenes. We defined the latter in an attempt to split the plot according to changes in character \textit{point-of-views}, the literary device used in novels to unveil the story. 

Although we now compare the comics and TV show using chapters and blocks, we must come back to issues and episodes to conduct our assessment, in order to allow a fair comparison with the performance previously presented for the other matching methods. We do so by considering that a novel chapter matching a comic chapter or a block matches the whole issue or episode. Following the same principle, when matching \textit{Comics vs. TV Show}, we match chapters and blocks, but consider issues and episodes to compute the performance.

\begin{table}[htb!]
    \centering
    \caption{Performance obtained when using \textit{structure}-based representations and the \textit{commensurate} narrative units to tackle the narrative matching task, expressed in terms of F1-score. Only the best results over all possible configurations are shown}
    \input{tables/narr_match/perf_structural_blocks_best}
    \label{tab:structural_matching_blocks_best}
\end{table}

Table~\ref{tab:structural_matching_blocks_best} shows the best alignment results using the commensurate narrative units, while Table~\ref{tab:perf_structural} in the Supplementary Material shows the results for all possible configurations. This scheme strongly increases matching performance when using the Smith--Waterman alignment algorithm, with gains of $9.4$, $2.5$ and $10.4$ points for the \textit{Novels vs. Comics}, \textit{Novels vs. TV Show} and \textit{Comics vs. TV Show} pairs respectively. This performance gain is further confirmed over the \texttt{U2} time period for the \textit{Novels vs. TV Show} pair with a large gain of $17.9$ F1 (see Section~\ref{sec:ApdxNarrMatchNovelsTVShowU2} in the Supplementary Material). Interestingly, even though we extract TV show \textit{blocks} automatically, which may induce errors, using these for matching still leads to better performance than with full episodes. Our results show that ensuring the \textit{scale} of the narrative units used for alignment are comparable is important for performance.

\subsection{Hybrid Method}
\label{sec:AlignCombo}
In this section, we strive to combine our textual and structural methods, in order to assess their complementarity. For each pair of media, we adopt a direct approach that consists in computing a new \textit{hybrid} similarity matrix $\mathbf{S}_{h}$, based on the structural and similarity matrices, respectively noted $\mathbf{S}_{s}$ and $\mathbf{S}_{t}$. We first rescale $\mathbf{S}_{s}$ and $\mathbf{S}_{t}$ separately using min-max normalisation: 
\begin{equation}
    \mathbf{S}'_{\star} = \frac{\mathbf{S}_{\star} - \min(\mathbf{S}_{\star})}{\max(\mathbf{S}_{\star}) - \min(\mathbf{S}_{\star})},
\end{equation}
where $\mathbf{S}_\star$ denotes $\mathbf{S}_{s}$ or $\mathbf{S}_{t}$. We then combine the resulting matrices using a weighted sum:
\begin{equation}
    \mathbf{S}_{h} = \alpha \mathbf{S}'_{s} + (1 - \alpha) \mathbf{S}'_{t},
\end{equation}
where $\alpha$ is a parameter controlling the relative importance of text vs. structure. As for our other parameters, we tune $\alpha$ for each media pair using the other two pairs as a development set.

While this combination method is pretty simple, note that we performed some additional exploratory experiments to combine textual and structural similarity, but that our attempts failed to improve over the results we present in this article. We experimented with training several machine-learning models to compute $\mathbf{S}_h$ from $\mathbf{S}_s$ and $\mathbf{S}_t$ instead of simply summing them. We also tried to perform early fusion by extracting embeddings from our dynamic character networks and combining them with SentenceBERT or TF-IDF vectors, and then experimented with multiple machine learning model to obtain $S_c$ from the resulting representation.

As in Section~\ref{sec:AlignStruct}, we experiment with matching using the text-constrained narrative units (Section~\ref{sec:AlignHybridOrig}) as well as the commensurate narrative units (Section~\ref{sec:AlignHybridAlter}).

\subsubsection{Text-Constrained Narrative Units}
\label{sec:AlignHybridOrig}
Table~\ref{tab:combined_matching} shows the results obtained when applying our hybrid method on the text-constrained narrative units already used in Sections~\ref{sec:AlignText} and~\ref{sec:AlignStructOrig}. As we experiment with all possible configurations of structural and textual matching, we only report the best performance for the sake of simplicity. We provide the full results in Table~\ref{tab:perf_combined} of the Supplementary Material.

\begin{table}[htb!]
    \centering
    \caption{Performance obtained when using \textit{hybrid} representations and the \textit{text-constrained} narrative units to tackle the narrative matching task, expressed in terms of F1-score. Only the best results across configurations are shown}
    \input{tables/narr_match/perf_combined_best}
    \label{tab:combined_matching}
\end{table}

We find that combining information from textual and structural matching increases performance for the \textit{Novels vs. Comics} pair ($+4.4$ F1), but fails to improve performance for the \textit{Novels vs. TV Show} ($-1.7$ F1) and \textit{Comics vs. TV Show} ($-2.2$ F1) pairs.

\subsubsection{Commensurate Narrative Units}
\label{sec:AlignHybridAlter}
As highlighted in Section~\ref{sec:AlignStructAlter}, performing structural matching on narrative units of comparable sizes can strongly increase performance. Therefore, we now want to conduct hybrid matching by combining text with the commensurate units from Section~\ref{sec:AlignStructAlter}. Our combination method requires to sum the textual and structural similarity matrices, and therefore, we need the matrices to be of the same dimension. It is not the case though: as already explained, the text-constrained units have a larger scale, and the corresponding similarity matrices are consequently smaller. Moreover, it is not possible to build textual representations at a smaller scale: for the TV show, episodes summaries can not easily be cut to correspond to the underlying blocks, and comics issues summaries cannot easily be split to apply to underlying chapters. Therefore, we propose to artificially \textit{extend} the textual similarity matrix in order to match the dimension of its structural counterpart. We do so by duplicating the rows and columns corresponding to a text-constrained narrative unit (e.g. TV episode) as many times as the number of commensurate units it contains (e.g. blocks).

\begin{table}[htb!]
    \centering
    \caption{Performance obtained when using \textit{hybrid} representations and the \textit{commensurate} narrative units to tackle the narrative matching task, expressed in terms of F1-score. Only the best results across configurations are shown}
    \input{tables/narr_match/perf_combined_blocks_best}
    \label{tab:combined_matching_alter}
\end{table}

Table~\ref{tab:combined_matching_alter} shows the best results of hybrid matching on commensurate narrative units, while the results for all configurations are available in the Supplementary Material (Table~\ref{tab:perf_combined}). As with purely structural matching, working with these narrative units greatly increases performance. Compared with hybrid matching on text-constrained units, we observe gains of $+11.1$, $+8.1$ and $+11.3$ F1 points on the \textit{Novels vs. Comics}, \textit{Novels vs. TV Show} and \textit{Comics vs. TV Show} pairs respectively. Compared with purely structural matching using commensurate units, we also observe improvements across the board: $+6.2$ F1 point for the \textit{Novels vs. Comics} pair, $+4$ points for the \textit{Novels vs. TV Show} pair and $+2.1$ points for the \textit{Comics vs. TV Show} pair. Overall, matching using hybrid similarity on the commensurate narrative units leads to the best performance.

\subsection{Takeaways}
\label{sec:AlignConclu}
\begin{figure}[htb!]
    \centering
        \includegraphics[width=1\textwidth]{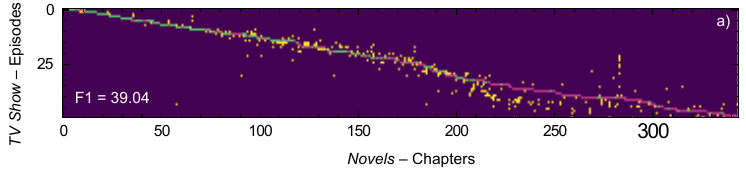}
        \includegraphics[width=1\textwidth]{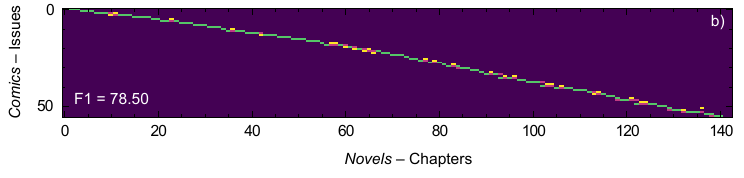}
        \includegraphics[width=1\textwidth]{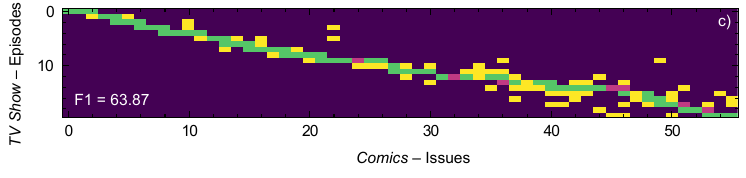}
    \caption{Best performing alignment for all pairs of media, from top to bottom: a) \textit{Novels} vs. \textit{TV Show}, b) \textit{Novels} vs. \textit{Comics}, and c)\textit{Novels} vs. \textit{Comics}. Green denotes true positives, red false positives, yellow false negatives and purple true negatives}
    \label{fig:best-alignments}
\end{figure}

The way we formalise the narrative matching task make it difficult: we perform many-to-many matching and use F1 as a metric, meaning any mismatch is strictly penalised. Even so, our best results ($78.50$ and $63.87$ F1 for the \textit{Novels vs. Comics} and \textit{Comics vs. TV Show} pairs, see Figure~\ref{fig:best-alignments} shows that our proposed alignment method can obtain good performance. Through our experiments, we are able to derive several important insights valuable to the application of the method. First, we note that structure-based matching using dynamic instant character networks outperforms text-based matching, highlighting the usefulness of networks for the task. To our knowledge, this is the first time that character networks are used for narrative alignment. We also show that combining text-based and structure-based similarities can yield better performance than using either alone. Furthermore, we demonstrate the importance of aligning stories using a comparable narrative scale, as taking commensurate narrative units into account consistently improves our results.

The results on the \textit{Novels vs. TV Show} pair over the \texttt{U5} period  are, however, more lackluster ($39.04$ F1 at best). We attribute this lower performance to the plot divergence between the original novels and their TV show adaptation, which increased over time. As a confirmation, results restricted to period \texttt{U2} are much better, with a top F1-score of $64.65$ (see Figure~\ref{fig:best_matching_tvshow-novels_U2} and Table~\ref{tab:best_matching_tvshow-novels_U2} of the Supplementary Material). As shown in Figure~\ref{fig:tvshow-novels_best_perf_through_time}, the matching performance progressively decreases as the TV show progresses for the structural, textual and hybrid similarities alike. Plot divergence is difficult to tackle for the Smith--Waterman algorithm, originally developed to align molecular sub-sequences: it assumes that parts of the sequences it tries to align are ordered similarly, but this assumption is challenged on the \textit{Novels vs. TV Show} pair. As seen in Figure~\ref{fig:best-alignments}, the blocks constituting the later seasons are ordered completely differently from their chapter counterparts, and some chapters are not even adapted, leading to low performance.

\begin{figure}[htb!]
    \centering
    \includegraphics{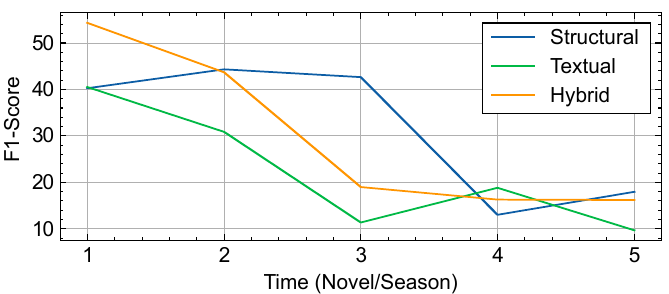}
    \caption{Performance of the best configurations of narrative alignment over seasons for the \textit{Novels vs. TV Show} pair, expressed in F1-score. As the TV show progressively diverges from the novels, the matching performance starts to decrease. F1-score for last two seasons does not exceed $20$}
    \label{fig:tvshow-novels_best_perf_through_time}
\end{figure}

\section{Conclusion}
\label{sec:Conclusion}
In this article, we propose a method based on dynamic character networks to analyse adaptations of a same story across media. This framework is meant to be used to derive insights on how the adaptation process affects the rendition of the plot and characters of a story, depending on the constraints of the target medium. We applied it to adaptations across media of the fantasy series \textit{A Song of Ice and Fire}. We obtained a corpus of three adaptations, including the original novels and two adaptations in the form of comics and TV show. We extracted several types of character networks from these raw data to model the three adaptations. We then focused on two research questions. 

The first was to determine whether such structural models allow matching characters from one adaptation to the other. Based on our results obtained with state-of-the-art Graph Matching methods, it appears that character interactions alone are not sufficient to reliably perform such task over \textit{all} characters. However, the performance is much better when focusing on a narrower set of the most \textit{important} characters, which hints at a stronger inter-medium similarity for this category of vertices. Moreover, adding character-related information such as sex or affiliation greatly improves the results, too. We conducted a centrality study that confirmed these observations.

Our second research question was to determine the possibility of matching the plots of the adaptations themselves, from one medium to the other. We formalised this task as a many-to-many matching problem, in which each narrative unit of one medium can be associated to one or several narrative units of the other medium. We proposed a method based on the computation of an inter-medium similarity matrix, which is then used to estimate a matching matrix. We considered several ways of building the similarity and matching matrices. We experimented with text- and structure-based dynamic representations of the adaptations. According to our results, the structure of the dynamic networks leads to much better performance than the textual representations. Combining them does improve our results, though, which indicates that they are complementary. Our results also show the importance of selecting narrative units of comparable scale to model the adaptations. Finally, our experiments provide objective elements that support the subjective perception of the audience regarding the divergence between the original novels and the TV show adaptation starting novel/season five and onwards.

%

Our work can be extended in several ways. First, some of our proposed methods could be improved. For the character matching task, the results obtained with the cumulative networks suggest that considering the plot dynamics is promising to increase performance. Concerning the narrative matching task, we hypothesise that performing an early fusion of text and structure by proposing an appropriate representation learning method could help to improve our results. 
Second, we plan on using this unique corpus to address other research questions. In particular, an interesting point concerns the position of women in the adaptations, and how the TV show differs from the novels.
Third, it would be interesting to tackle the same problems on other adaptations across media, in order to study how this affects the performance. A number of other works of fiction were the object of such adaptations: \textit{Harry Potter}, \textit{The Witcher}, \textit{Dune}, etc. However, this would require a significant annotation work, which is why our corpus is unique. 
The fourth perspective concerns the different tasks proposed in the literature that require to match character networks. For example, some authors compare character networks to real-world ones in order to assess the level of realism of stories~\citep{Stiller2003, Gleiser2007, Kenna2016, gramsch2017medieval}; others want to automatically distinguish original works from adaptations~\cite{Chaturvedi2018, Chowdhury2019}. These comparisons are typically conducted over static graphs: our proposed methods could be used in this context to take into account the dynamics of the stories.
Fifth and finally, an application in which we are most interested in is the analysis of historical and mythological texts. As mentioned earlier, works comparing hagiography and history display different network properties. The social networks of epics and sagas have also been compared (for example see~\cite{mac2012universal}), however the data used here is frequently from medieval manuscripts recorded after centuries of oral transmission. The final recorded version may have significant differences from the original which we no longer have access too.  \emph{A Song of Ice and Fire} and two of its adaptations are ``closed systems''. By determining differences in character roles, plots, and social networks, insights can be made into how much these can deviate given centuries of oral tradition.

\backmatter
\bmhead{Acknowledgements}
We thank Jeffrey Lancaster for constituting his dataset about the \textit{Games of Thrones} TV show and sharing it online, as well as for his feedback. We also thank Tanzir Pial, who kindly provided us with explanations about his implementation and use of the Smith--Waterman algorithm in~\cite{Pial2023a, Pial2023b}.

\bibliography{got_article.bib}

\newpage
\newgeometry{top=1.5cm, bottom=1.5cm, left=1cm, right=1cm}
\color{black!60!blue}
\pagecolor{black!1}
\renewcommand*{\theHsection}{SM\arabic{section}}%
\renewcommand{\thesection}{SM\arabic{section}}
\setcounter{section}{0}
\renewcommand{\thetable}{SM\arabic{table}}
\setcounter{table}{0}
\renewcommand{\thefigure}{SM\arabic{figure}}
\setcounter{figure}{0}
\begin{center}
    \noindent\Huge{Supplementary Material} \\[0.2cm]
    \noindent\makeatletter\Large{\@title}\makeatother\\[0.2cm]
    \noindent\artauthors
\end{center}

\section{Dataset and Descriptive Analysis}
\label{sec:ApdxDescrAnal}
This section aims at providing more details regarding the dataset and its preparation. Section~\ref{sec:ApdxDescrAnalNarr} is dedicated to the raw data and the adaptations themselves, whereas Section~\ref{sec:ApdxDescrAnalNets} focuses on the networks that we extract from these adaptations.

\subsection{Adaptations}
\label{sec:ApdxDescrAnalNarr}
In this section, we give more information regarding the three adaptations studied in the main article (Section~\ref{sec:ApdxDescrAnalNarrOrga}), and the way we constitute the character sets used in our experiments (Section~\ref{sec:ApdxDescrAnalNarrChars}). 

\subsubsection{Organisation}
\label{sec:ApdxDescrAnalNarrOrga}
\begin{figure}[htb!]
    \centering
    \includegraphics[width=1\textwidth]{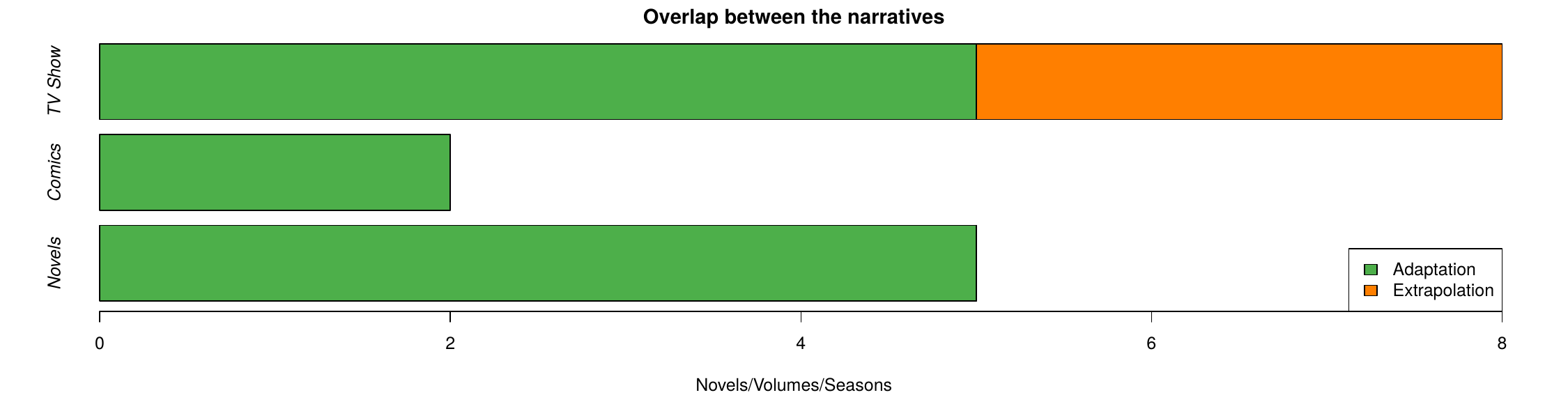}
    \caption{Overlap between the three considered adaptations, in terms of books (for the novels), volumes (for the comics), and seasons (for the TV show)}
    \label{fig:NarrOverlap}
\end{figure}

The original material is constituted of five published books out of a total of seven envisioned novels:
\begin{enumerate}
    \item \textit{A Game of Thrones} (1996)
    \item \textit{A Clash of Kings} (1998)
    \item \textit{A Storm of Swords} (2000)
    \item \textit{A Feast for Crows} (2005)
    \item \textit{A Dance with Dragons} (2011)
    \item \textit{The Winds of Winter} (forthcoming)
    \item \textit{A Dream of Spring} (planned)
\end{enumerate}
A few draft chapters of the last two novels have also been published online. A part of these were integrated in the TV show, in addition to the first five novels. The showrunner also had access to unpublished material (and to the author). By comparison, the comics aim to be a straightforward adaptation of the first two novels~\cite{AWoIaF2021}. Table~\ref{tab:NarrStats} shows how the three adaptations break down into various types of narrative units, whereas Figure~\ref{fig:NarrOverlap} shows the overlap between the three adaptations, in terms of their largest narrative units: books, volumes, and seasons.

\begin{table}[htb!]
    \caption{Numbers of narrative units for all three adaptations studied in the main article}
    \label{tab:NarrStats}%
    \begin{tabular}{l r r r}
        \toprule
        Narrative unit & \textit{Novels} & \textit{Comics} & \textit{TV Show} \\
        \midrule
        Scenes   & -- & 1,437 & 4,165 \\
        Blocks   & -- & -- & 739 \\
        Chapters & 344 & 143 & -- \\
        Issues/Episodes & -- & 56 & 73 \\
        Books/Volumes/Seasons & 5 & 2 & 8 \\
        \botrule
    \end{tabular}
\end{table}

\subsubsection{Characters}
\label{sec:ApdxDescrAnalNarrChars}
In order to normalise the character names, we first scrape a \textit{main list} of names from the Wiki mentioned in the main article, and dedicated to the novels: \textit{A Wiki of Ice and Fire}\footnote{\url{https://awoiaf.westeros.org/index.php/Main_Page}} (AWoIaF). We then list the names appearing in all three adaptations, and automatically match them to similar names in the main list. Next, the many remaining names are matched manually. The TV show contains a number of additional characters compared to the novels. We complement the main list based on the other Wiki mentioned in the main article, \textit{Game of Thrones Wiki}\footnote{\url{https://gameofthrones.fandom.com/wiki/Game_of_Thrones_Wiki}} (GoTW), which is dedicated to the TV show.

In the process, we identify and delete a few spurious characters, e.g. the names of some actors and staff in the \textit{TV Show} dataset, and some expressions not matching any character in the \textit{Novels} dataset. We also identify some characters appearing twice under different names (e.g. \textit{The Three-Eyed Raven} and \textit{The Last Greenseer}). Sometimes, different versions of the same character are explicitly distinguished (e.g. normal vs. young, or live vs. wight): we merge them for the sake of consistency.

After this, the main list contains a total of 3,863 characters (some of them appearing in other media than those we consider in this work): 3,778 characters from the AWoIaF Website, and 85 additional characters retrieved from the GoTW Website. In the end, we can leverage this list to produce conversion maps specific to each medium, allowing to normalise all character names.

\begin{table}[htb!]
    \caption{Lists of the 20 most important characters (\texttt{top-20}) for periods \texttt{U2} (left) and \texttt{U5} (right), by decreasing order of importance. The scores correspond to normalised numbers of occurrences (cf. Section~\ref{sec:DataChar} in the main article). Stars indicate the four characters that are present only in the left or only the right table}
    \label{tab:CharacterList}%
    \begin{tabular}{l l l r r r r p{1cm} l r r r}
        \cmidrule{1-7}\cmidrule{9-12}
        Character & Affiliation & Sex & \textit{Nov.} & \textit{Com.} & \textit{TV} &  Mean &  & Character & \textit{Nov.} & \textit{TV} &  Mean \\
        \cmidrule{1-7}\cmidrule{9-12}
        Tyrion Lannister & House Lannister & M & 0.62 & 1.00 & 1.00 & 0.87 &  & Tyrion Lannister & 0.76 & 1.00 & 0.88 \\
        Eddard Stark & House Stark & M & 1.00 & 0.64 & 0.58 & 0.74 &  & Cersei Lannister & 0.86 & 0.76 & 0.81 \\
        Arya Stark & House Stark & F & 0.62 & 0.88 & 0.62 & 0.71 &  & Jon Snow & 0.57 & 0.89 & 0.73 \\
        Catelyn Stark & House Stark & F & 0.65 & 0.80 & 0.49 & 0.65 &  & Sansa Stark & 0.70 & 0.68 & 0.69 \\
        Jon Snow & House Stark & M & 0.50 & 0.78 & 0.60 & 0.63 &  & Arya Stark & 0.67 & 0.65 & 0.66 \\
        Sansa Stark & House Stark & F & 0.64 & 0.57 & 0.54 & 0.59 &  & Eddard Stark & 1.00 & 0.25 & 0.62 \\
        Bran Stark & House Stark & M & 0.63 & 0.63 & 0.40 & 0.56 &  & Jaime Lannister & 0.75 & 0.47 & 0.61 \\
        Cersei Lannister & House Lannister & F & 0.76 & 0.30 & 0.57 & 0.54 &  & Robb Stark & 0.89 & 0.30 & 0.59 \\
        Daenerys Targaryen & House Targaryen & F & 0.20 & 0.75 & 0.62 & 0.52 &  & Joffrey Baratheon & 0.77 & 0.41 & 0.59 \\
        Robb Stark & House Stark & M & 0.76 & 0.29 & 0.42 & 0.49 &  & Tywin Lannister* & 0.73 & 0.31 & 0.52 \\
        Joffrey Baratheon & House Baratheon & M & 0.68 & 0.24 & 0.44 & 0.45 &  & Daenerys Targaryen & 0.31 & 0.72 & 0.52 \\
        Theon Greyjoy & House Greyjoy & M & 0.25 & 0.42 & 0.52 & 0.39 &  & Stannis Baratheon & 0.71 & 0.31 & 0.51 \\
        Robert I Baratheon & House Baratheon & M & 0.80 & 0.15 & 0.15 & 0.37 &  & Robert I Baratheon & 0.92 & 0.06 & 0.49 \\
        Petyr Baelish & House Baelish & M & 0.39 & 0.26 & 0.36 & 0.34 &  & Catelyn Stark & 0.67 & 0.32 & 0.49 \\
        Jaime Lannister & House Lannister & M & 0.68 & 0.11 & 0.21 & 0.33 &  & Bran Stark & 0.56 & 0.37 & 0.46 \\
        Sandor Clegane & House Lannister & M & 0.38 & 0.19 & 0.37 & 0.31 &  & Sandor Clegane & 0.42 & 0.31 & 0.37 \\
        Jorah Mormont & House Targaryen & M & 0.19 & 0.32 & 0.37 & 0.29 &  & Samwell Tarly* & 0.20 & 0.49 & 0.35 \\
        Stannis Baratheon & House Baratheon & M & 0.53 & 0.07 & 0.24 & 0.28 &  & Theon Greyjoy & 0.31 & 0.37 & 0.34 \\
        Renly Baratheon* & House Baratheon & M & 0.47 & 0.17 & 0.18 & 0.27 &  & Jorah Mormont & 0.20 & 0.47 & 0.33 \\
        Varys* & House Baratheon & M & 0.33 & 0.18 & 0.28 & 0.26 &  & Petyr Baelish & 0.35 & 0.30 & 0.33 \\
        \cmidrule{1-7}\cmidrule{9-12}
    \end{tabular}
\end{table}

Table~\ref{tab:CharacterList} provides the lists of the 20 most important characters, according to the method described in the main article, for periods \texttt{U2} (first two books, volumes, and seasons) and \texttt{U5} (first five books and seasons). Period \texttt{U5} concerns only the novels and TV show, since the comics only cover the first two books. The characters are almost all the same in both lists, albeit in a different order, except for Renly Baratheon and Varys (only in \texttt{U2}), and Samwell Tarly and Petyr Baelish (only in \texttt{U5}). In addition to the character names, both lists contain the scores used to rank the characters in each adaptation and overall, which correspond to max-normalised numbers of occurrences. Finally, the left list also shows the \textit{Sex} and \textit{Affiliation} attributes, later used to match the characters from one network to the other.

\subsection{Networks}
\label{sec:ApdxDescrAnalNets}
\begin{figure}[tbh!]
    \centering
    \includegraphics[height=5.57cm]{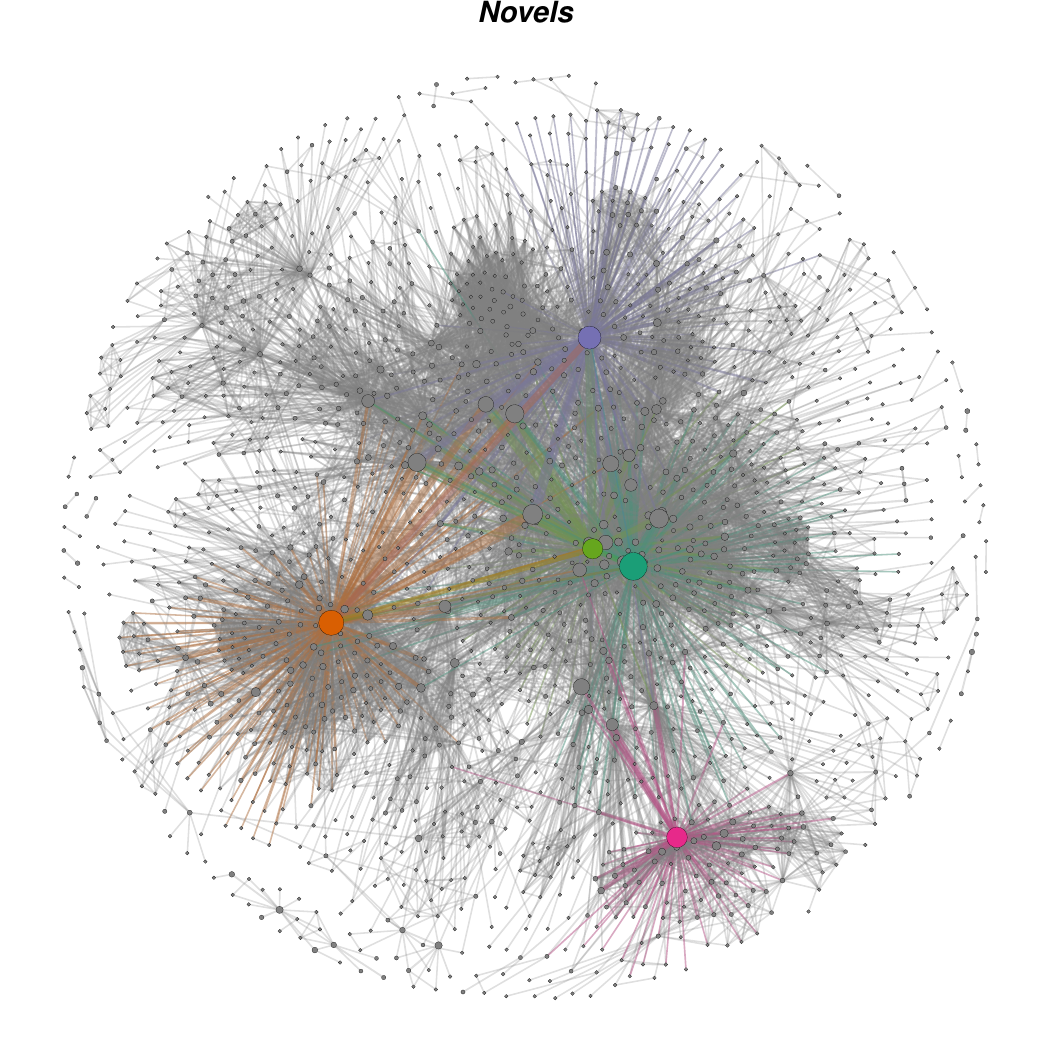}
    \hfill
    \includegraphics[height=5.57cm]{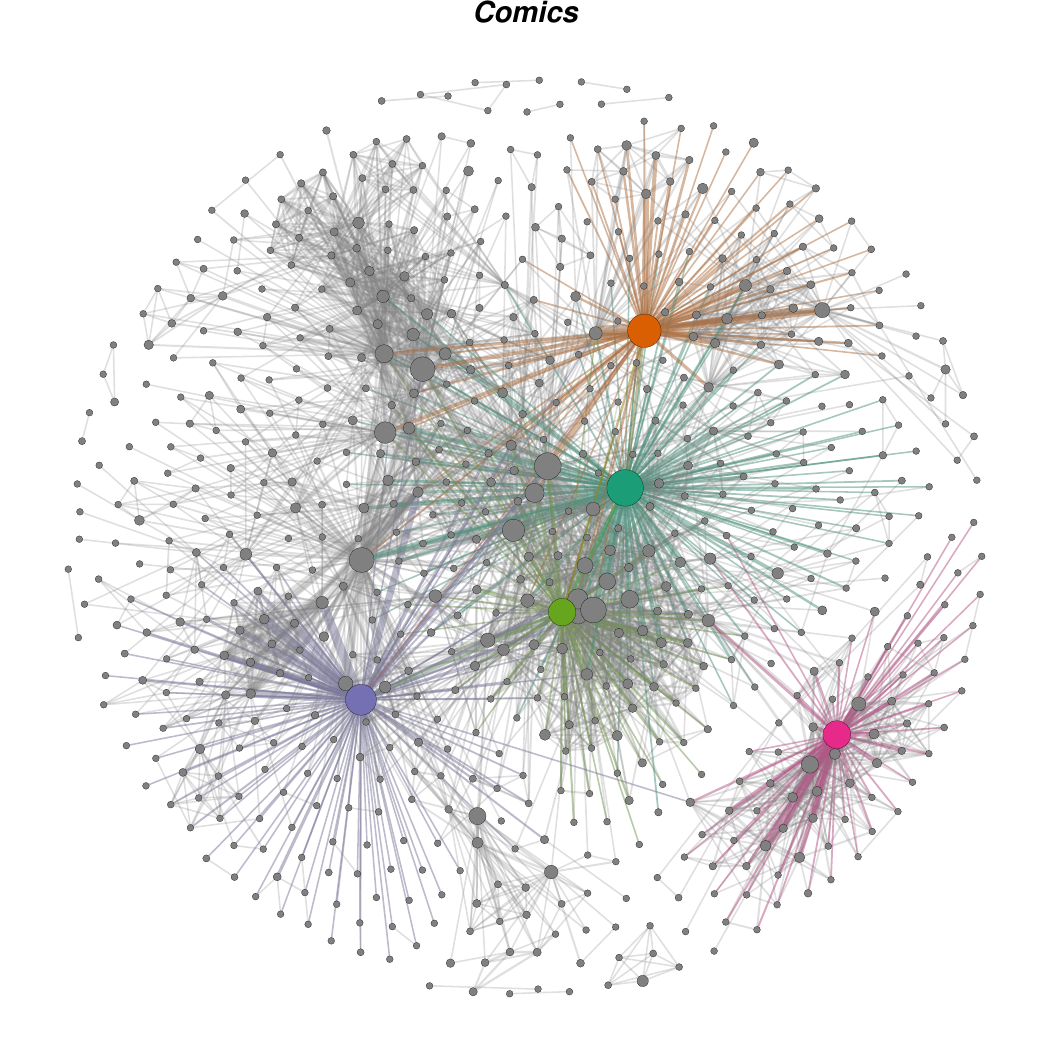}
    \hfill
    \includegraphics[height=5.57cm]{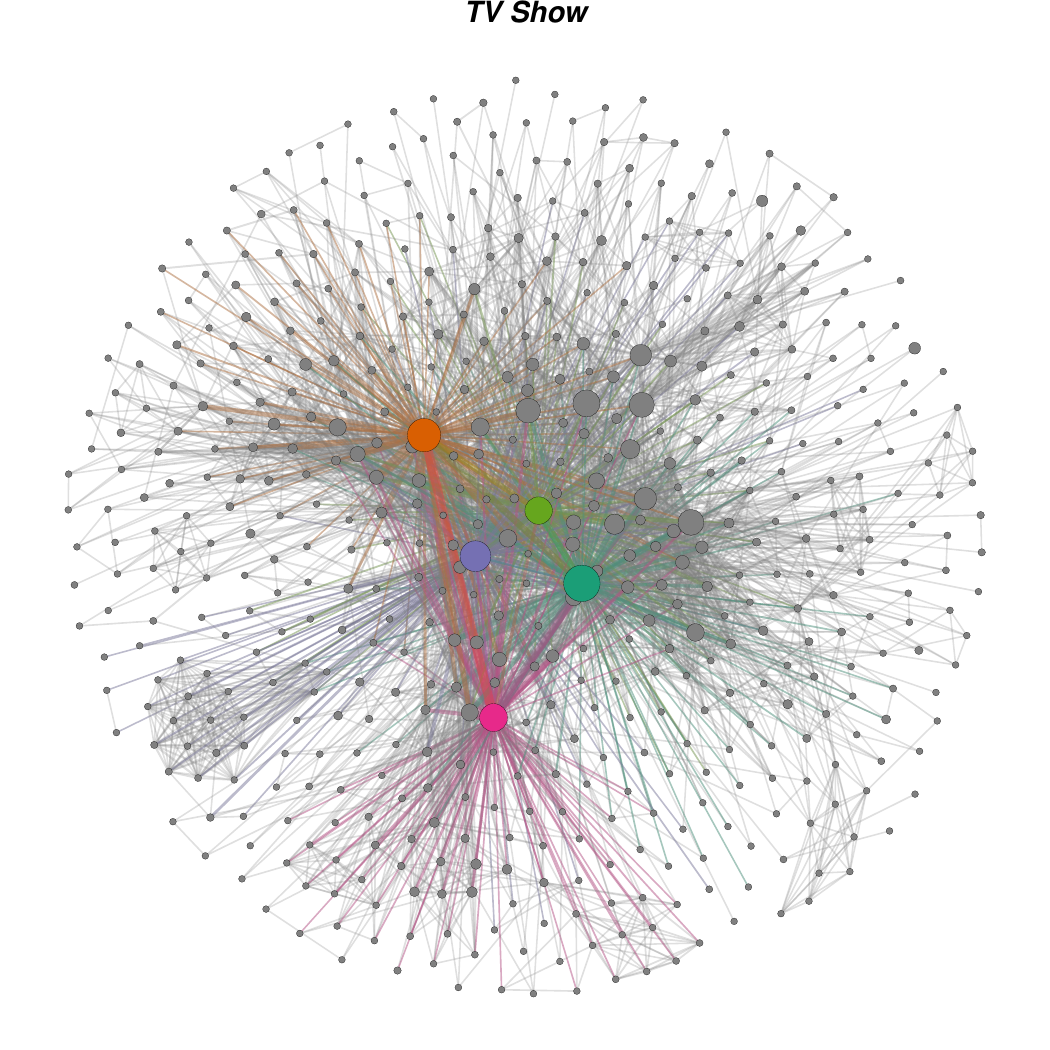}
    \hfill
    \includegraphics[height=5.57cm]{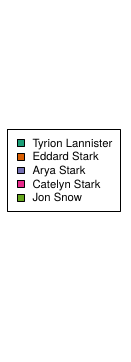}
    \caption{Static networks including all the characters, for all three adaptations, for the longest period they cover. The 5 most important characters are highlighted in colour}
    \label{fig:NetsAll}
\end{figure}

Figure~\ref{fig:NetsAll} shows the static character network obtained for each adaptation, when considering the largest period it covers: five books for the novels, two volumes for the comics, and eight seasons for the TV show. The 5 most important characters are represented in colour, and vertex size reflects the importance score (see Table~\ref{tab:CharacterList}). Edge thickness corresponds to the number of interactions over the considered period. We used the Fruchterman--Reingold method~\cite{Fruchterman1991} to provide similar layouts and ease visual comparison. The \textit{Novels} network gives the impression of being denser, but this is not the case: as shown by Table~\ref{tab:networks} from the main article, the TV Show network is clearly the densest. 

\begin{figure}[tbh!]
    \centering
    \includegraphics[width=1\textwidth]{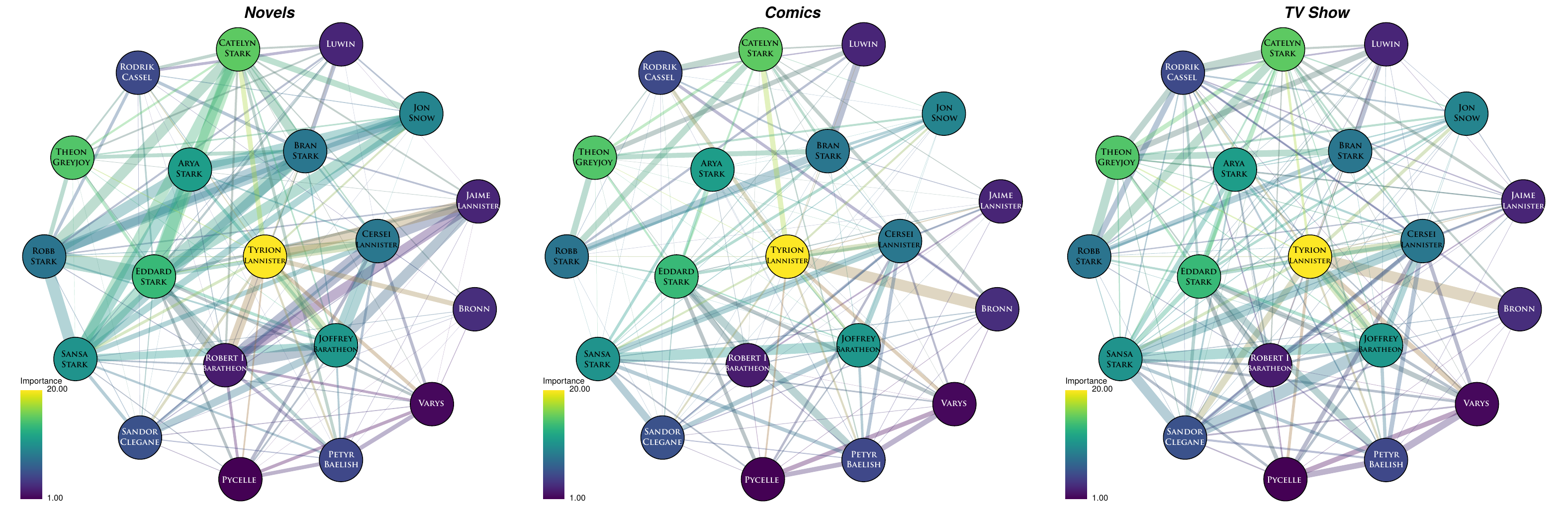}
    \caption{Static networks limited to the \texttt{top20} character set, for all three adaptations, for period \texttt{U2} (first two novels, volumes, and seasons). The vertex layout is fixed for all three graphs}
    \label{fig:NetsTop20}
\end{figure}

Figure~\ref{fig:NetsTop20} represents, for each adaptation, the subnetwork of the 20 most important characters overall, for period \texttt{U2} (i.e. first two novels, volumes, and seasons). As in Figure~\ref{fig:NetsAll}, the layout is based on the Fruchterman--Reingold method~\cite{Fruchterman1991}. However, this time we fix the layout across networks, as the characters are exactly the same (by construction) for all three of them. Vertex colour represents character importance.

Table~\ref{tab:NetsTopoCharsets} shows the same topological measure as in Table~\ref{tab:networks} from the main article, but for the three different character sets (and not \textit{all} of them). The increasing filtering of characters obviously affects the number of vertices, but also increases the density. This means that the more important characters are, the more they tend to be interconnected. This also shows in the assortativity, which tends to increase (even if it remains quite low), the average clustering coefficient, which also increases, and the modularity, which decreases, indicating that the important characters are more tightly knit.  

\begin{table}[htb!]
    \caption{Network statistics for each of the time-periods, giving the number of vertices $n$ and edges $L$, the density $\delta$, the mean degree $\langle k \rangle$, the average shortest path length $\langle \ell \rangle$, the average clustering coefficient $\langle C \rangle$, the degree assortativity $r$, and the modularity $Q$. This table shows the statistics for each character set: by comparison, Table~\ref{tab:networks} from the main article deals with \textit{all} available characters}
    \label{tab:NetsTopoCharsets}%
    \begin{tabular}{p{2cm} p{2cm} p{2.5cm} r r r r r r r r r r}
        \toprule
        Characters & Period & Adaptation & $n$ & $L$ & $\delta$ & $\langle k \rangle $& $ \langle \ell \rangle $& $\langle C \rangle$ & $r$ & $Q$\\
        \midrule
        \texttt{named} & \texttt{U2} & Novels & 731 & 4,959 & 0.019 & 13.57 & 2.98 & 0.60 & $0.00$ & 0.53 \\
         &  & Comics & 324 & 2,039 & 0.039 & 12.59 & 2.88 & 0.72 & $-0.03$ & 0.68 \\
         &  & TV Show & 167 & 1,185 & 0.085 & 14.19 & 2.83 & 0.76 & $0.11$ & 0.58 \\
        \cmidrule{2-11}
         & \texttt{U5} & Novels & 1,877 & 13,859 & 0.008 & 14.77 & 3.32 & 0.58 & $-0.02$ & 0.58 \\
         &  & TV Show & 285 & 2,090 & 0.052 & 14.67 & 2.71 & 0.75 & $-0.01$ & 0.61 \\
        \cmidrule{2-11}
         & \texttt{U8} & TV Show & 348 & 2,974 & 0.049 & 17.09 & 2.51 & 0.76 & $-0.07$ & 0.42 \\
        \midrule
        \texttt{common} & \texttt{U2} & Novels & 123 & 907 & 0.121 & 14.75 & 2.38 & 0.68 & $0.12$ & 0.45 \\
         &  & Comics & 123 & 699 & 0.093 & 11.37 & 2.67 & 0.68 & $0.06$ & 0.62 \\
         &  & TV Show & 123 & 850 & 0.113 & 13.82 & 2.69 & 0.76 & $0.16$ & 0.54 \\
        \cmidrule{2-11}
         & \texttt{U5} & Novels & 216 & 1,859 & 0.080 & 17.21 & 2.46 & 0.67 & $0.03$ & 0.47 \\
         &  & TV Show & 216 & 1,534 & 0.066 & 14.20 & 2.65 & 0.76 & $0.05$ & 0.59 \\
        \cmidrule{2-11}
         & \texttt{U8} & TV Show & 153 & 1,330 & 0.114 & 17.39 & 2.28 & 0.74 & $-0.02$ & 0.42 \\
        \midrule
        \texttt{top-20} & \texttt{U2} & Novels & 20 & 127 & 0.668 & 12.70 & 1.41 & 0.79 & $-0.08$ & 0.01 \\
         &  & Comics & 20 & 111 & 0.584 & 11.10 & 1.28 & 0.73 & $0.13$ & 0.13 \\
         &  & TV Show & 20 & 123 & 0.647 & 12.30 & 1.20 & 0.82 & $0.15$ & 0.09 \\
        \cmidrule{2-11}
         & \texttt{U5} & Novels & 20 & 127 & 0.668 & 12.70 & 1.34 & 0.83 & $-0.07$ & 0.01 \\
         &  & TV Show & 20 & 117 & 0.616 & 11.70 & 1.44 & 0.88 & $0.22$ & 0.33 \\
        \cmidrule{2-11}
         & \texttt{U8} & TV Show & 20 & 151 & 0.795 & 15.10 & 1.21 & 0.92 & $-0.19$ & 0.00 \\
        \botrule
    \end{tabular}
\end{table}

\begin{figure}[tbh!]
    \centering
    \includegraphics[width=1\textwidth]{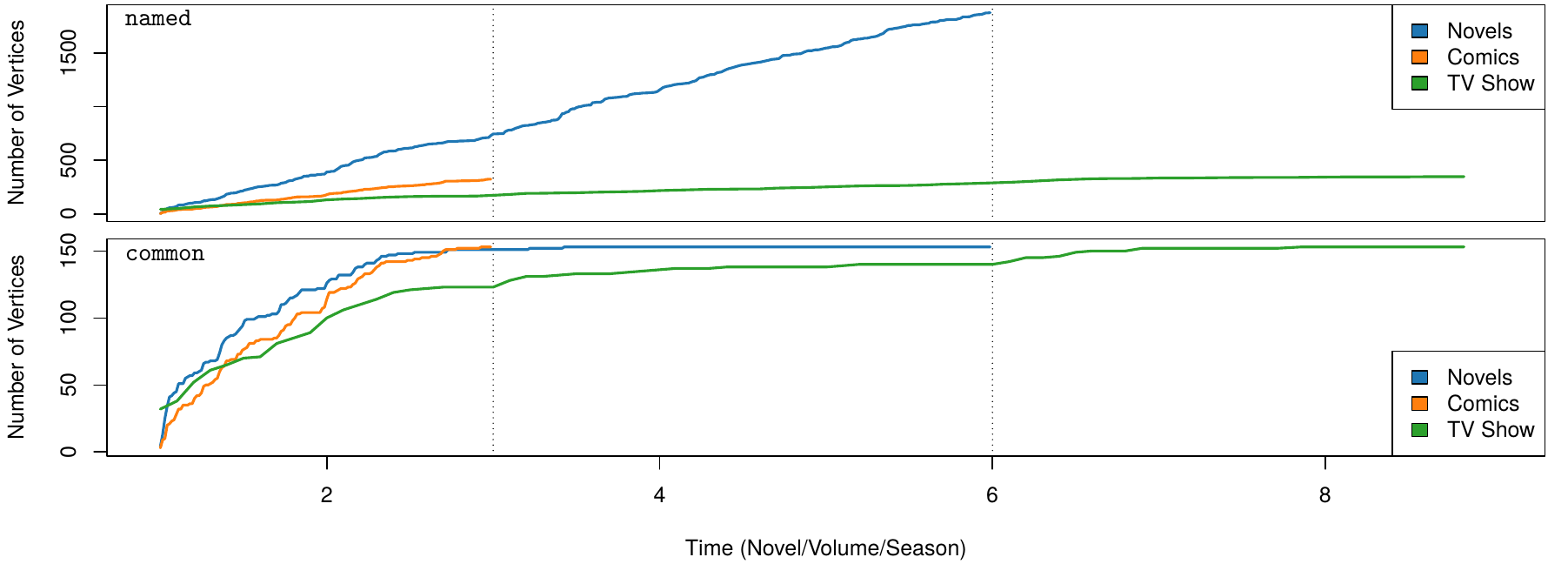}
    \caption{Evolution of the number of vertices in each adaptation, for the \texttt{named} (top) and \texttt{common} (bottom) character sets. Table~\ref{fig:topology} from the main article provides the same plot, but for all existing characters}
    \label{fig:EvolStatsVertex}
\end{figure}

Figure~\ref{fig:EvolStatsVertex} shows the evolution of the number of vertices (i.e. characters) in each adaptation, over books/volumes/seasons. By comparison, Figure~\ref{fig:topology} from the main article includes all characters (named or not), and shows that 1) their number increases linearly for all adaptations, 2) faster for novels and comics than for the TV show, and 3) the growth rate is similar for novels and comics. When considering only \texttt{named} characters in Figure~\ref{fig:EvolStatsVertex}, we see that that the first two observations still holds, but not the third one: comics exhibit a smaller growth rate. This is because the annotations for this medium include a large proportion of unnamed characters (typically, standing in the background and only witnessing a scene). When focusing on characters that are \texttt{common} to the three media, all of them exhibit a sublinear growth of the number of vertices, and a very similar evolution. 

\begin{figure}[tbh!]
    \centering
    \includegraphics[width=1\textwidth]{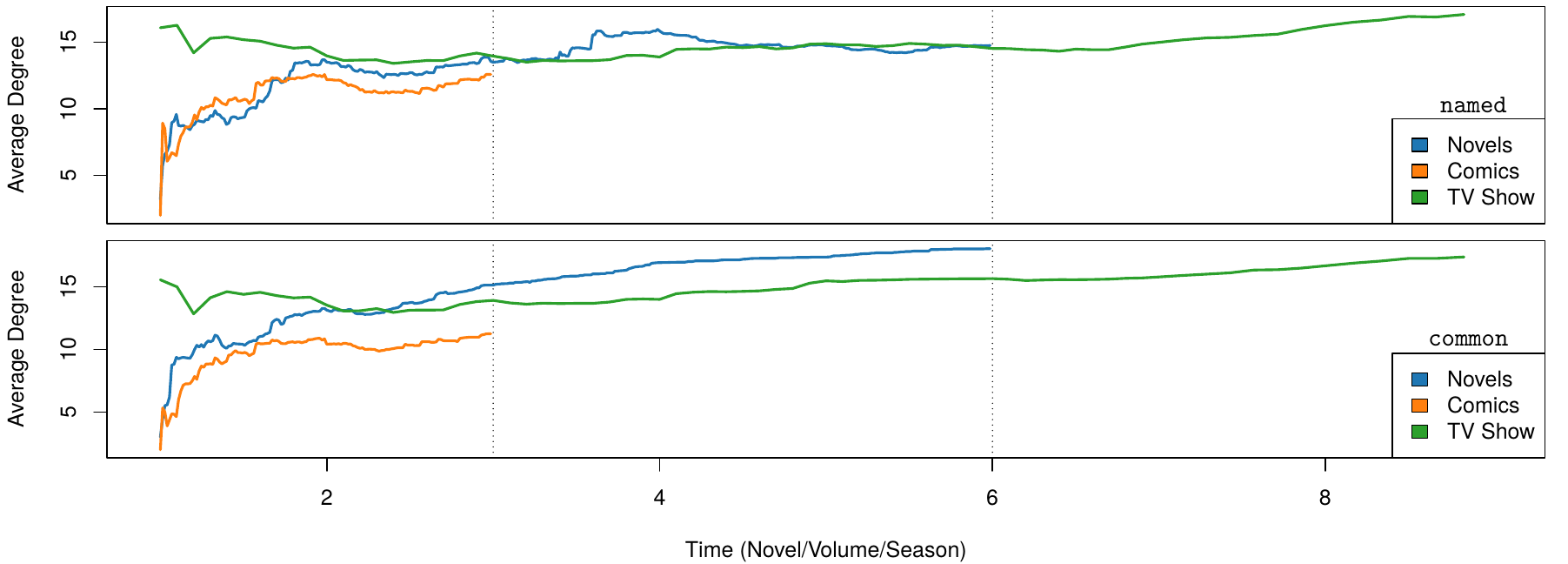}
    \caption{Evolution of the average degree in each adaptation, for the \texttt{named} (top) and \texttt{common} (bottom) character sets. Table~\ref{fig:topology} from the main article provides the same plot, but for all existing characters}
    \label{fig:EvolStatsDegree}
\end{figure}

Figure~\ref{fig:EvolStatsDegree} shows the evolution of the average degree over time. By comparison, Figure~\ref{fig:topology} from the main article includes all characters, and shows that it is relatively stable in the TV show, whereas the novels and comics are characterised by an increase during the first seasons, before stabilising too. These observations holds when focusing on \texttt{named} and \texttt{common} characters (Figure~\ref{fig:EvolStatsDegree}). Moreover, novels and comics appear to be very similar on this aspect.

\section{Character Comparison}
\label{sec:ApdxCharComp}
This section aims at providing additional results for Section~\ref{sec:CharComp} from the main article. Like in the main article, we first consider the GM problem used to match characters between adaptations (Section~\ref{sec:ApdxCharCompMatch}), before turning to a centrality analysis of the characters (Section~\ref{sec:ApdxCharCompCentr}).

\subsection{Graph Matching}
\label{sec:ApdxCharCompMatch}
In this section, we present the comprehensive results obtained when applying the five selected state-of-the-art GM methods to our networks. By comparison, the main article (Section~\ref{sec:CharCompMatch}) focuses only on the best methods, for the sake of concision.

\subsubsection{Basic Methods}
\label{sec:ApdxCharCompMatchBasic}
Table~\ref{tab:MatchingResultsCentering} shows the results obtained when applying all methods to all pairs of networks, without using any seeds or vertex attributes, with and without adjacency matrix centring. When applying the percolation method, which requires at least one seed, we use the most important character.

When considering all available \texttt{named} characters, there is no unique best approach, as their performance vary over network pairs. The best results are obtained when comparing the comics and TV show. Centring does not seem to affect the results much.

\begin{table}[htb!]
    \caption{Graph matching results obtained for all methods when using no seeds and no attributes. The table highlights how centring affects the performance, which is expressed in terms of correctly matched for each pair of networks, considering the three character sets used throughout the paper: all named characters (\texttt{named}), only those common to both compared networks (\texttt{common}), and only the 20 most important characters (\texttt{top-20})}
    \label{tab:MatchingResultsCentering}%
    \begin{tabular}{p{3cm}@{~}p{3cm} r@{~}r@{~}r r@{~}r@{~}r r@{~}r@{~}r}
        \toprule
        Method & Centring & \multicolumn{3}{c}{\textit{Novels} vs. \textit{Comics}} & \multicolumn{3}{c}{\textit{Novels} vs. \textit{TV Show}} & \multicolumn{3}{c}{\textit{Comics} vs. \textit{TV Show}} \\
         &  & \texttt{named} & \texttt{common} & \texttt{top-20} & \texttt{named} & \texttt{common} & \texttt{top-20} & \texttt{named} & \texttt{common} & \texttt{top-20} \\
        \midrule
        Convex      & Yes & 0.13\% & 3.42\% & 10.00\% & 0.00\% &  2.16\% &  0.00\% & 0.28\% & 16.67\% & 10.00\% \\
        Indefinite  & Yes & 0.00\% & 1.03\% &  5.00\% & 0.00\% & \textbf{10.07\%} &  0.00\% & 0.00\% &  4.55\% & \textbf{50.00\%} \\
        Concave     & Yes & 0.00\% & 2.05\% & \textbf{20.00\%} & 0.00\% &  1.44\% &  0.00\% & 0.28\% & \textbf{21.97\%} & 10.00\% \\
        Percolation & Yes & 0.13\% & 2.40\% & 15.00\% & 0.13\% &  2.88\% & 20.00\% & 0.57\% &  5.30\% & 10.00\% \\
        Umeyama     & Yes & 0.13\% & 0.00\% & 10.00\% & 0.00\% &  0.72\% &  5.00\% & 0.00\% &  0.76\% & 10.00\% \\
        \midrule
        Convex      &  No & 0.40\% & 3.08\% & 10.00\% & 0.00\% &  0.72\% &  5.00\% & \textbf{5.10\%} & 16.67\% & 25.00\% \\
        Indefinite  &  No & 0.26\% & 1.03\% &  5.00\% & 0.00\% &  1.44\% &  0.00\% & 2.83\% & 10.61\% & 35.00\% \\
        Concave     &  No & \textbf{0.53\%} & 2.05\% &  5.00\% & 0.13\% &  2.16\% &  5.00\% & 4.82\% & 21.21\% & 30.00\% \\
        Percolation &  No & \textbf{0.53\%} & \textbf{3.77\%} & \textbf{20.00\%} & \textbf{0.26\%} &  3.60\% & \textbf{25.00\%} & 0.85\% & 10.61\% & 15.00\% \\
        Umeyama     &  No & 0.00\% & 0.00\% & 15.00\% & 0.00\% &  0.00\% &  0.00\% & 0.00\% &  0.76\% &  5.00\% \\
        \botrule
    \end{tabular}
\end{table}

Focusing only on \texttt{common} characters (i.e. named characters appearing in \textit{both} compared graphs) systematically improves the performance. This allows avoiding padding the smaller graphs with many isolates in order to get same-sized graphs, which makes the problem much easier. The \textit{Convex} method produces the best performance overall, with \textit{Concave} a close second. Again, the best match is between the \textit{Comics} and \textit{TV Show} networks. It seems that centring does not affect performance in our case, or even makes it worse, so we do not use it in the rest of our experiments.

Finally, the performance is even better when focusing only on the 20 most important characters (\texttt{top-20}). This could be due to the networks being more similar in terms of interconnections between these specific characters, or this could be just because of the smaller number of vertices.

\subsubsection{Adaptive Seeds}
\label{sec:ApdxCharCompMatchAdaptive}
As explained in the main article, the adaptive seeds approach is an iterative method consisting in first applying one of the previous algorithms as before, and selecting the first few best matches according to some heuristic measure. These are then used as \textit{hard} seeds in the next iteration (as if they were ground truth). By using an increasing number of seeds at each iteration, the matching is supposed to get better and better. Alternatively, it is possible to use the previous best matches as \textit{soft} seeds, which allow including some uncertainty by allowing multiple matching.

\begin{table}[htb!]
    \caption{Graph matching results obtained for all methods when using adaptive seeds, no attributes, and no centring. The table compares the use of \textit{hard} (top part) and \textit{soft} (bottom part) adaptive seeds. Performance is expressed as in Table~\ref{tab:MatchingResultsCentering}}
    \label{tab:MatchingResultsAdaptive}%
    \begin{tabular}{p{3cm}@{~}p{3cm} r@{~}r@{~}r r@{~}r@{~}r r@{~}r@{~}r}
        \toprule
        Method & Adaptive & \multicolumn{3}{c}{\textit{Novels} vs. \textit{Comics}} & \multicolumn{3}{c}{\textit{Novels} vs. \textit{TV Show}} & \multicolumn{3}{c}{\textit{Comics} vs. \textit{TV Show}} \\
         & Seed & \texttt{named} & \texttt{common} & \texttt{top-20} & \texttt{named} & \texttt{common} & \texttt{top-20} & \texttt{named} & \texttt{common} & \texttt{top-20} \\
        \midrule
        Convex      & Hard & 0.40\% &  6,16\% & \textbf{15,00\%} & \textbf{1.45\%} &  0,00\% &  5,00\% & 12.36\% & \textbf{34,85\%} & 10,00\% \\
        Indefinite  & Hard & 0.13\% &  0,00\% &  0,00\% & 0.13\% & 11,51\% &  0,00\% & 2.81\% &  4,55\% &  0,00\% \\
        Concave     & Hard & 1.32\% &  5,14\% & 10,00\% & 1.32\% & 10,79\% &  5,00\% & 14.89\% & 28,03\% & 10,00\% \\
        Percolation & Hard & \textbf{4.35\%} & \textbf{25,34\%} &  0,00\% & 0.00\% &  2,88\% &  0,00\% & 1.97\% &  0,76\% & \textbf{50,00\%} \\
        Umeyama     & Hard & 0.00\% &  0,00\% & \textbf{15,00\%} & 0.00\% &  0,00\% &  0,00\% & 0.00\% &  0,76\% &  5,00\% \\
        \midrule
        Convex      & Soft & 0.40\% &  1,71\% &  0,00\% & 0.79\% & \textbf{12,23\%} &  5,00\% & \textbf{15.73\%} & 30,30\% &  0,00\% \\
        Indefinite  & Soft & 0.26\% &  2,40\% &  0,00\% & 0.13\% &  5,76\% &  0,00\% & 1.12\% &  0,00\% &  0,00\% \\
        Concave     & Soft & 0.40\% &  2,74\% & 10,00\% & 0.92\% &  9,35\% & 15,00\% & 9.83\% & 21,97\% & 10,00\% \\
        Percolation & Soft & 0.40\% &  4,79\% & \textbf{15,00\%} & 0.66\% &  6,47\% & \textbf{25,00\%} & 3.09\% &  8,33\% & 15,00\% \\
        Umeyama     & Soft & 0.00\% &  0,00\% & \textbf{15,00\%} & 0.00\% &  0,00\% &  0,00\% & 0.00\% &  0,76\% &  5,00\% \\
        \botrule
    \end{tabular}
\end{table}

Table~\ref{tab:MatchingResultsAdaptive} shows the results obtained when applying the adaptive seeds method using all the previous algorithms. Considering \textit{hard} seeds, compared to the seedless approach, there is a clear improvement for the \textit{Percolation} algorithm when applied to the \textit{Novels} vs. \textit{Comics}, and for the \textit{Indefinite} algorithm when applied to the \textit{Comics} vs. \textit{TV Show}. No method dominates the others, as each best score for a network pair was obtained through a different algorithm. Using \textit{soft} seeds instead leads to lesser results, and the best methods are not the same. The best performance for the \textit{Novels} vs. \textit{Comics} networks, obtained with the \textit{Concave} method, increases, though.

\begin{figure}[htb!]
    \centering
    \includegraphics[width=0.32\textwidth]{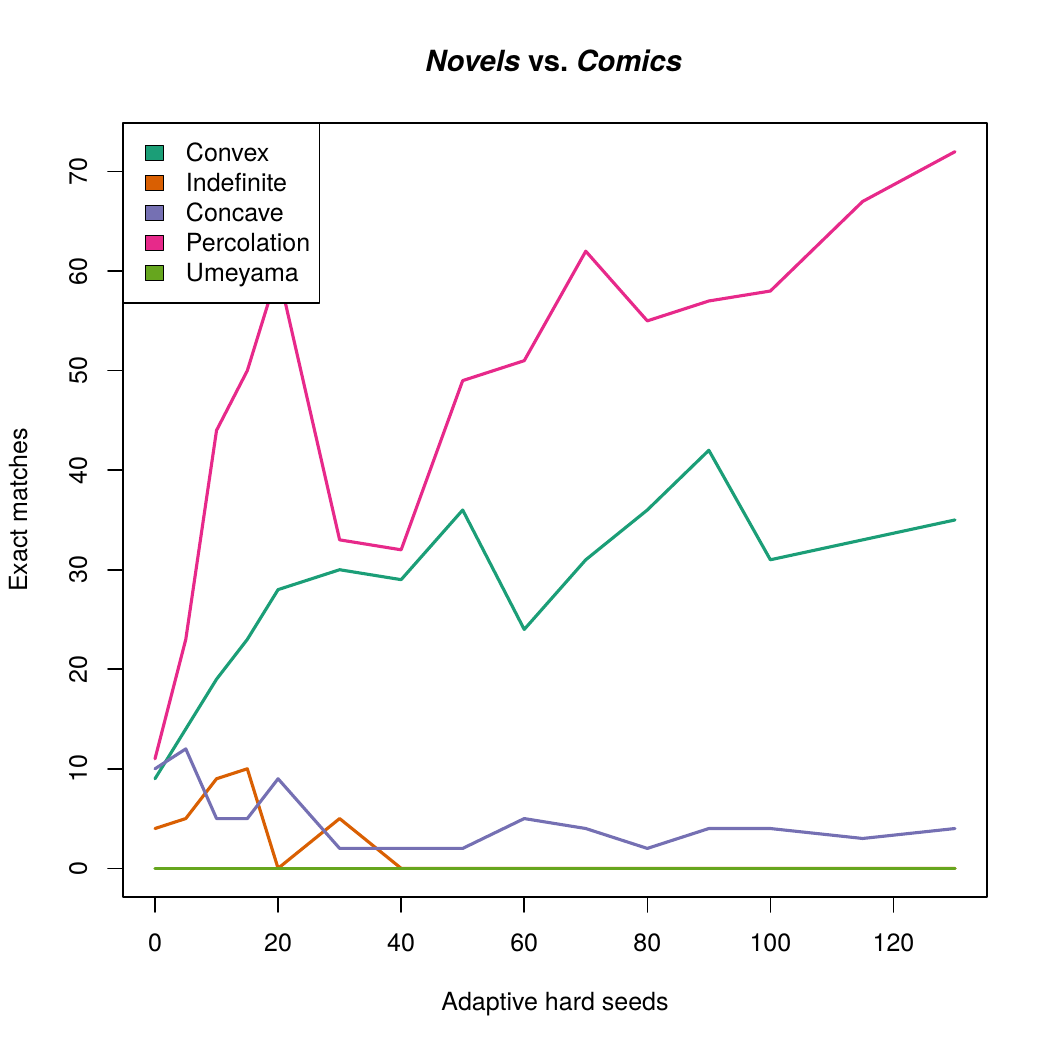}
    \hfill
    \includegraphics[width=0.32\textwidth]{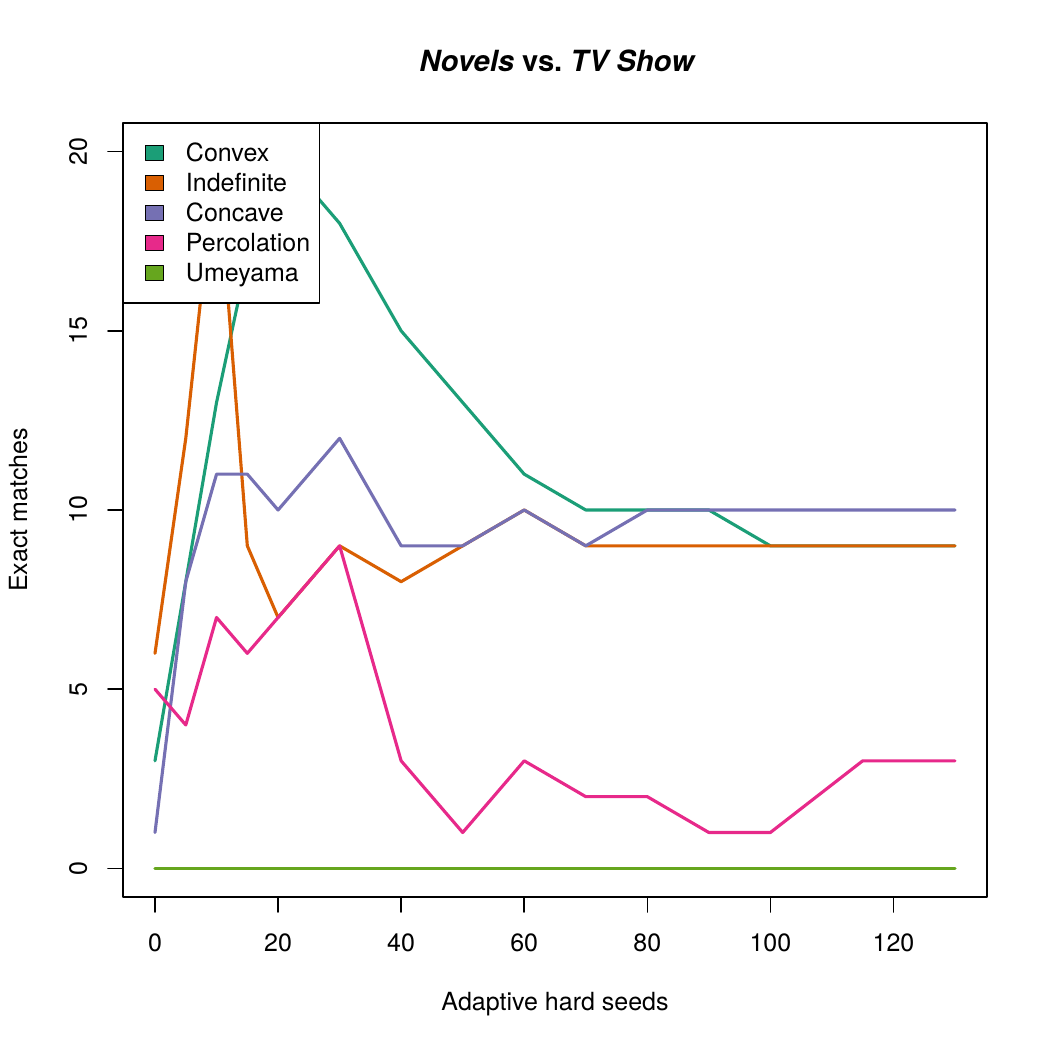}
    \hfill
    \includegraphics[width=0.32\textwidth]{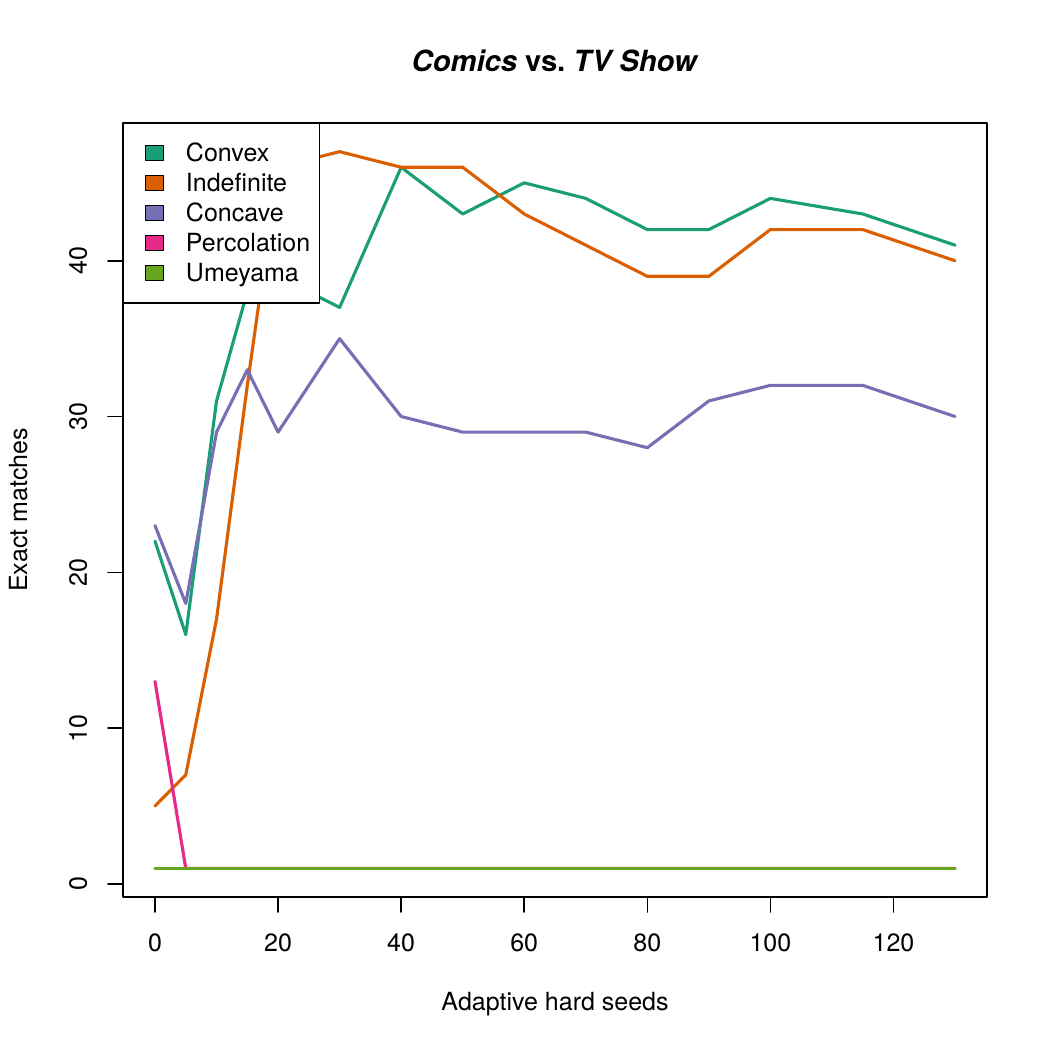}\\
    \includegraphics[width=0.32\textwidth]{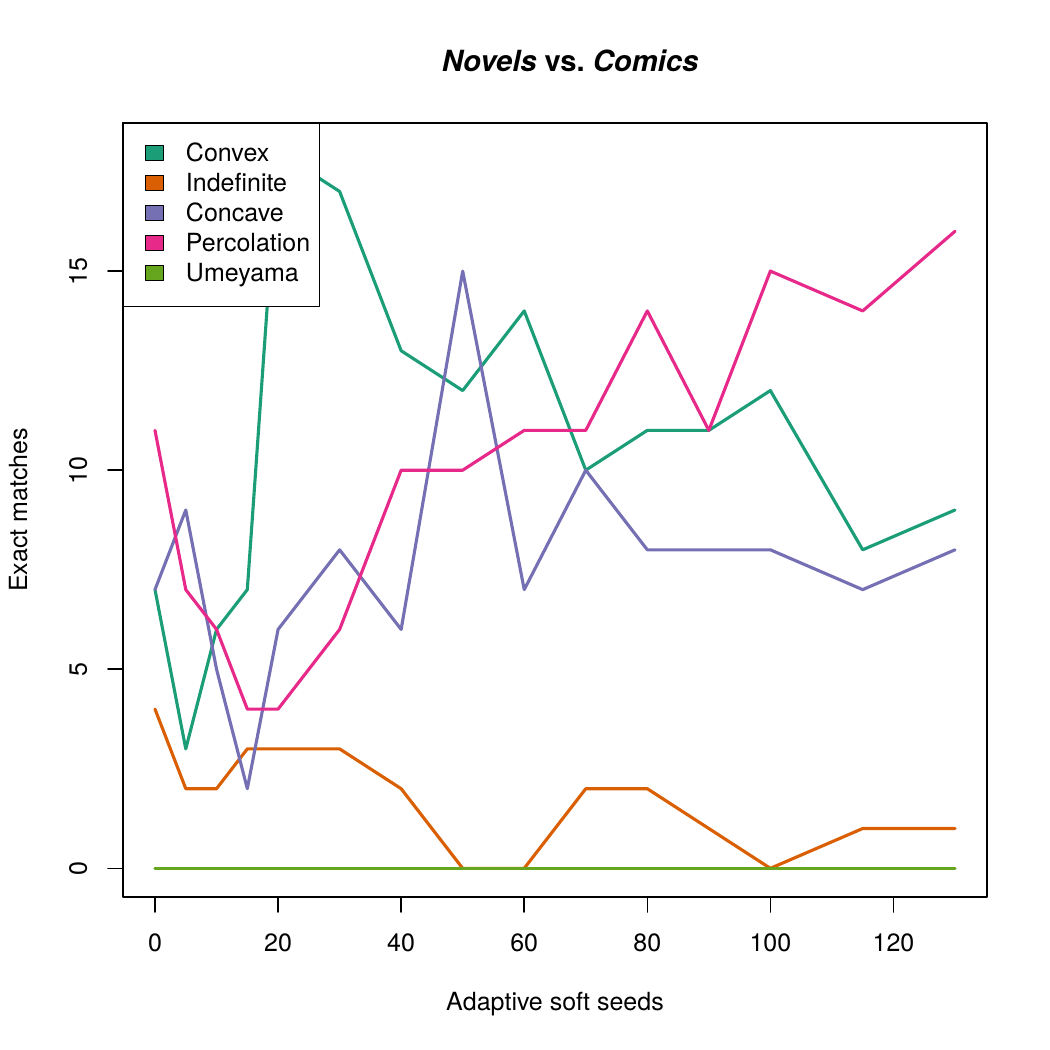}
    \hfill
    \includegraphics[width=0.32\textwidth]{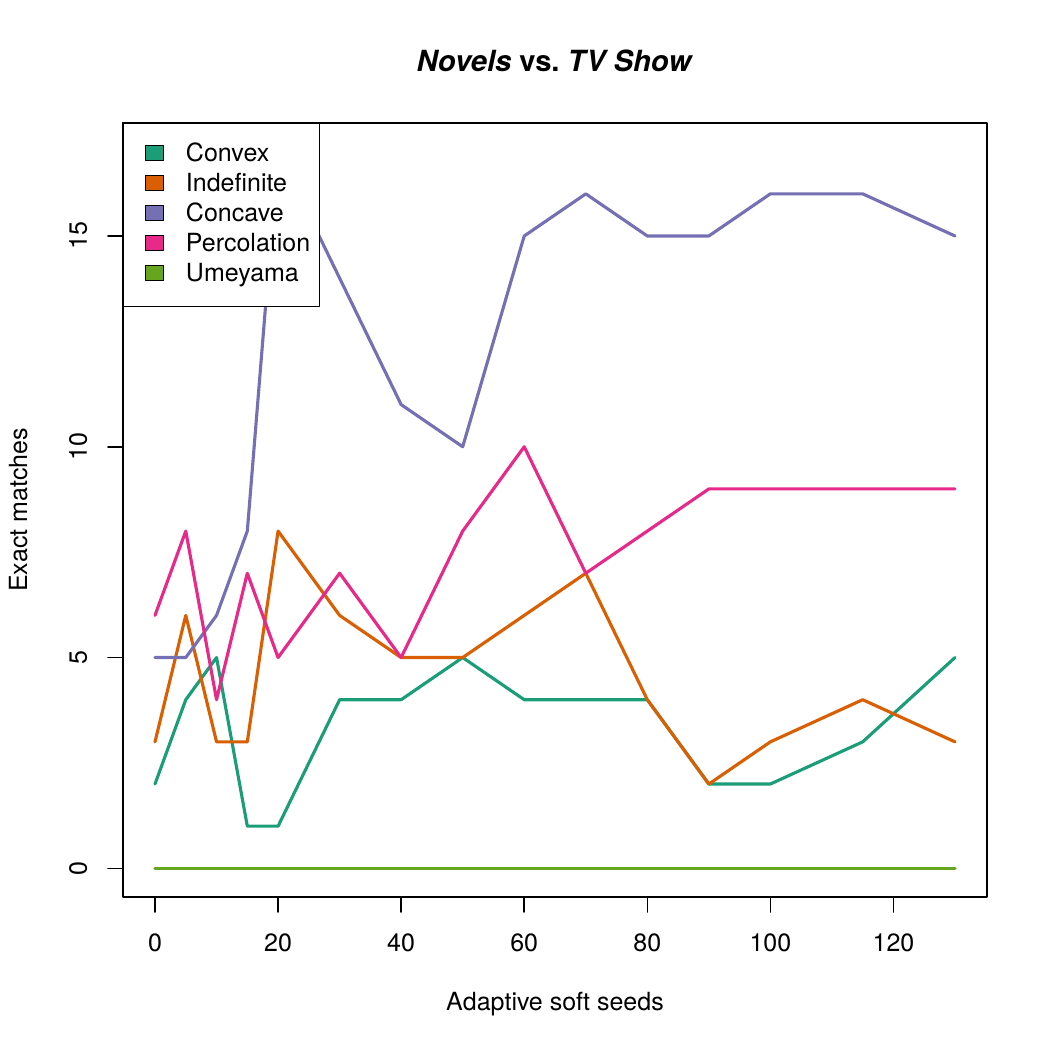}
    \hfill
    \includegraphics[width=0.32\textwidth]{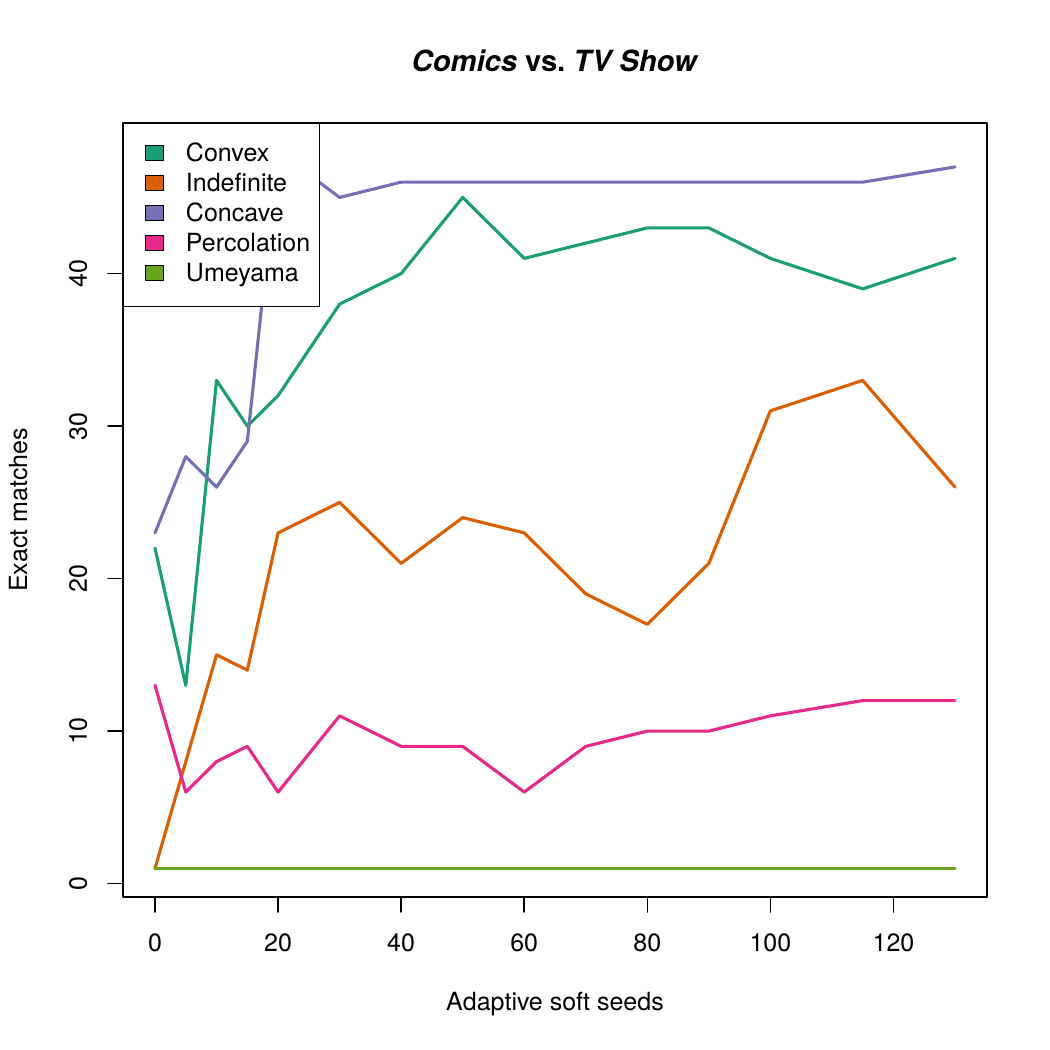}\\
    \caption{Evolution of the number of matches as a function of the number of \textit{hard} (top row) and \textit{soft} (bottom row) adaptive seeds, leading to the results shown in the top part of Table~\ref{tab:MatchingResultsAdaptive}}
    \label{fig:MatchingResultsAdaptive}
\end{figure}

Figure~\ref{fig:MatchingResultsAdaptive} shows how the number of exact matches evolves as a function of the number of adaptive seeds. The rightmost points in the plots correspond to the values from Table~\ref{tab:MatchingResultsAdaptive}. It shows that no algorithm dominates the others on more than one comparison. Moreover, the performance is not always improved by increasing the number of seeds: for instance, in the \textit{Novels} vs. \textit{TV Show} comparison, the \textit{Convex} algorithm correctly matches 20 characters when using 22 seeds, but its performance is only 9 with 130 seeds. 

We also experiment with what we call adaptive \textit{temporal} seeds, which takes time into account. Instead of iteratively applying the vertex matching algorithms using an increasing number of seeds on the same static network, we do so on the series of time slices constituting the dynamic networks, using the matches from the previous time slice as the seeds of the next one. We only focus on the \textit{Novels} vs. \textit{Comics} comparison, as they have the same temporal scale (chapters), and use the cumulative networks in order to be sure to find the previous best matched characters in the next time slice. We focus on the hard seeds, which seem to work better on the static networks. The method does not perform well for our dataset, with all scores almost zero (not shown here).

\subsubsection{Ground Truth Seeds}
\label{sec:ApdxCharCompMatchSeeds}
As explained in the main article, we relax the problem by leveraging hard seeds from the ground truth. We consider using the 5, 10 and 15 most important characters as seeds. Table~\ref{tab:MatchingResultsGT} presents the obtained performances.

\begin{table}[htb!]
    \caption{Graph matching results obtained for all methods when using ground truth seeds, no attributes, and no centring. Performance is expressed as in Tables~\ref{tab:MatchingResultsCentering} \&~\ref{tab:MatchingResultsAdaptive}}
    \label{tab:MatchingResultsGT}%
    \begin{tabular}{l@{~}r r@{~}r@{~}r r@{~}r@{~}r r@{~}r@{~}r}
        \toprule
        Method & Seeds & \multicolumn{3}{c}{\textit{Novels} vs. \textit{Comics}} & \multicolumn{3}{c}{\textit{Novels} vs. \textit{TV Show}} & \multicolumn{3}{c}{\textit{Comics} vs. \textit{TV Show}} \\
         &  & \texttt{named} & \texttt{common} & \texttt{top-20} & \texttt{named} & \texttt{common} & \texttt{top-20} & \texttt{named} & \texttt{common} & \texttt{top-20} \\
        \midrule
        Convex      &  5 & 3.99\% & 10.80\% &  60.00\% & 2.39\% & 20.15\% & 26.67\% & 12.36\% & 38.58\% &  86.67\% \\
        Indefinite  &  5 & 3.06\% &  7.67\% &  20.00\% & 1.99\% & 20.15\% & 13.33\% & 10.34\% & 36.22\% &  80.00\% \\
        Concave     &  5 & 4.26\% & 14.63\% &  60.00\% & 2.25\% & 20.90\% & 26.67\% & 12.07\% & 35.43\% &  86.67\% \\
        Percolation &  5 & 0.53\% &  8.01\% &  26.67\% & 1.06\% &  5.22\% & 40.00\% &  4.31\% & 18.90\% &  46.67\% \\
        Umeyama     &  5 & 0.00\% &  0.00\% &  13.33\% & 0.00\% &  0.00\% &  0.00\% &  0.00\% &  1.57\% &   0.00\% \\
        \midrule
        Convex      & 10 & 6.43\% & 20.92\% &  80.00\% & 2.40\% & 27.13\% & \textbf{60.00\%} & 10.79\% & 38.52\% & \textbf{100.00\%} \\
        Indefinite  & 10 & 5.09\% & 12.06\% &  40.00\% & 2.27\% & 27.91\% & \textbf{60.00\%} &  9.91\% & \textbf{42.62\%} & \textbf{100.00\%} \\
        Concave     & 10 & 6.29\% & 23.40\% &  80.00\% & 2.54\% & 27.13\% & \textbf{60.00\%} & 11.66\% & 40.16\% & \textbf{100.00\%} \\
        Percolation & 10 & 2.01\% &  9.93\% &  60.00\% & 0.53\% & 11.63\% & 40.00\% &  6.71\% & 20.49\% &  50.00\% \\
        Umeyama     & 10 & 0.00\% &  0.00\% &  10.00\% & 0.00\% &  0.00\% & 10.00\% &  0.00\% &  0.82\% &  20.00\% \\
        \midrule
        Convex      & 15 & \textbf{7.01\%} & 22.02\% & \textbf{100.00\%} & 3.49\% & \textbf{33.87\%} & \textbf{60.00\%} & 10.36\% & 35.90\% & \textbf{100.00\%} \\
        Indefinite  & 15 & 6.87\% & 24.55\% & \textbf{100.00\%} & 2.28\% & 29.84\% & \textbf{60.00\%} &  8.28\% & 39.32\% & \textbf{100.00\%} \\
        Concave     & 15 & \textbf{7.01\%} & \textbf{26.35\%} & \textbf{100.00\%} & \textbf{3.76\%} & 33.06\% & \textbf{60.00\%} & \textbf{12.95\%} & 39.32\% & \textbf{100.00\%} \\
        Percolation & 15 & 5.39\% & 18.41\% &  60.00\% & 1.48\% & 13.71\% & \textbf{60.00\%} &  5.62\% & 17.09\% &  60.00\% \\
        Umeyama     & 15 & 0.00\% &  0.00\% &  40.00\% & 0.00\% &  0.00\% & \textbf{60.00\%} &  0.00\% &  0.85\% &  20.00\% \\
        \botrule
    \end{tabular}
\end{table}

Increasing the number of seeds increases the performance, for (almost) all network pairs and all matching methods. It is worth noting that \textit{Comics} vs. \textit{TV Show}, the network pair whose performance is the best without seeds, gets the smallest improvement when increasing the seed number. As observed before, the performance is systematically the lowest for the \texttt{named} character set, and the higher for the \texttt{top-20} character set.

\subsubsection{Vertex Attributes}
\label{sec:ApdxCharCompMatchAttr}
We relax the problem differently by leveraging some vertex attributes to help the methods matching the characters. The algorithms implemented in \texttt{iGraphMatch} take an optional vertex similarity matrix as input, which can be used to leverage additional knowledge. In particular, it makes it possible to indirectly consider vertex attributes. A simple way of doing so in the case of categorical attributes is to put a 1 if the characters have the same attribute value, and 0 otherwise. We experiment with attributes sex, character affiliation, and both.

\begin{table}[htb!]
    \caption{Graph matching results obtained for all methods when using vertex attributes, no seeds, and no centring. Performance is expressed as in Tables~\ref{tab:MatchingResultsCentering}--\ref{tab:MatchingResultsGT}}
    \label{tab:MatchingResultsAttr}%
    \begin{tabular}{l@{~}l r@{~}r@{~}r r@{~}r@{~}r r@{~}r@{~}r}
        \toprule
        Method & Attr. & \multicolumn{3}{c}{\textit{Novels} vs. \textit{Comics}} & \multicolumn{3}{c}{\textit{Novels} vs. \textit{TV Show}} & \multicolumn{3}{c}{\textit{Comics} vs. \textit{TV Show}} \\
         &  & \texttt{Named} & \texttt{Common} & \texttt{Top-20} & \texttt{named} & \texttt{common} & \texttt{top-20} & \texttt{named} & \texttt{common} & \texttt{top-20} \\
        \midrule
        Convex      &  Sex &  0.13\% &  2.74\% &  5.00\% & 0.13\% &  0.72\% &  0.00\% &  0.28\% &  3.79\% &  10.00\% \\
        Indefinite  &  Sex &  0.00\% &  3.77\% & 10.00\% & 0.13\% &  5.76\% &  5.00\% &  0.28\% &  5.30\% &  80.00\% \\
        Concave     &  Sex &  0.00\% &  2.05\% & 15.00\% & 0.00\% &  2.16\% &  0.00\% &  0.00\% &  3.03\% &  10.00\% \\
        Percolation &  Sex &  0.40\% &  3.42\% & 10.00\% & 0.66\% &  7.91\% & 15.00\% &  0.85\% & 16.67\% &  35.00\% \\
        Umeyama     &  Sex &  0.13\% &  0.00\% & 15.00\% & 0.00\% &  0.72\% &  0.00\% &  0.00\% &  2.27\% &   5.00\% \\
        \midrule
        Convex      & Aff. &  0.00\% &  0.00\% &  0.00\% & 0.00\% &  0.00\% &  5.00\% &  0.00\% &  0.00\% &   0.00\% \\
        Indefinite  & Aff. &  9.51\% & 65.75\% & 60.00\% & 3.43\% & 63.31\% & 40.00\% &  9.07\% & 81.06\% &  80.00\% \\
        Concave     & Aff. &  0.00\% &  0.00\% &  0.00\% & 0.00\% &  0.72\% &  0.00\% &  0.00\% &  0.00\% &  10.00\% \\
        Percolation & Aff. & 14.93\% & 57.88\% & 15.00\% & 6.32\% & \textbf{64.75\%} & 20.00\% & 15.58\% & 58.33\% &  80.00\% \\
        Umeyama     & Aff. &  0.13\% &  1.03\% &  5.00\% & 0.13\% &  3.60\% &  0.00\% &  0.00\% &  0.76\% &   5.00\% \\
        \midrule
        Convex      & Both &  0.00\% &  0.00\% &  0.00\% & 0.00\% &  0.00\% &  0.00\% &  0.00\% &  0.00\% &   0.00\% \\
        Indefinite  & Both &  9.51\% & \textbf{73.97\%} & \textbf{65.00\%} & 3.43\% & \textbf{64.75\%} & \textbf{55.00\%} &  8.50\% & \textbf{83.33\%} & \textbf{100.00\%} \\
        Concave     & Both &  0.00\% &  0.00\% &  0.00\% & 0.00\% &  0.00\% &  0.00\% &  0.00\% &  0.00\% &  10.00\% \\
        Percolation & Both & \textbf{16.12\%} & 56.85\% & 10.00\% & \textbf{6.46\%} & 54.68\% & 20.00\% & \textbf{15.86\%} & 61.36\% &  80.00\% \\
        Umeyama     & Both &  0.13\% &  0.68\% &  5.00\% & 0.13\% &  3.60\% &  0.00\% &  0.00\% &  0.76\% &   5.00\% \\
        \botrule
    \end{tabular}
\end{table}

Table~\ref{tab:MatchingResultsAttr} shows the performance obtained without any seed or centring. Leveraging the attributes clearly improves the results for the \textit{Indefinite}, \textit{Concave} and \textit{Percolation} algorithms, especially the former. The effect is much stronger for affiliation than for sex, but the latter still helps to improve the results a little bit. As before, focusing only on the characters which are common to both compared networks improves performance, and even more so for the \texttt{top-20} character set.

Interestingly, using ground truth seeds in addition to attributes only marginally improves the results (not shown here). This suggests that both types of additional information are equally relevant for the character matching task.

\subsubsection{Vertex Similarity}
\label{sec:ApdxCharCompMatchSim}
\begin{figure}[htb!]
    \centering
    \includegraphics[width=0.32\textwidth]{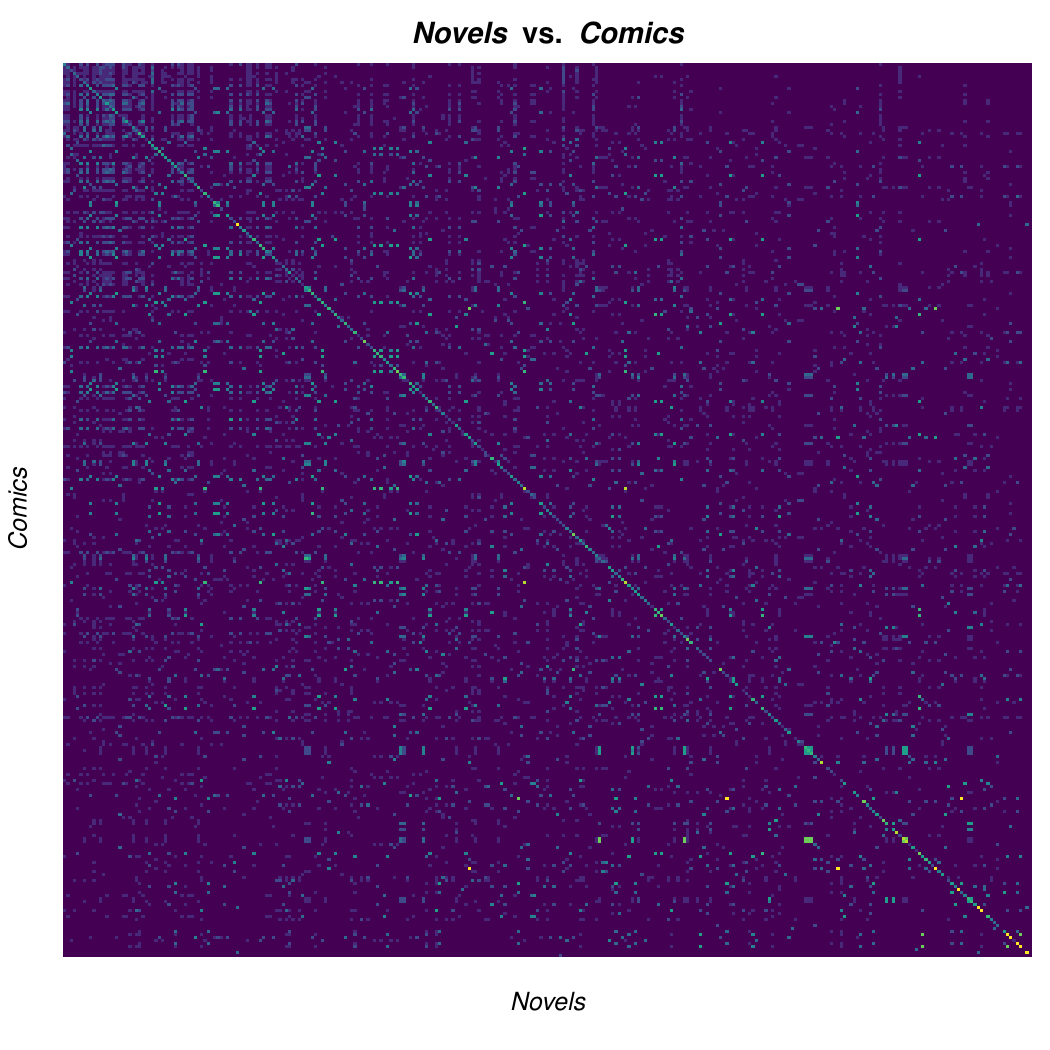}
    \hfill
    \includegraphics[width=0.32\textwidth]{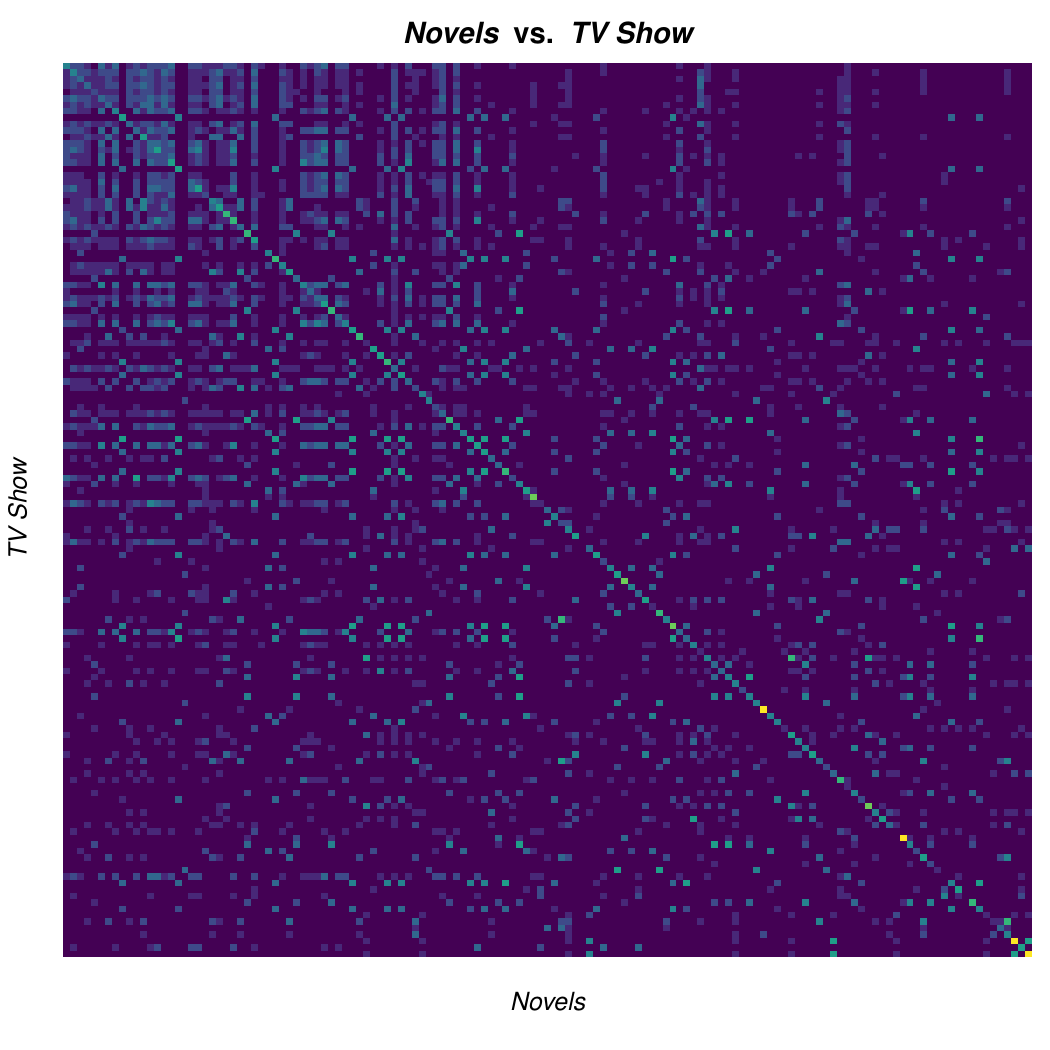}
    \hfill
    \includegraphics[width=0.32\textwidth]{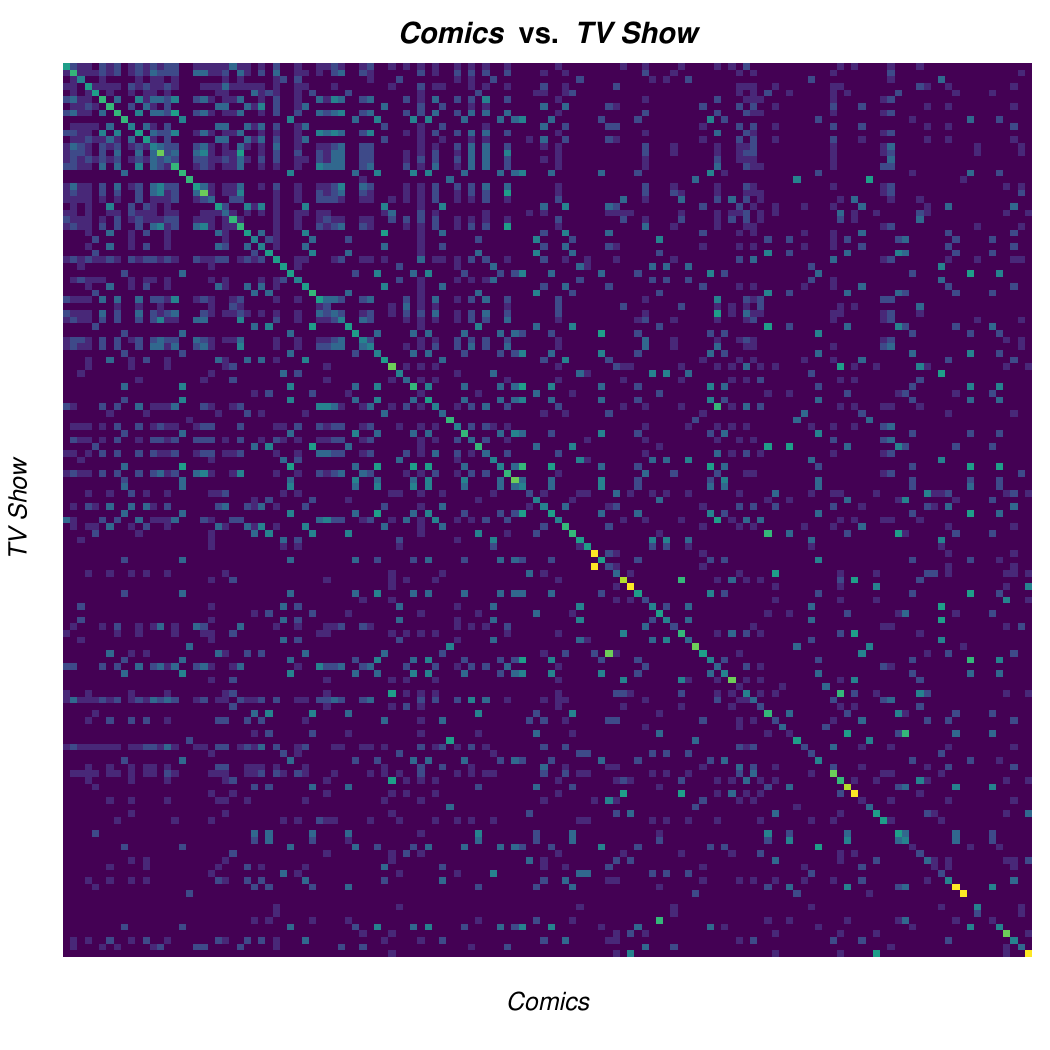}
    \caption{Similarity matrices obtained with Ru\v{z}i\v{c}ka's similarity, for the \texttt{common} character set and period \texttt{U2}. By comparison, in Figure~\ref{fig:MatchingResultsSimTop}, the main article focuses on \texttt{top-20}, i.e. the 20 most important characters}
    \label{fig:MatchingResultsSimAll}
\end{figure}

As in the main article, we use Ru\v{z}i\v{c}ka's similarity to measure the similarity between two characters based on their respective neighborhoods. Figure~\ref{fig:MatchingResultsSimAll} shows the similarity matrices obtained for all characters, when focusing only on the \texttt{common} character set. Each matrix corresponds to a specific pair of adaptations. It is important to stress that these matrices are not symmetric, as the compared pairs of characters belong to two distinct networks. Therefore, the similarity between a character $v_1$ in the novels and a character $v_2$ in the comics is not necessarily equal to that between $v_2$ in the novels and $v_1$ in the comics. 

In all three matrices, the diagonal appears clearly, i.e. the similarity between both instances of the same character is high. However, a number of off-diagonal cells are also highlighted, which means that there are potential mismatches. It is even more the case when considering all \texttt{named} characters (not shown here). The main article provides similar matrices focusing on \texttt{top-20}, the 20 most important characters (Figure~\ref{fig:MatchingResultsSimTop}), showing that the number of off-diagonal high values (i.e. mismatches) appears lower for this subset of characters.

\begin{figure}[htb!]
    \centering
    \includegraphics[width=0.32\textwidth]{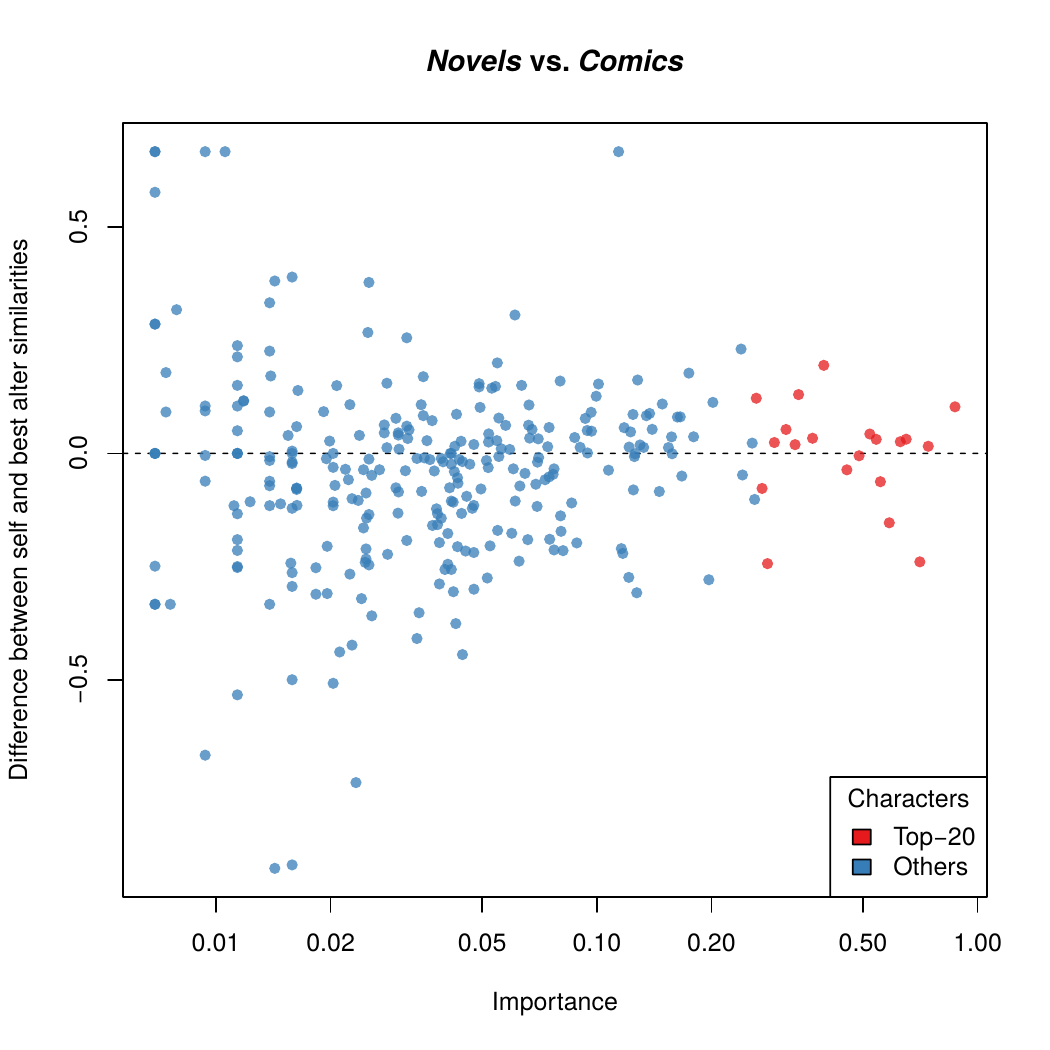}
    \hfill
    \includegraphics[width=0.32\textwidth]{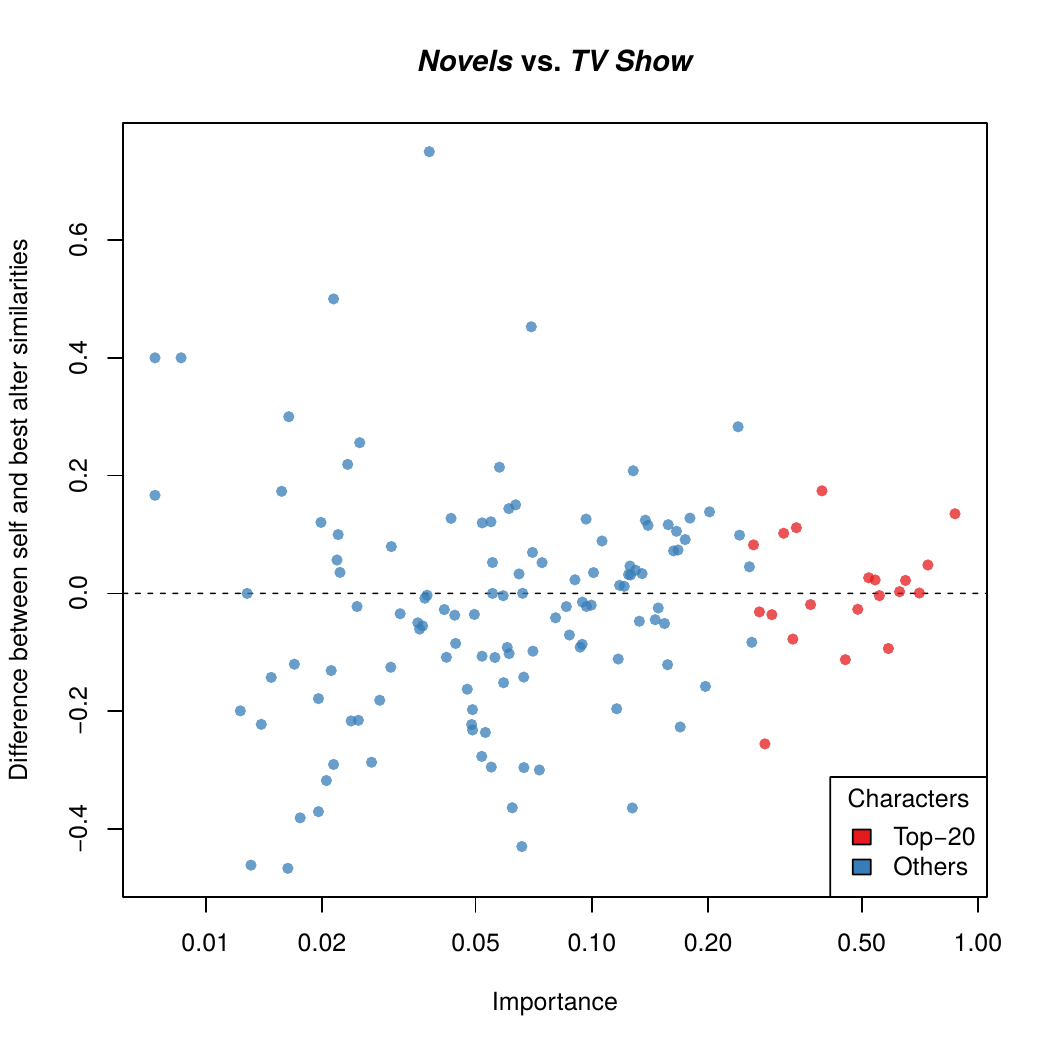}
    \hfill
    \includegraphics[width=0.32\textwidth]{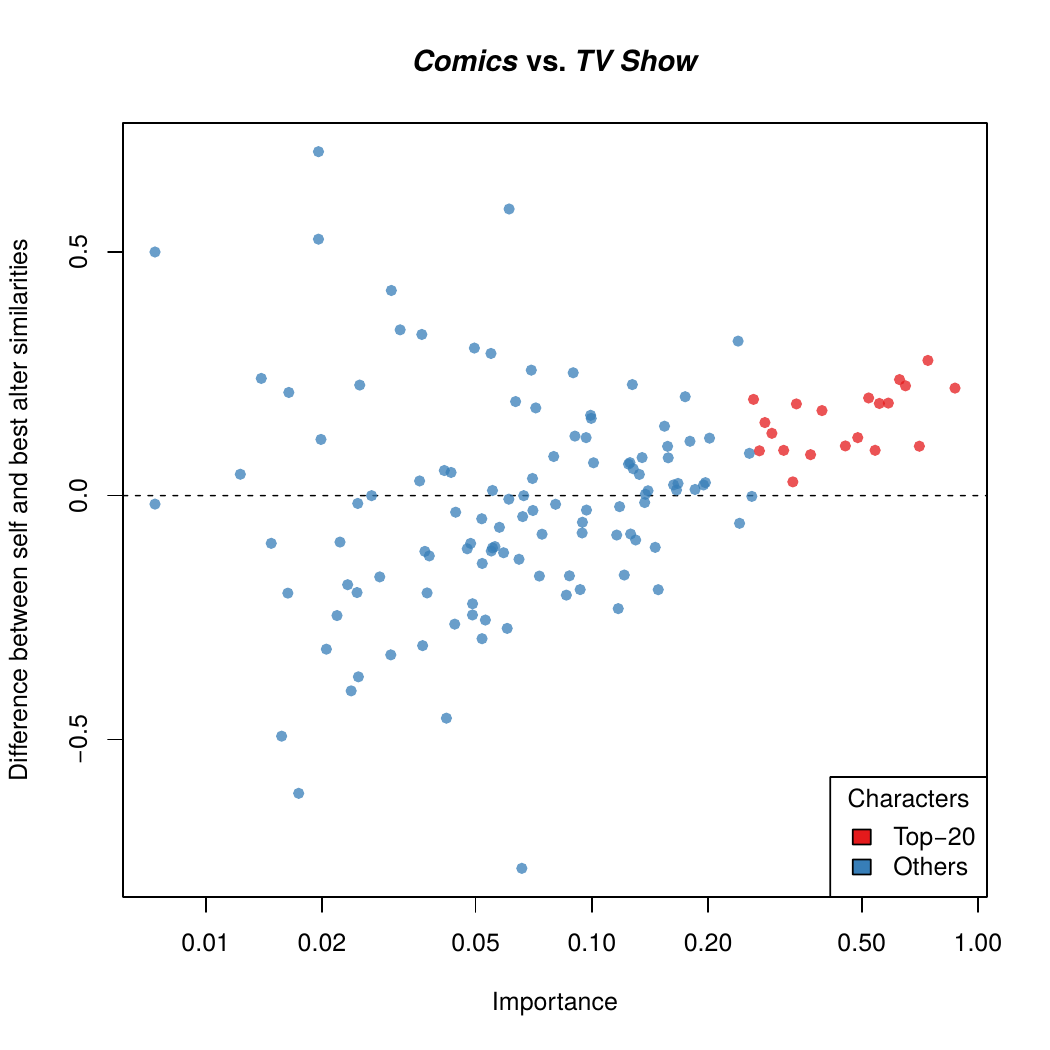}
    \caption{Similarity difference between, on one side, a character and itself, and, on the other side, the same character and the most similar alternate character, as a function of character importance. The considered character set is \texttt{common}. Notice the logarithmic scale on the $y$-axis}
    \label{fig:MatchingResultsSimVsImp}
\end{figure}

We call \textit{self-similarity} the similarity between a character in a given network, and the same character in a different network. The \textit{best alter-similarity} is the similarity between the same character and the best alternate character in the second network. Consequently, the difference between these quantities indicates whether the match is correct (positive value, i.e. the best match is the same character) or not (negative value, i.e. the best match is another character). Figure~\ref{fig:MatchingResultsSimVsImp} shows this difference as a function of character importance. The most important characters are located on the right side of the plots, the 20 most important characters being represented in red. These \texttt{top-20} characters tend to be located in the positive half (i.e. they are correctly matched), especially for the \textit{Comics} vs. \textit{TV Show} comparison. The correlation between these variables is intermediary (0.36) but significant ($p<.001$) for \textit{Comic} vs. \textit{TV Show}, according to Spearman's coefficient. On the contrary, there is no significant correlation for both other pairs of adaptations (\textit{Novels} vs. \textit{Comics}, and \textit{Novels} vs. \textit{TV Show}).

\subsubsection{Characters of Interest}
\label{sec:ApdxCharCompMatchChars}
Based on the fan Wikis\footref{ftn:awoiaf}\textsuperscript{,}\footref{ftn:wow} mentioned in Section~\ref{sec:DataAttr} of the main article, we elaborate two list of characters of particular interest~\cite{AWoIaF2022, WoW2023, WoW2023a}. The first one contains \textit{pairs} of characters that have a different name in the novels and TV show, but are known to correspond to the same individual, sometimes with various additional differences due to the adaptation process. Unlike the \texttt{named} character set, \texttt{common} and \texttt{top-20} do not contain the pair of characters of interest, by construction. For this reason, in the first case the computation of character similarity is exactly as described in Section~\ref{sec:CharCompMatchSim} of the main article, but the two other character sets require adjusting the procedure. This is done by forcing the inclusion of the listed characters of interest in \texttt{common} and \texttt{top-20}.

\begin{table}[htb!]
    \caption{Similarity difference between, on one side, pairs of characters of interest from the Novels and TV Show, and, on the other side, the best alternative. Positive values correspond to correct matches}
    \label{tab:SpecCharSimSubst}%
    \begin{tabular}{p{3cm} p{3cm} p{0.2cm} r r r p{0.2cm} r r r}
        \toprule
        \textit{Novels} & \textit{TV Show} & & \multicolumn{3}{r}{Period \texttt{U2}} & & \multicolumn{3}{r}{Period \texttt{U5}} \\
        Character & Character & & \texttt{named} & \texttt{common} & \texttt{top-20} & & \texttt{named} & \texttt{common} & \texttt{top-20} \\
        \cmidrule{1-2}\cmidrule{4-6}\cmidrule{8-10}
        Jeyne Westerling & Talisa Stark && - & - & - && -0.04 & -0.21 & 0.55 \\
        Vargo Hoat & Locke && - & - & - && 0.01 & 0.01 & 0.10 \\
        Cleos Frey & Alton Lannister && 0.02 & 0.03 & 0.12 && -0.03 & -0.06 & 0.05 \\
        Asha Greyjoy & Yara Greyjoy && 0.06 & 0.21 & 0.84 && 0.04 & 0.19 & 0.45 \\
        Robert Arryn & Robin Arryn && 0.02 & -0.15 & -0.10 && 0.06 & 0.09 & 0.08 \\
        Jhogo & Kovarro && -0.13 & -0.17 & 0.48 && -0.14 & -0.18 & 0.53 \\
        Grazdan mo Eraz & Razdal mo Eraz && - & - & - && -0.15 & 0.15 & 0.21 \\
        Grazdan mo Ullhor & Greizhen mo Ullhor && - & - & - && -0.04 & 0.01 & 0.21 \\
        Stalwart Shield & White Rat && - & - & - && 0.00 & -0.13 & 0.00 \\
        \botrule
    \end{tabular}
\end{table}

Table~\ref{tab:SpecCharSimSubst} shows the result of the matching process based on Ru\v{z}i\v{c}ka's similarity. Each row corresponds to a pair of characters, and shows the difference between, on one side, the similarity between the characters of the considered pair, and, on the other side, the best alternative. Consequently, positive values correspond to correct matches. Some characters only enter the story after the second novel or season, which is why some values are missing. As remarked before, we obtain better performance when focusing on narrower character sets. Note that in the present case, \texttt{top-20} means that we compare the two character of interest based only on the relationships with the 20 most important characters. In the case of period \texttt{U5}, it is worth stressing that all matches are correct when using \texttt{top-20}.

\begin{table}[htb!]
    \caption{Similarity difference between, on one side, pairs of characters of interest from the Novels and TV Show, and, on the other side, the best alternative. Positive values correspond to correct matches}
    \label{tab:SpecCharSimMerge}%
    \begin{tabular}{p{2cm} p{2cm} p{2cm} p{0.2cm} r r r p{0.2cm} r r r}
        \toprule
        \multicolumn{2}{l}{\textit{Novels}} & \textit{TV Show} & & \multicolumn{3}{r}{Period \texttt{U2}} & & \multicolumn{3}{r}{Period \texttt{U5}} \\
        Character 1 & Character 2 & Character & & \texttt{named} & \texttt{common} & \texttt{top-20} & & \texttt{named} & \texttt{common} & \texttt{top-20} \\
        \cmidrule{1-3}\cmidrule{5-7}\cmidrule{9-11}
        Dirk & Clubfoot Karl & Karl && - & - & - && -0.07 & -0.05 & 0.05 \\
        Gendry & Edric Storm & Gendry && 0.08 & 0.17 & 0.12 && 0.07 & 0.18 & 0.21 \\
        \botrule
    \end{tabular}
\end{table}

The second list contains \textit{triples} of characters: two from the novels and one from the TV show. The TV Show character is known to result from the merging of both novels characters. In certain cases, the TV show character bears the same name as one of the novels characters: this means that the TV Show character corresponds mainly to their homonym from the novels, but also includes a significant amount of the remaining character's characteristics and/or plot. We originally identified 7 such triples, but only the two shown in Table~\ref{tab:SpecCharSimMerge} are usable: some of the concerned characters do not appear at all in our annotations (as already mentioned, these are relatively minor characters). Like before, it appears that the character set has a strong influence on the result, and both cases are successfully matched for period \texttt{U5} using \texttt{top-20}.

\subsection{Centrality Analysis}
\label{sec:ApdxCharCompCentr}
This section aims at complementing Section~\ref{sec:CharCompCentr} from the main article, by providing additional results related to character centrality.

\subsubsection{Centrality Correlation}
\label{sec:ApdxCharCompCentrCorr}
\begin{figure}[htb!]
    \centering
    \includegraphics[width=0.32\textwidth]{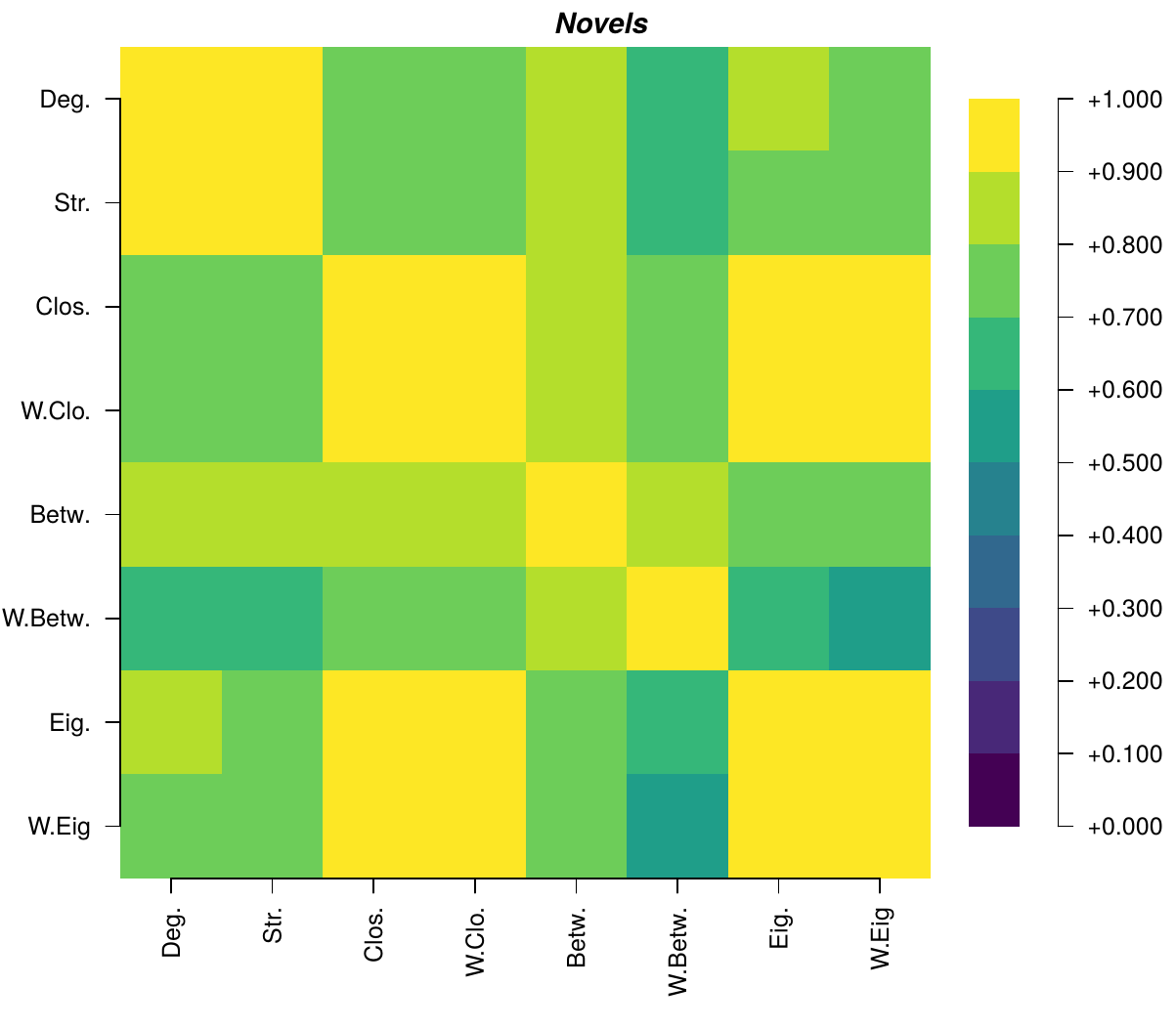}
    \hfill
    \includegraphics[width=0.32\textwidth]{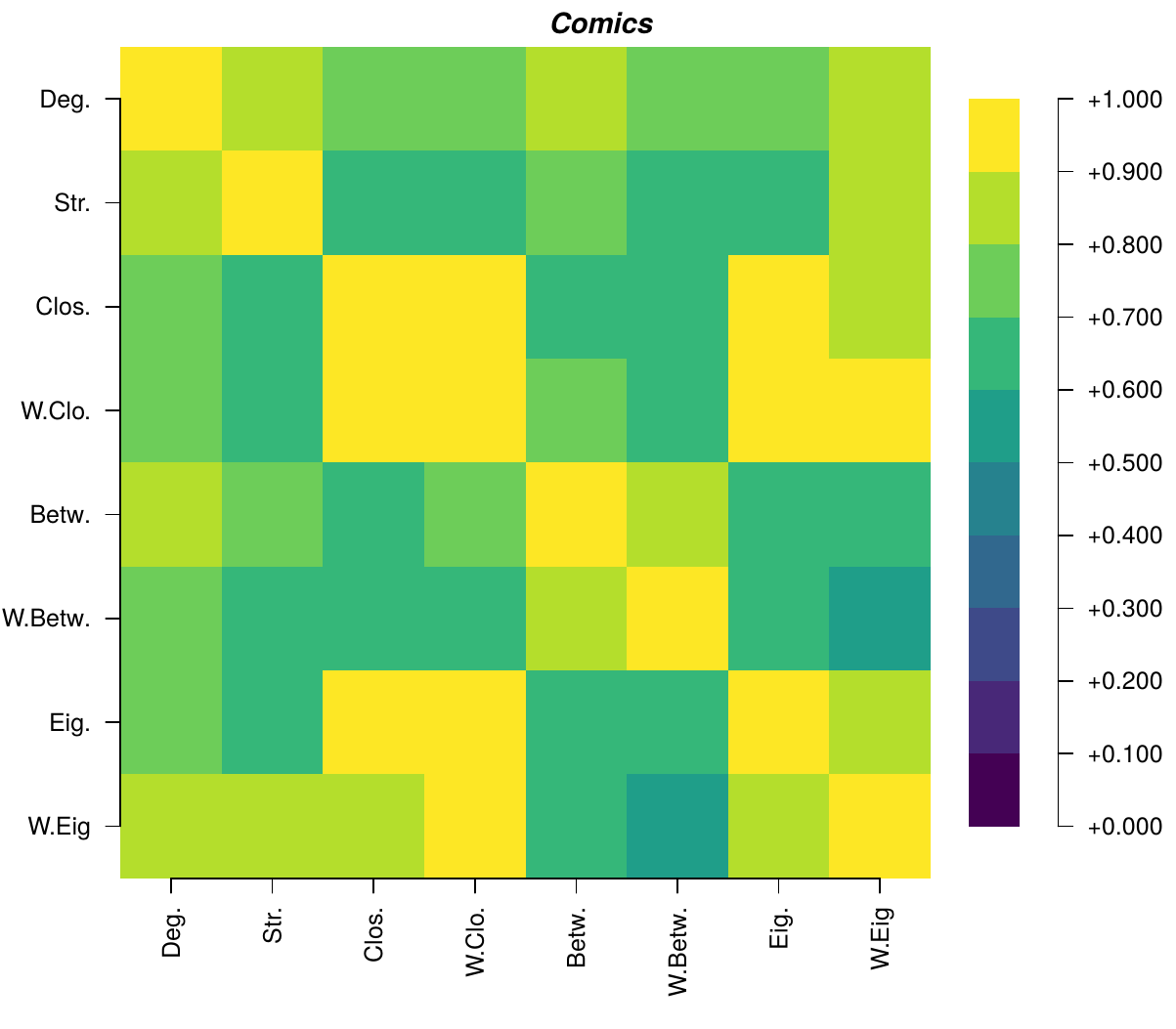}
    \hfill
    \includegraphics[width=0.32\textwidth]{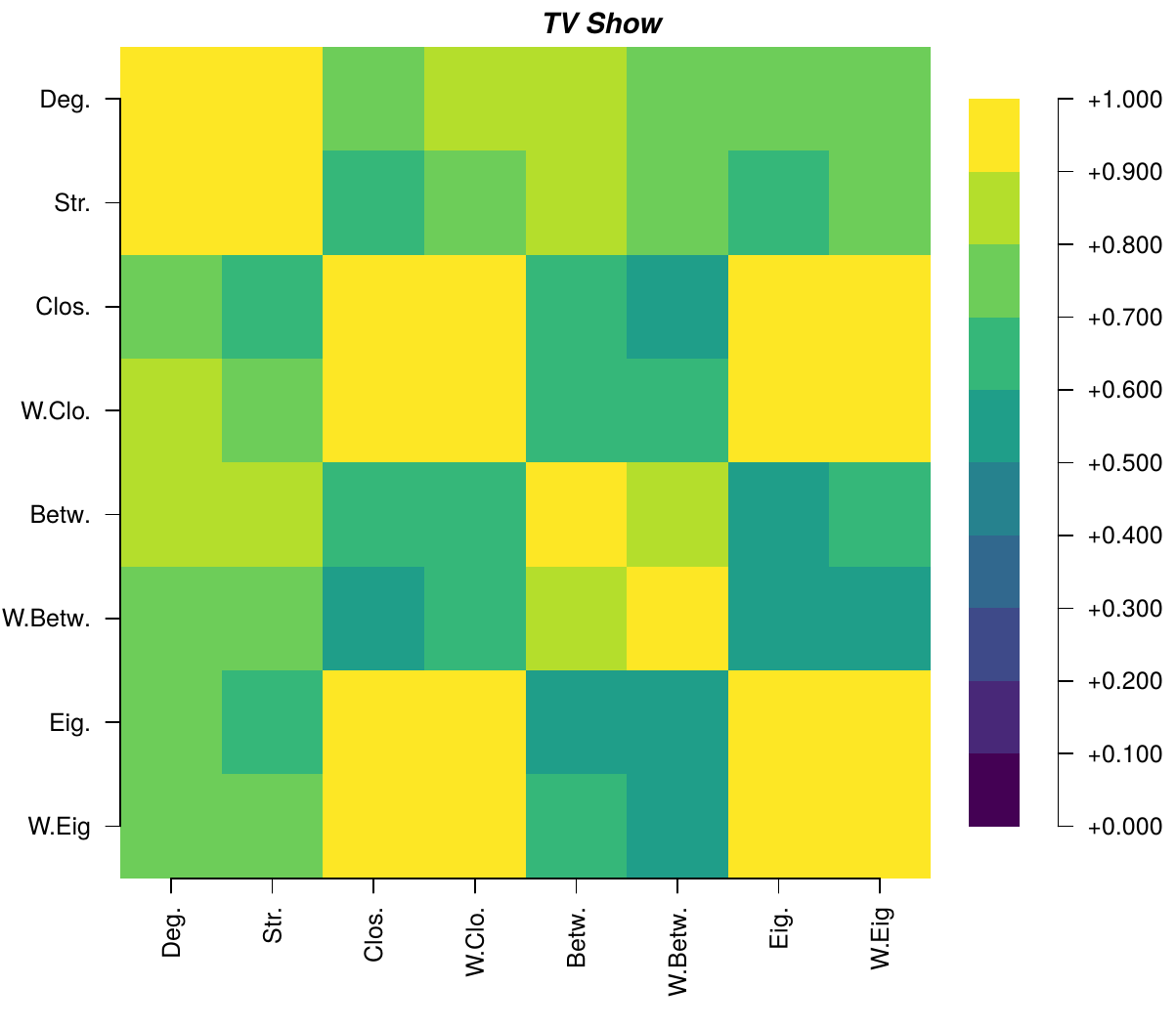}
    \caption{Spearman's correlation between the selected centrality metrics, for each of the three adaptations, considering the \texttt{common} character set. By comparison, Figure~\ref{fig:VtxMatch_CentrNamedCorr} from the main article shows the matrices for \texttt{named} and \texttt{top-20} character sets}
    \label{fig:VtxMatch_CentrNamedCorr_Comm}
\end{figure}

Figure~\ref{fig:VtxMatch_CentrNamedCorr} from the main article shows only the centrality metric correlations for the \texttt{named} and \texttt{top-20} character sets. Figure~\ref{fig:VtxMatch_CentrNamedCorr_Comm} provides the same view for \texttt{common} characters. For all adaptations, we get results very similar to those obtained for the TV show in Figure~\ref{fig:VtxMatch_CentrNamedCorr}. This is because the TV show annotations contain much fewer unnamed characters, and the plot is much more focused on the core characters present in all three adaptations. Consequently, using only common characters does not change much for this adaptation, whereas it makes both other more similar to the TV show.

\begin{figure}[htb!]
    \centering
    \includegraphics[width=0.32\textwidth]{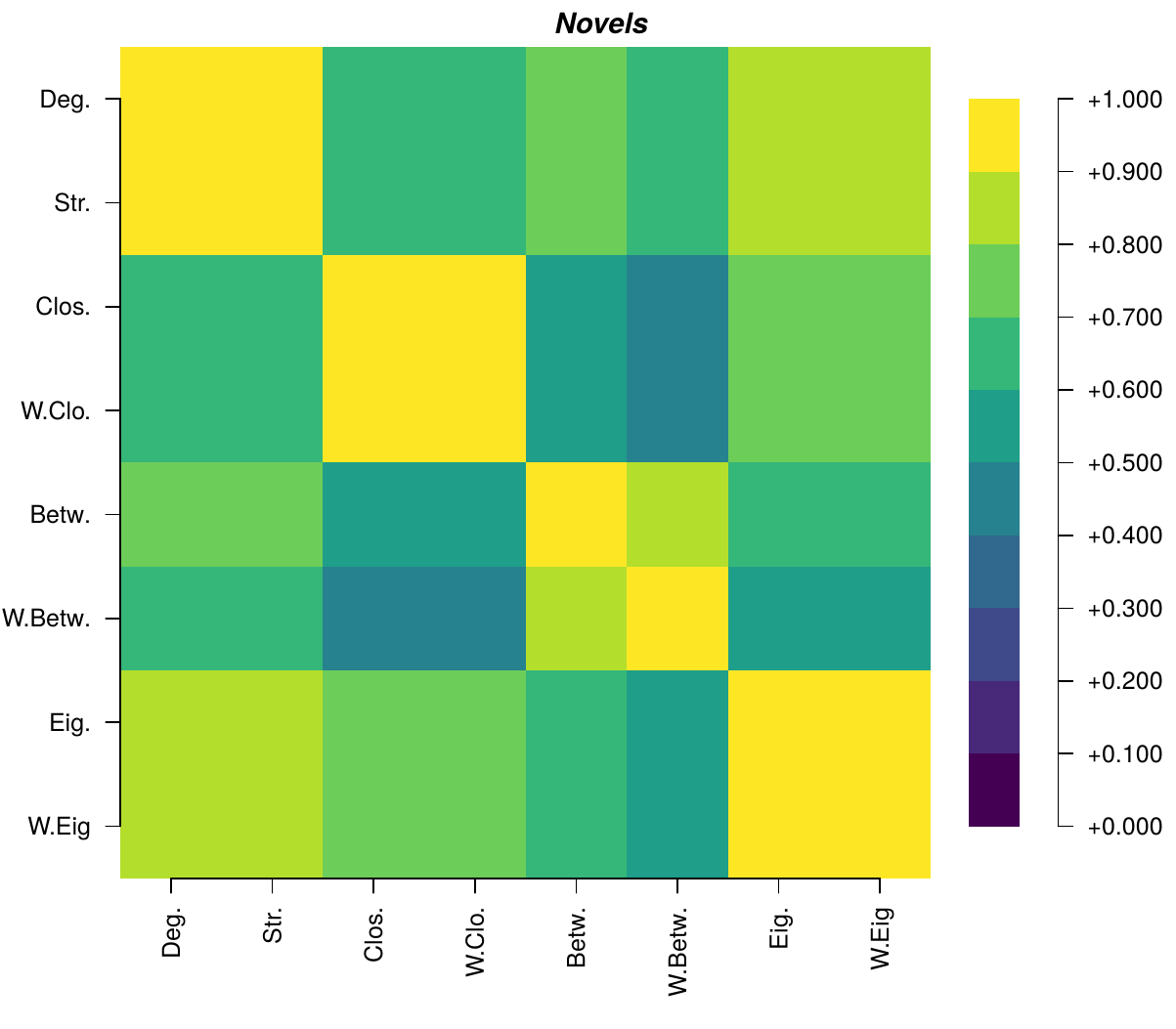}
    \hfill
    \includegraphics[width=0.32\textwidth]{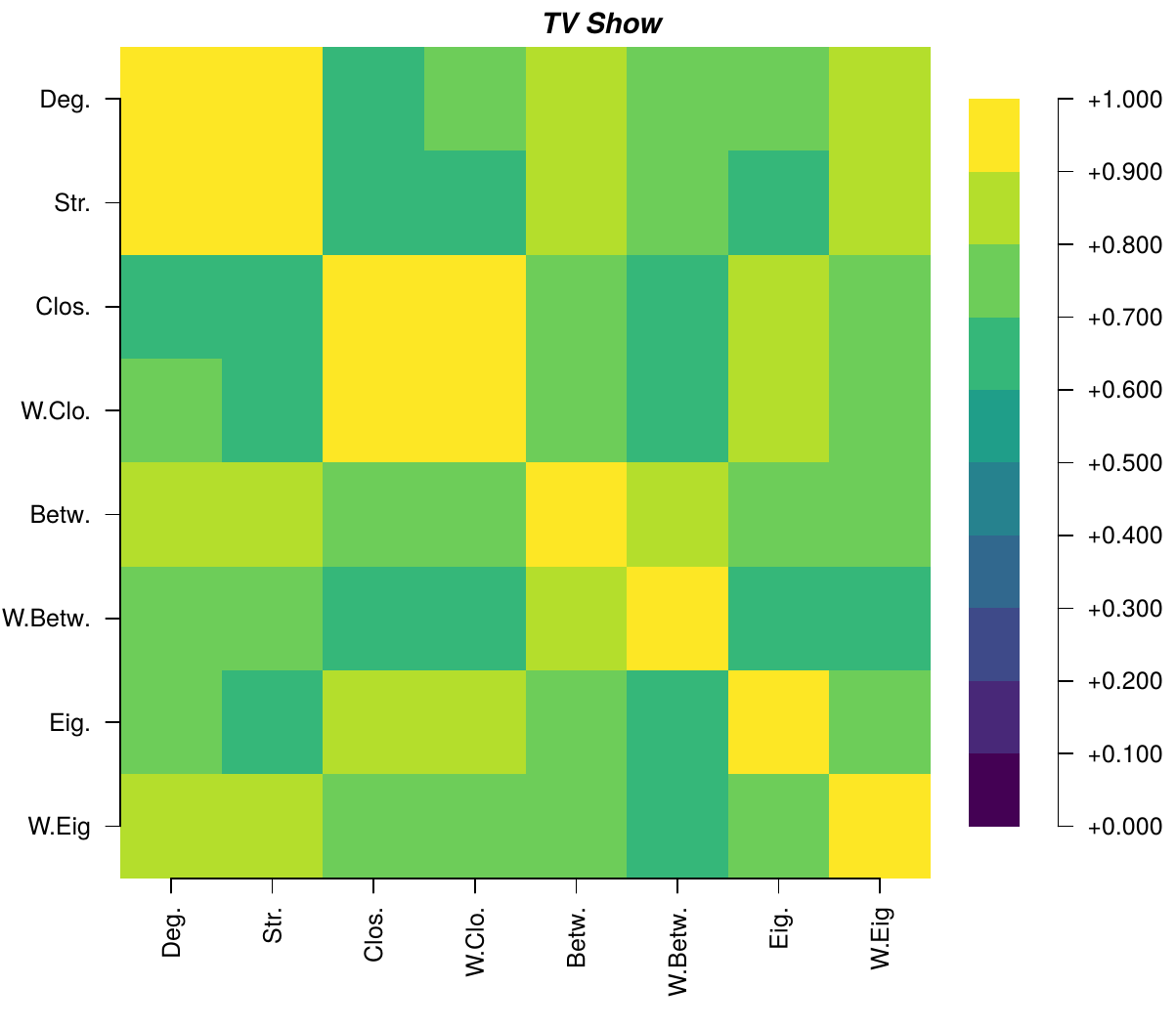}
    \hfill
    \includegraphics[width=0.32\textwidth]{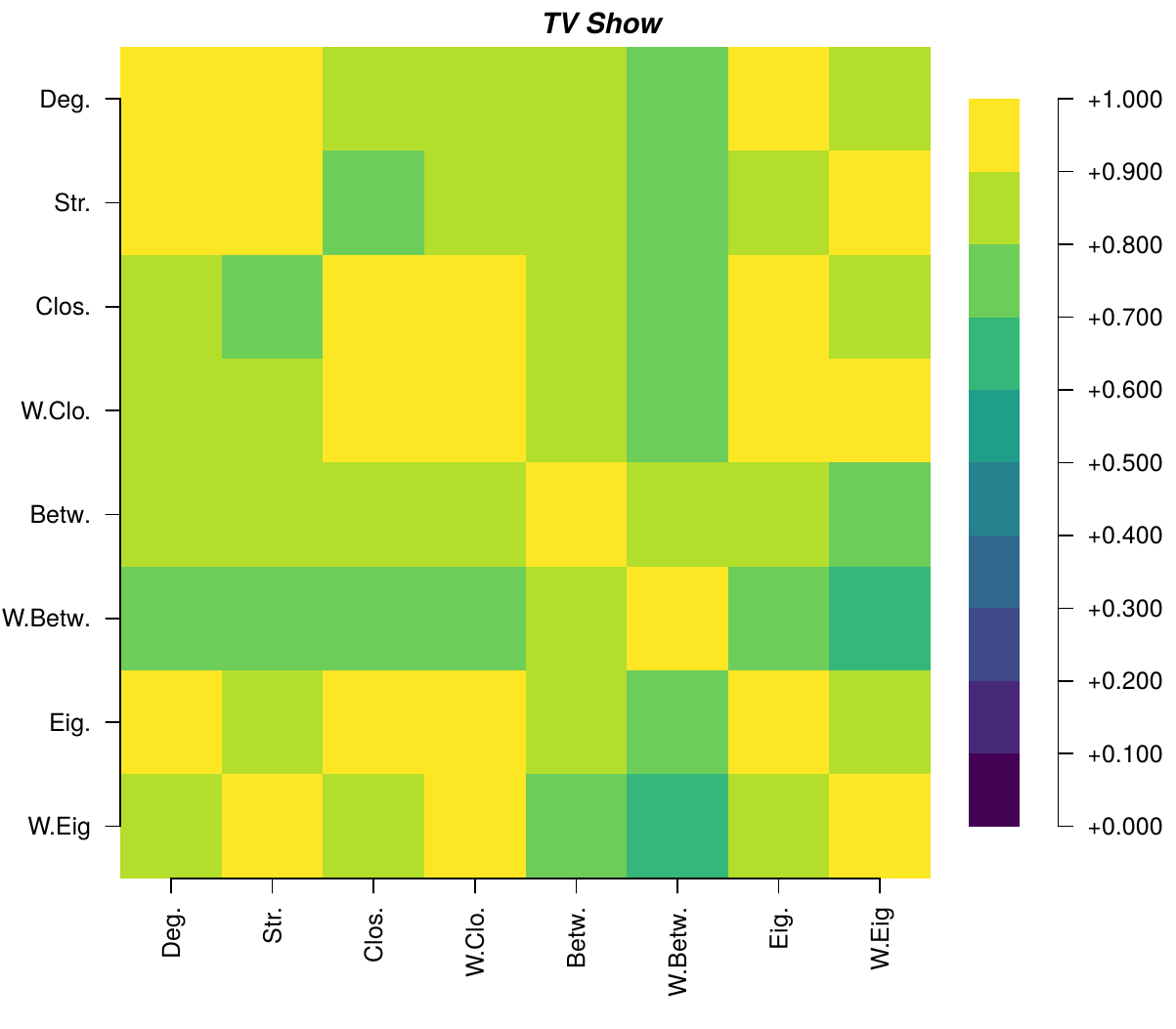}
    \includegraphics[width=0.32\textwidth]{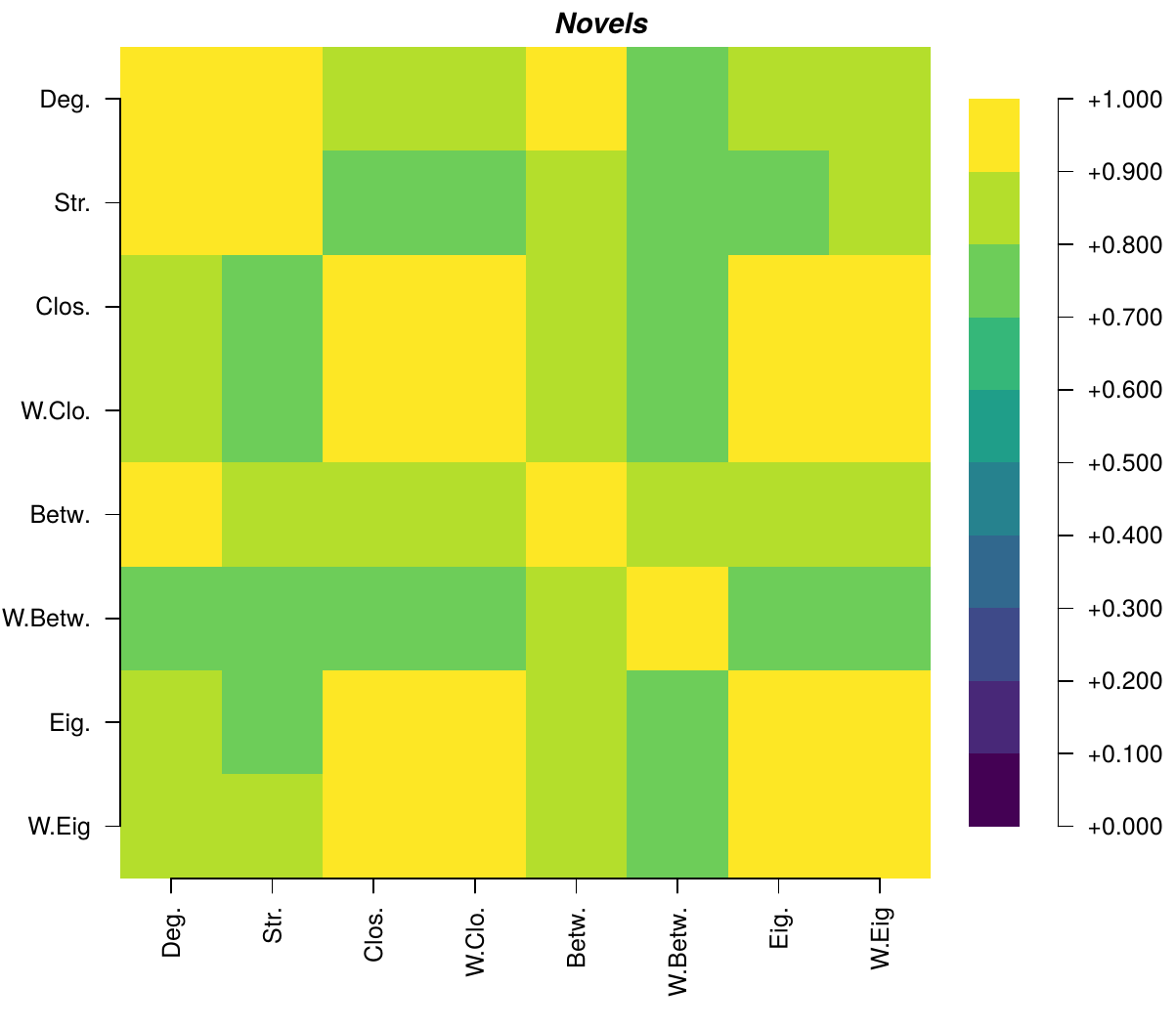}
    \hfill
    \includegraphics[width=0.32\textwidth]{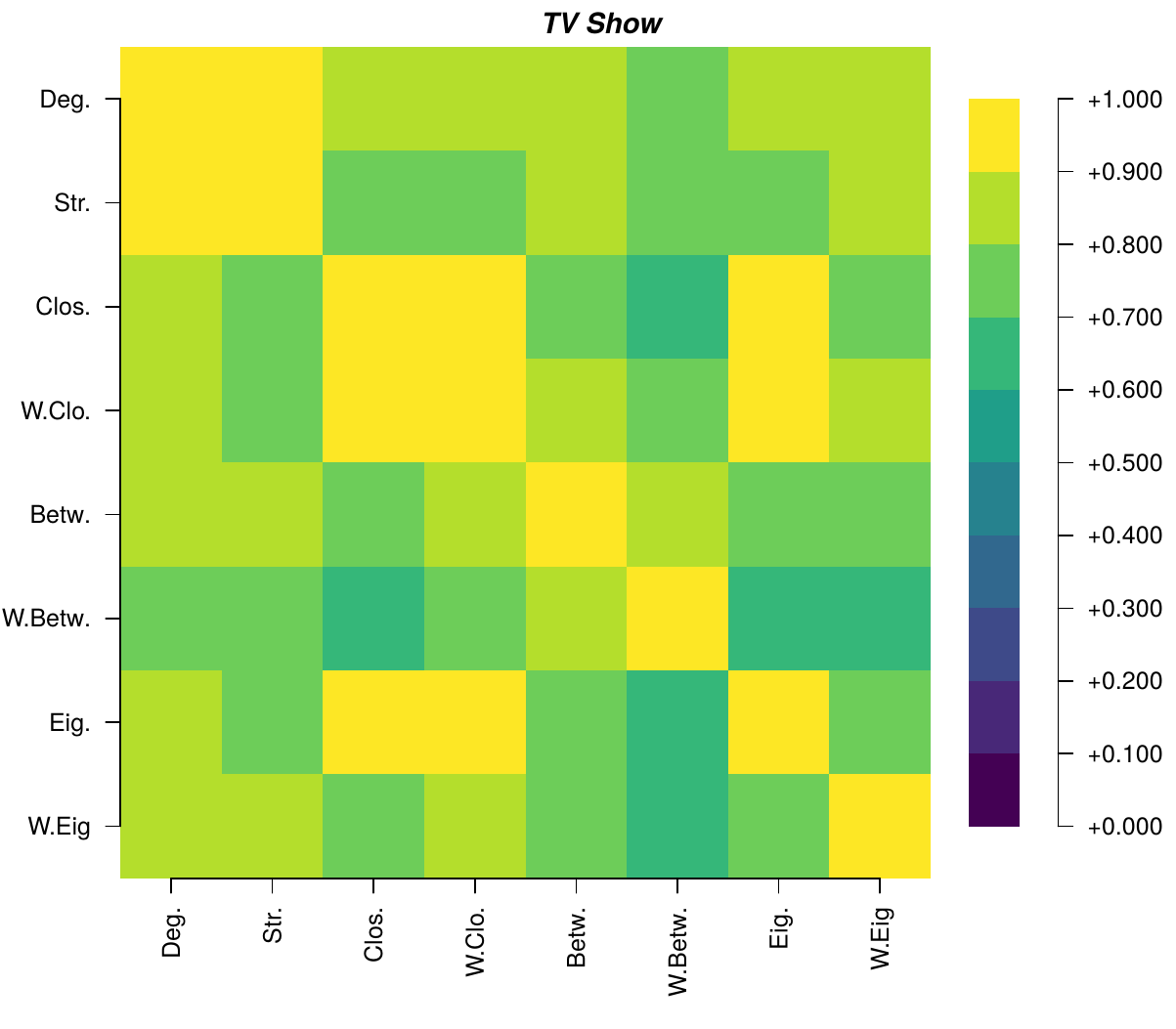}
    \hfill
    \includegraphics[width=0.32\textwidth]{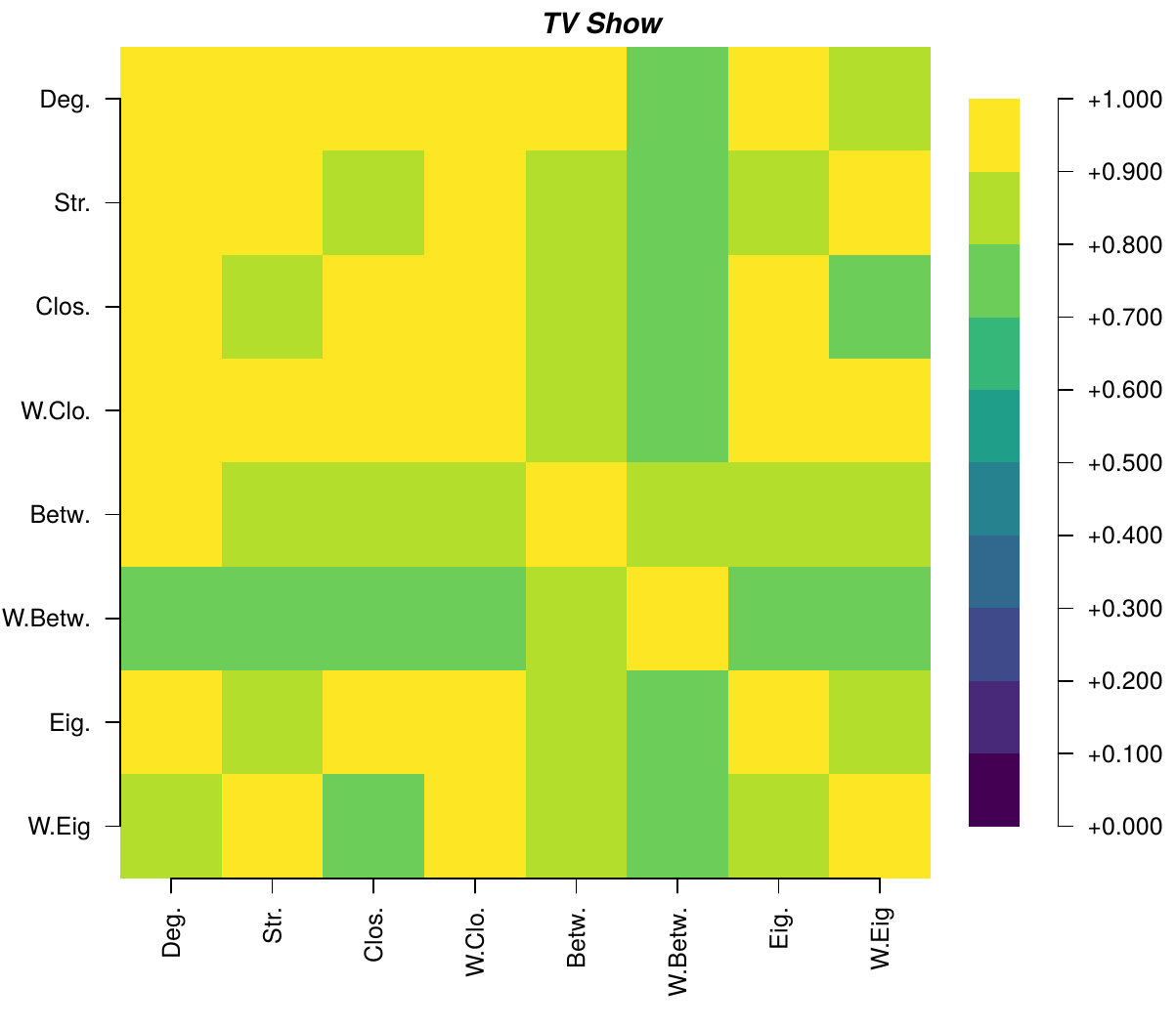}
    \includegraphics[width=0.32\textwidth]{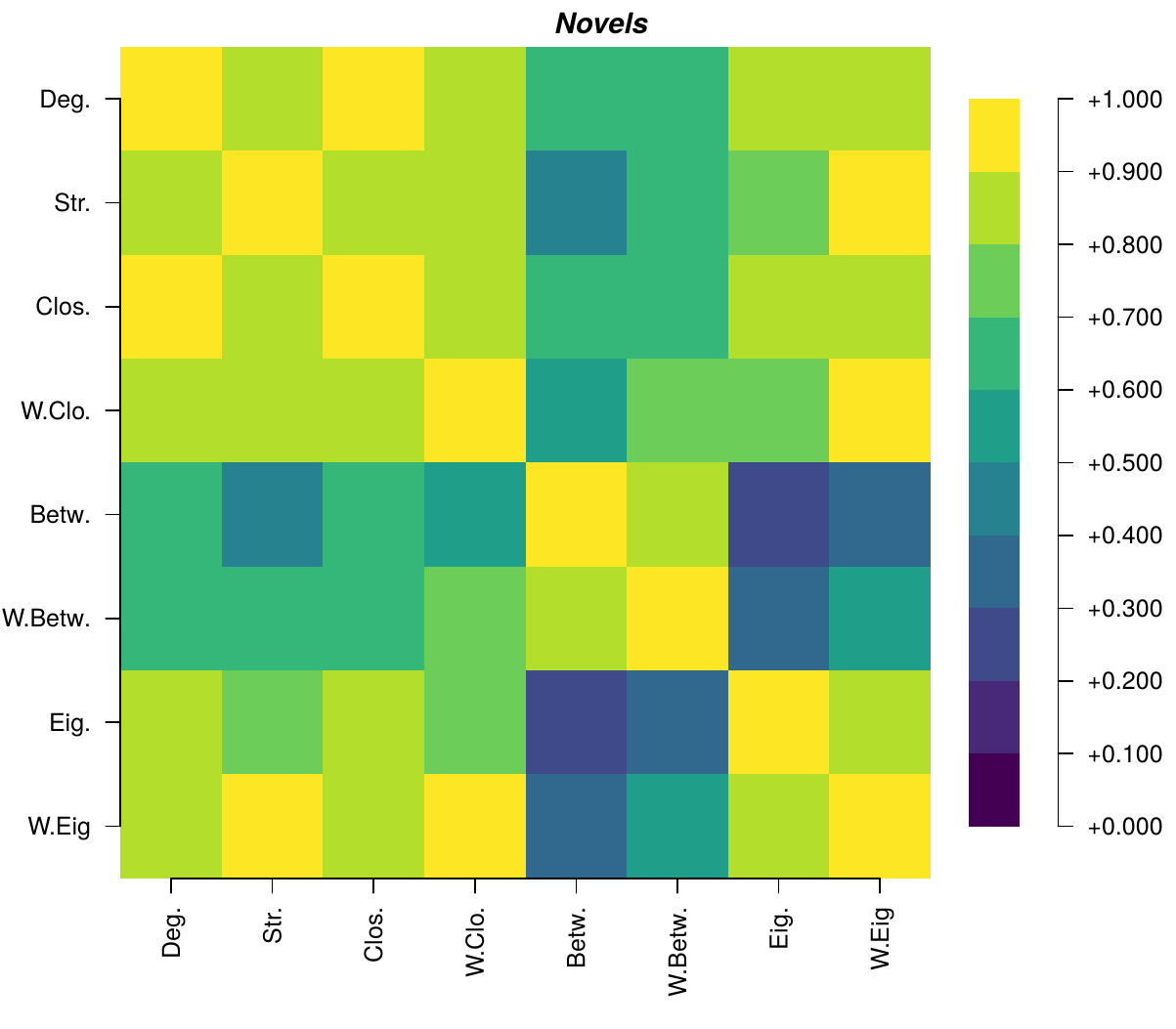}
    \hfill
    \includegraphics[width=0.32\textwidth]{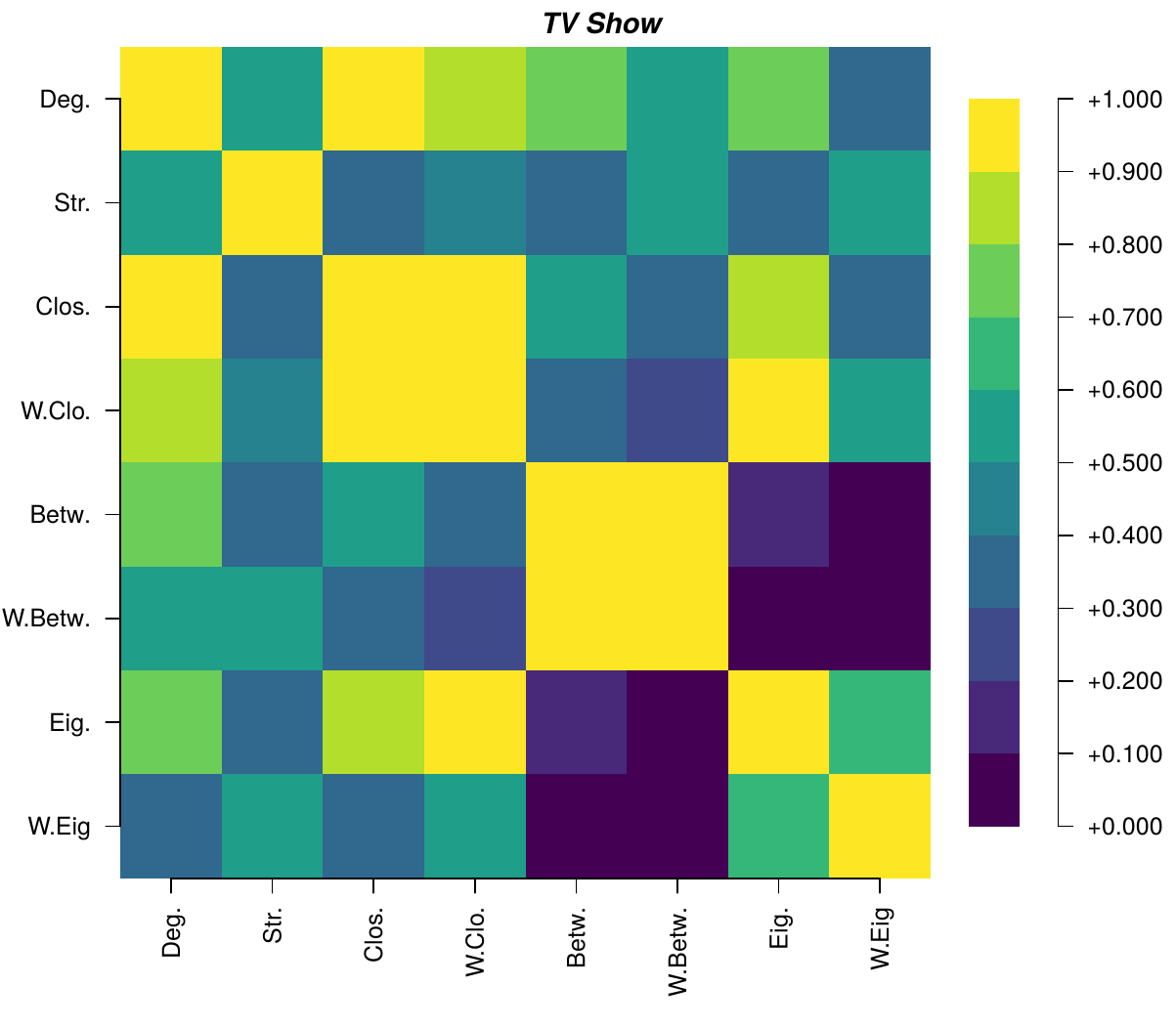}
    \hfill
    \includegraphics[width=0.32\textwidth]{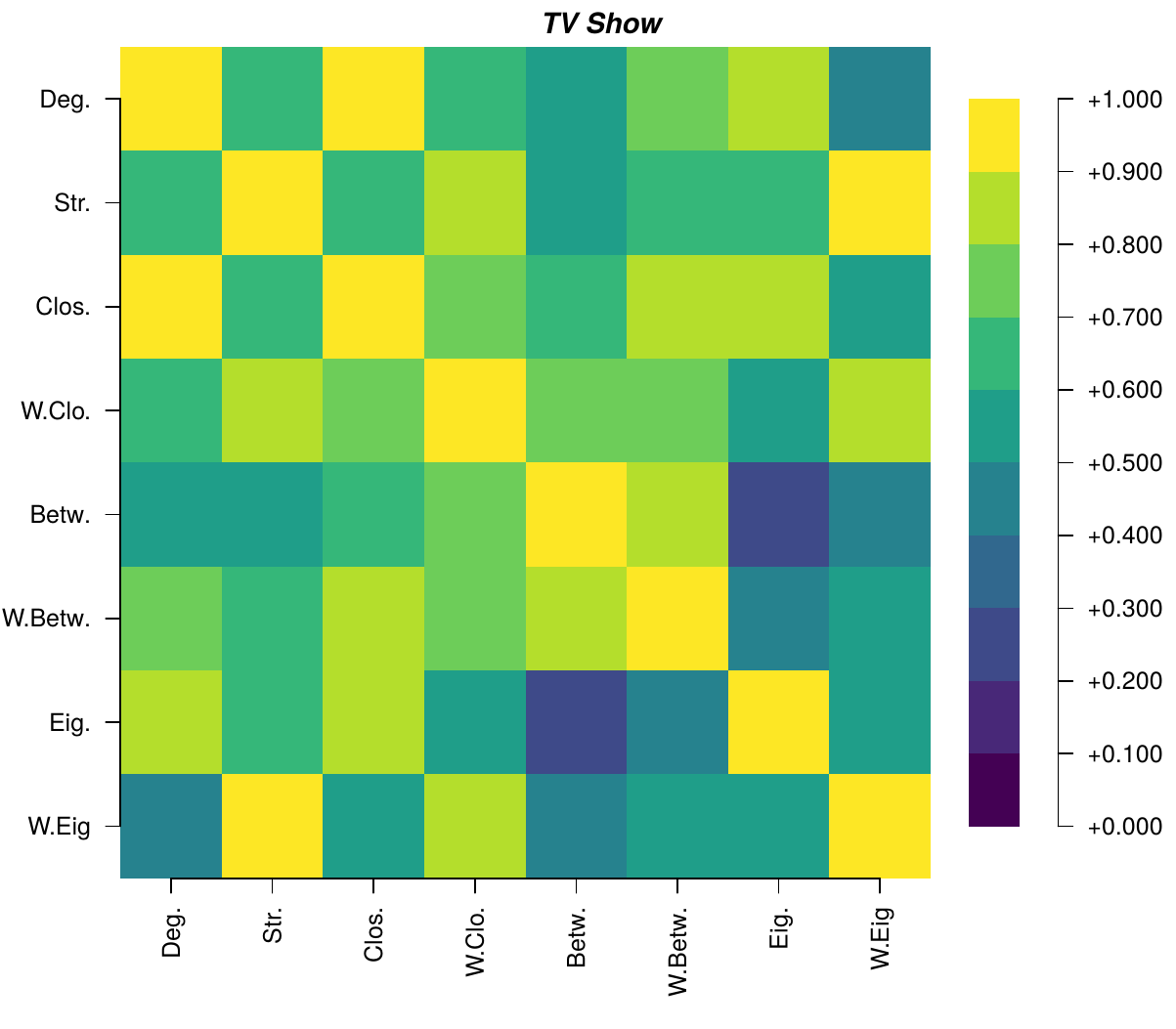}
    \caption{Spearman's correlation between the selected centrality metrics, for the first 5 books of the novels (left column), the first 5 seasons of the TV Show (centre column) and all 8 seasons of the TV Show (right column). Rows differ in the considered character set: \texttt{named} (top row), \texttt{common} (centre row), and \texttt{top-20} (bottom row). By comparison, Figures~\ref{fig:VtxMatch_CentrNamedCorr} and~\ref{fig:VtxMatch_CentrNamedCorr_Comm} focus on period \textit{U2}, i.e. the first two books, volumes, and seasons}
    \label{fig:VtxMatch_CentrNamedCorr_Ext}
\end{figure}

Figure~\ref{fig:VtxMatch_CentrNamedCorr_Ext} shows the centrality correlation matrices obtained when considering more than two books or seasons: first five books of the novels (left column in the figure), first five seasons of the TV show (centre column), and all eight seasons of the TV Show (right column). For the novels, using five books instead of two only noticeably affect the \texttt{top-20} matrix, which exhibits clearer blocks that gather the degree, closeness, and Eigencentrality, whereas the betweenness is isolated. For the first five seasons of the TV show, we observe a higher general correlation level, which increases even more when considering all eight seasons. This could be due to the densification of the network. 

\subsubsection{Centrality Profiles}
\label{sec:ApdxCharCompCentrProf}
In~\cite{Silva2023}, Silva \textit{et al}. adopt an  approach similar to ours, with a corpus of Portuguese-language literary works, except they only use \textit{unweighted} centrality metrics, and the $k$-means method to perform the cluster analysis. They identify the following clusters, which we note $C'$ for convenience:
\begin{itemize}
    \item $C'_1$: medium betweenness;  high Eigenvector, closeness, and degree. These are important figures with a significant impact on the story.
    \item $C'_2$: low betweenness and degree; medium closeness; high Eigenvector. These are characters with a high but local importance.
    \item $C'_3$: low betweenness, Eigenvector, and degree; medium closeness. These are secondary characters that act as mediators or connectors between different subplots or social groups.
    \item $C'_4$: low betweenness, Eigenvector, closeness, and degree. These are minor characters.
\end{itemize}

Table~\ref{tab:CentrComp} provides a visual summary of these clusters, as well as the ones identified in Section~\ref{sec:CharCompCentrProf} of the main article. It is not straightforward to match these clusters, because ours are not as precisely defined in terms of centrality scores: certain metrics cover a range of values, e.g. the degree scores in $C_1$ (novels) range from low to medium. Still, it appears that, for all three adaptations, the first cluster is similar to $C'_4$ (minor characters), while the second (third for the TV show) can be matched to $C_1$ (major characters).

\begin{table}[htb!]
    \caption{Summary of the classes of centrality identified through clustering for the three adaptations considered in this article and shown in Figure~\ref{fig:VtxMatch_CentrCommClust}, and those studied by Silva \textit{et al}.~\cite{Silva2023}}
    \label{tab:CentrComp}%
    \begin{tabular}{p{4cm} p{2cm} p{2.5cm} p{2.5cm} p{2.5cm} p{2.5cm}}
        \toprule
        Adaptation & Cluster & Degree & Eigenvector & Betweenness & Closeness \\
        \midrule
        \textit{Novels}    & $C_1$ & \cellcolor{orange!50!red!50} Low / Medium & \cellcolor{orange!50!red!50} Low / Medium & \cellcolor{red!50} Low & \cellcolor{orange!50!red!50} Low / Medium \\
         & $C_2$ & \cellcolor{green!50} High   & \cellcolor{green!50} High   & \cellcolor{orange!50} Medium   & \cellcolor{green!50} High   \\
        \cline{2-6}
        \textit{Comics}    & $C_1$ & \cellcolor{orange!50!red!50} Low / Medium & \cellcolor{orange!50!red!50} Low / Medium & \cellcolor{red!50} Low      & \cellcolor{orange!50} Medium \\
         & $C_2$ & \cellcolor{green!50} High   & \cellcolor{green!50} High   & \cellcolor{orange!50} Low / High & \cellcolor{green!50} High   \\
        \cline{2-6}
        \textit{TV Show}   & $C_1$ & \cellcolor{red!50} Low    & \cellcolor{red!50} Low    & \cellcolor{red!50} Low     & \cellcolor{orange!50!red!50} Low / Medium \\
         & $C_2$ & \cellcolor{green!50!orange!50} Medium / High   & \cellcolor{green!50!orange!50} Medium / High   & \cellcolor{red!50} Low      & \cellcolor{green!50!orange!50} Medium / High \\
         & $C_3$ & \cellcolor{green!50} High   & \cellcolor{green!50} High   & \cellcolor{green!50} High     & \cellcolor{green!50} High   \\
        \midrule
        Silva \textit{et al}.~\cite{Silva2023} 
         & $C'_1$ & \cellcolor{green!50} High & \cellcolor{green!50} High & \cellcolor{orange!50} Medium & \cellcolor{green!50} High   \\
         & $C'_2$ & \cellcolor{red!50} Low  & \cellcolor{green!50} High & \cellcolor{red!50} Low    & \cellcolor{orange!50} Medium \\
         & $C'_3$ & \cellcolor{red!50} Low  & \cellcolor{red!50} Low  & \cellcolor{red!50} Low    & \cellcolor{orange!50} Medium \\
         & $C'_4$ & \cellcolor{red!50} Low  & \cellcolor{red!50} Low  & \cellcolor{red!50} Low    & \cellcolor{red!50} Low    \\
        \botrule
    \end{tabular}
\end{table}

\begin{figure}[htb!]
    \centering
    \includegraphics[width=1\textwidth]{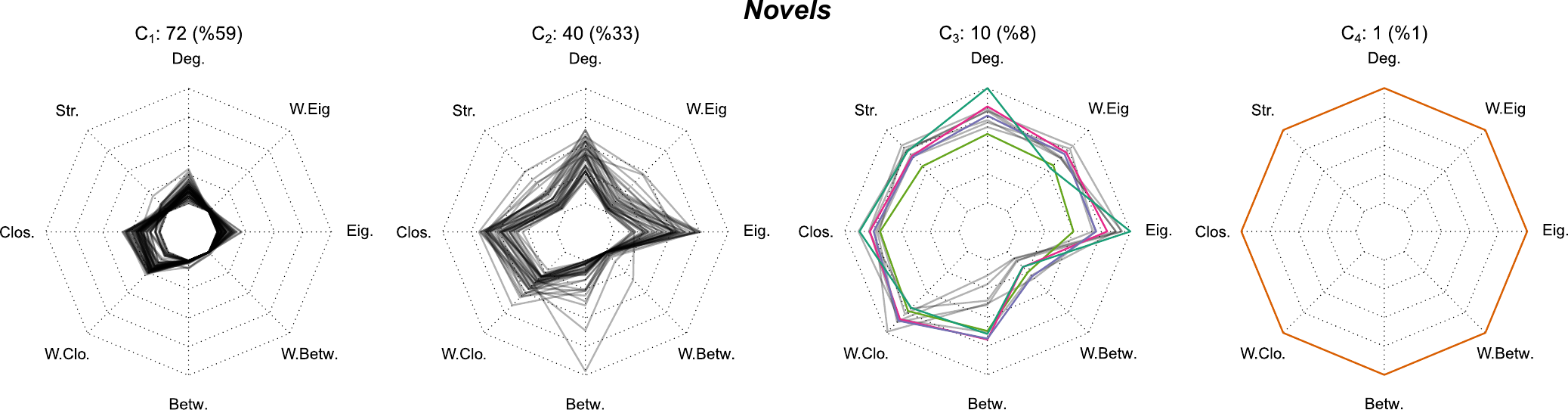}
    \\
    \includegraphics[width=1\textwidth]{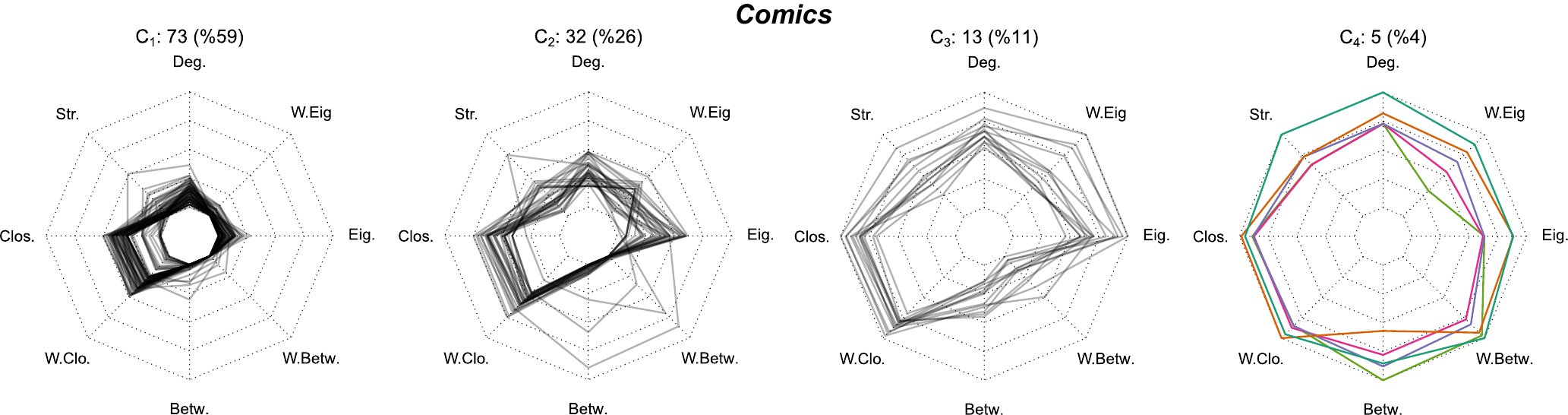}
    \\
    \includegraphics[width=1\textwidth]{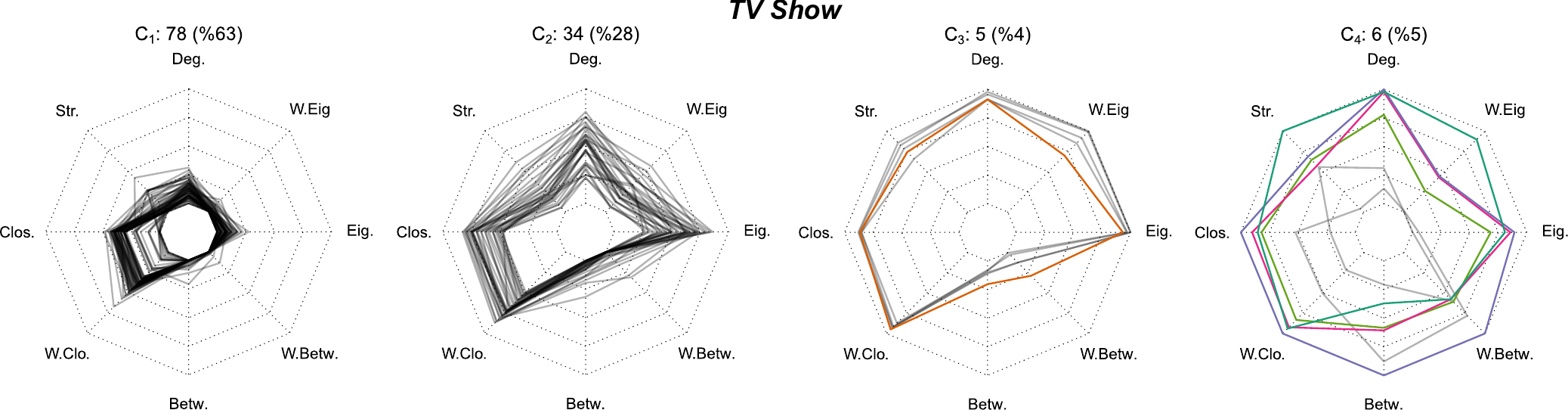}
    \caption{Centrality profiles of the \texttt{common} character set, for all three adaptations, when fixing the number of clusters to four. The five most important characters are represented in color, as in Figures~\ref{fig:VtxMatch_CentrCommRadar} and~\ref{fig:VtxMatch_CentrCommClust} from the main article}
    \label{fig:VtxMatch_CentrCommClustK4}
\end{figure}

In order to get more comparable clusters, we select the dendrogram cuts that correspond to $k=4$ clusters, as shown in Figure~\ref{fig:VtxMatch_CentrCommClustK4}. These are not the best cuts, which are discussed in the main article (cf. Figure~\ref{fig:VtxMatch_CentrCommClust}). But they allow a direct comparison between, on the one hand, the three adaptations, and on the other hand, the results of Silva \textit{et al}.~\cite{Silva2023}.

When considering four clusters, the adaptations exhibit relatively similar classes of centrality. Clusters $C_1$ gather minor characters, with low to medium closeness, and low degree, Eigenvector centrality, and betweenness. These clusters match Silva \textit{et al}.'s $C'_3$ and $C'_4$. Clusters $C_2$ contain minor characters that are a bit more central, exhibiting higher levels of closeness, degree, and Eigenvector centrality, while their betweenness stays low. This case is not covered in Silva \textit{et al}.'s typology. Clusters $C_3$ are constituted of major characters with high closeness, degree, and Eigenvector centrality, but low to medium betweenness, which matches Silva \textit{et al}.'s $C'_1$. Finally, clusters $C_4$ contain the few major characters that are central in terms of all four metrics, and not covered in Silva's \textit{et al}. typology.

\begin{table}[htb!]
    \caption{Summary of the classes of centrality identified through clustering ($k=4$) for the three adaptations considered in this article and shown in Figure~\ref{fig:VtxMatch_CentrCommClustK4}, and those studied by Silva \textit{et al}.~\cite{Silva2023}}
    \label{tab:CentrCompK4}%
    \begin{tabular}{p{4cm} p{2cm} p{2.5cm} p{2.5cm} p{2.5cm} p{2.5cm}}
        \toprule
        Adaptation & Cluster & Degree & Eigenvector & Betweenness & Closeness \\
        \midrule
        ASoIaF \& GoT & $C_1$ & \cellcolor{red!50} Low & \cellcolor{red!50} Low & \cellcolor{red!50} Low & \cellcolor{orange!50!red!50} Low / Medium \\
         & $C_2$ & \cellcolor{orange!50} Medium & \cellcolor{orange!50} Medium & \cellcolor{red!50} Low & \cellcolor{green!50} High \\
         & $C_3$ & \cellcolor{green!50} High & \cellcolor{green!50} High & \cellcolor{orange!50} Medium & \cellcolor{green!50} High \\
         & $C_4$ & \cellcolor{green!50} High & \cellcolor{green!50} High & \cellcolor{green!50} High & \cellcolor{green!50} High \\
        \midrule
        Silva \textit{et al}.~\cite{Silva2023} & $C'_1$ & \cellcolor{green!50} High & \cellcolor{green!50} High & \cellcolor{orange!50} Medium & \cellcolor{green!50} High   \\
         & $C'_2$ & \cellcolor{red!50} Low  & \cellcolor{green!50} High & \cellcolor{red!50} Low    & \cellcolor{orange!50} Medium \\
         & $C'_3$ & \cellcolor{red!50} Low  & \cellcolor{red!50} Low  & \cellcolor{red!50} Low    & \cellcolor{orange!50} Medium \\
         & $C'_4$ & \cellcolor{red!50} Low  & \cellcolor{red!50} Low  & \cellcolor{red!50} Low    & \cellcolor{red!50} Low    \\
        \botrule
    \end{tabular}
\end{table}

To be complete, we have to mention the work of Mas\'{i}as \textit{et al.}~\cite{Masias2017} and Pang \textit{et al.}~\cite{Pang2023}. They adopt a similar approach when identifying classes of characters by clustering them based on their centrality. However, their results are not comparable to ours, because the methods they used are too different. First, they do not use standard centrality metrics, but rather some parametric weighted versions, which allow handling two types of weights at once. Second, they study the clusters in terms of their stability relative to these parameters. Third, they do not normalise the metric values to account for scale differences. Nevertheless, it is worth mentioning that they typically identify 3 clusters corresponding to 3 increasing levels of centrality. They do not discuss each metric separately, so we assume that these levels concern all metrics indiscriminately.

\section{Narrative Matching}
This section provides additional results for Section~\ref{sec:Align} of the main article. First, Section~\ref{sec:ApdxNarrMatchBestConf} highlights combinations of parameters leading to the best matching performance, using both \textit{text-constrained} and \textit{commensurate} narrative units. Then, Section~\ref{sec:ApdxNarrMatchSim} shows the similarity matrices corresponding to the best matching performance for the \textit{text-constrained} units. Section~\ref{sec:ApdxNarrMatchStruct} and~\ref{sec:ApdxNarrMatchComb} present results for all possible configurations of structural and hybrid matching. Section~\ref{sec:ApdxNarrMatchNovelsTVShowU2} discusses the results of narrative matching for the \textit{Novels vs. TV Show} pair over period \texttt{U2}, as opposed to \texttt{U5} in the main article. Finally, Section~\ref{sec:ApdxNarrMatchCumNets} displays the results of structural matching using cumulative dynamic character networks, whereas we present results with instant networks in the main article.

\subsection{Best Configurations}
\label{sec:ApdxNarrMatchBestConf}
The top part of Table~\ref{tab:best-alignment-config} shows the best configurations of narrative matching using text-constrained narrative units, while its bottom part deals with the best configurations using commensurate narrative units. Overall, the best performance is obtained using commensurate units and hybrid similarity.

\begin{table}[htb!]
    \centering
    \caption{Best configurations for the task of narrative matching using the \textit{text-constrained} and \textit{commensurate} narrative units}
    \begin{tabular}{l l l l l l l l r}
        \toprule
         Narr. Unit & Adaptations Pair                  & Similarity & Repr. & Measure         & Character Set & Text Sim. & Alignment & F1 \\
         \midrule
         Text-Constr. & \textit{Novels vs. Comics}  & Hybrid     & Vertices          & Ru\v{z}i\v{c}ka & \texttt{common}     & \textit{tfidf}     & Smith--Waterman & $67.37$ \\
         & \textit{Novels vs. TV Show} & Structural & Edges          & Jaccard         & \texttt{common}     & --                 & Smith--Waterman & $32.63$ \\
         & \textit{Comics vs. TV Show} & Structural & Vertices          & Ru\v{z}i\v{c}ka & \texttt{common}     & --                 & Smith--Waterman & $51.40$ \\
         \midrule
         Commens. & \textit{Novels vs. Comics}  & Hybrid     & Edges          & Ru\v{z}i\v{c}ka  & \texttt{named}      & \textit{tfidf}     & Smith--Waterman & $78.50$ \\
         & \textit{Novels vs. TV Show} & Hybrid     & Vertices          & Ru\v{z}i\v{c}ka  & \texttt{common}     & \textit{tfidf}     & Smith--Waterman & $39.04$ \\
         & \textit{Comics vs. TV Show} & Hybrid     & Edges          & Jaccard          & \texttt{top20}      & \textit{sbert}     & Smith--Waterman & $63.87$ \\
         \bottomrule
    \end{tabular}
    \label{tab:best-alignment-config}
\end{table}

\subsection{Similarity Matrices}
\label{sec:ApdxNarrMatchSim}
\begin{figure}[htb!]
    \centering
    \includegraphics[width=1\textwidth]{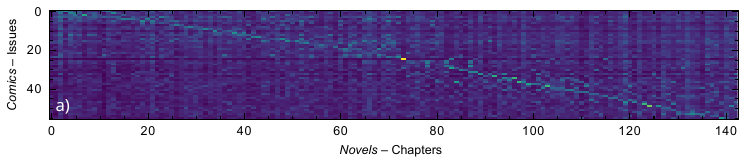}
    \includegraphics[width=1\textwidth]{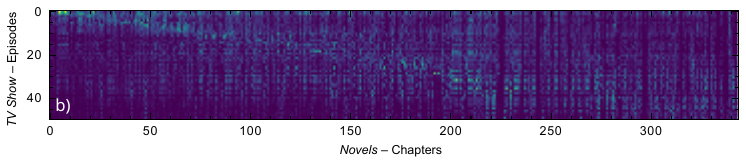}
    \includegraphics[width=1\textwidth]{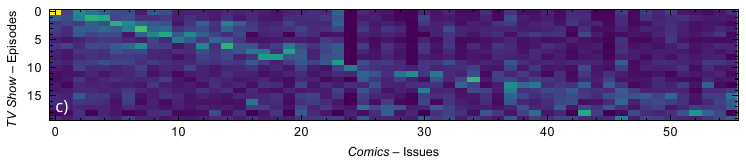}
    \caption{Similarity matrices that produce the best alignments for all pairs of media: a) \textit{Novels} vs. \textit{Comics}; b) \textit{Novels vs. TV Show}; c) \textit{Comics vs. TV Show}. Configurations with commensurate units are excluded, since matrices are of different shapes. See Table~\ref{tab:best-alignment-config} for the precise configurations}
    \label{fig:best-alignments-matrices}
\end{figure}

Figure~\ref{fig:best-alignments-matrices} shows the similarity matrices $\mathbf{S}$ that lead to the best alignment for each pair of media (except for configurations with blocks), as described in Table~\ref{tab:best-alignment-config}. These matrices exhibit patterns where an overall alignment can loosely be observed. These similarity matrices however are not sufficient to align narrative units, as we find that the performance of the simple thresholding alignment baseline is lower than that of the Smith--Waterman algorithm.

\subsection{Structural Matching}
\label{sec:ApdxNarrMatchStruct}
Table~\ref{tab:perf_structural} shows the results of structural alignment for all surveyed configurations. By comparison, Table~\ref{tab:perf_structural_best} in the main article only focuses on the best results.

\begin{table}[htb!]
    \centering
    \caption{Narrative matching performance, expressed in F1-score, for all configurations of \textit{structure}-based matching. The first column indicates the type of narrative unit used: text-constrained (\textit{Text-Constr.}) or commensurate (\textit{Commens.}). The \textit{Repr.} column indicates whether the similarity is computed over vertex or edge sets. The \textit{Measure} column indicates whether the similarity measure ignores weights (\textit{Jaccard}) or take them into account (\textit{Ru\v{z}i\v{c}ka})}
    \input{tables/narr_match/perf_structural}
    \label{tab:perf_structural}
\end{table}

We can conclude that the Smith--Waterman algorithm performs better than thresholding. In addition, the best results are obtained with the \texttt{common} character set, whereas \texttt{top-20} performs the worst. However, it is difficult to reach a conclusion for the other two parameters: compared objects (edges vs. vertices), and weight usage (Jaccard vs. Ru\v{z}i\v{c}ka).

\subsection{Hybrid Matching}
\label{sec:ApdxNarrMatchComb}
Table~\ref{tab:perf_combined} shows all the results of narrative matching using the \textit{hybrid}-based similarity for the \textit{text-constrained} and \textit{commensurate} narrative units respectively. When using the \textit{commensurate} units, hybrid similarity improves the results compared to structural similarity by between $8.1$ and $11.3$ F1 depending on the media pair. Overall, we obtain the best overall matching performance using the \textit{commensurate} units to perform hybrid matching.

\begin{table}[htb!]
    \centering
    \caption{Narrative matching performance, expressed in F1-score, for all configurations of \textit{hybrid}-based matching. The first column indicates the type of narrative unit used: text-constrained (\textit{Text-Constr.}) or commensurate (\textit{Commens.}). The \textit{Text Sim.} column indicates the textual similarity measure in usage. The \textit{Struct Repr.} column indicates whether the similarity is computed over vertex or edge sets. The \textit{Struct Measure} column indicates whether the similarity measure ignores weights (\textit{Jaccard}) or take them into account (\textit{Ru\v{z}i\v{c}ka})}
    \input{tables/narr_match/perf_combined}
    \label{tab:perf_combined}
\end{table}

\subsection{\textit{Novels} vs. \textit{TV Show} over \texttt{U2}}
\label{sec:ApdxNarrMatchNovelsTVShowU2}
In this section, we present the results of narrative matching for the \textit{Novels vs. TV Show} media pair over the \texttt{U2} time period. By comparison, the main article focuses on \texttt{U5}. Since the TV show diverges more and more from the novels across seasons, we expect that aligning adaptations over the \texttt{U2} period is easier than over the \texttt{U5} period.

\begin{figure}[htb!]
    \centering
    \includegraphics[width=1\textwidth]{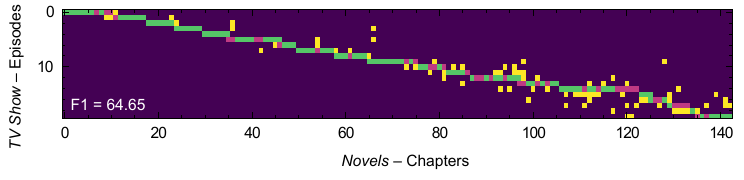}
    \caption{Best performing alignment for the \textit{Novels vs. TV Show} pair over time period \texttt{U2}}
    \label{fig:best_matching_tvshow-novels_U2}
\end{figure}

The best results over all configurations can be found in Table~\ref{tab:best_matching_tvshow-novels_U2}, while the best alignment can be found in Figure~\ref{fig:best_matching_tvshow-novels_U2}. We restrict ourselves to the best results per similarity due to the great number of possible configurations. As we expect, the performance is higher for the \texttt{U2} than the \texttt{U5} period, due to the divergence of the later seasons. We also observe that, as when aligning over \texttt{U5}, matching using commensurate units yields a positive performance boost, with a large increase of $17.9$ F1. Combining textual and structural matching increases performance ($+5.8$ F1), but less than taking commensurate into account.

\begin{table}[htb!]
    \centering
    \caption{Performance obtained on the \texttt{U2} time period of the \textit{Novels vs. TV Show} pair, expressed in terms of F1-score. Only the best results across configurations are shown. The mentions \textit{Text-Constrained} and \textit{Commensurate} refer to the corresponding narrative units, see Section~\ref{sec:AlignStruct} in the main article.}
    \input{tables/narr_match/tvshow-novels/perf_U2_best}
    \label{tab:best_matching_tvshow-novels_U2}
\end{table}





\subsection{Cumulative Networks}
\label{sec:ApdxNarrMatchCumNets}
Table~\ref{tab:matching_cumulative} presents results of structural narrative matching using \textit{cumulative} networks on the text-constrained narrative units. By comparison, the main article focuses on \textit{instant} networks.

\begin{table}[htb!]
    \centering
    \caption{Performance obtained when using \textit{structural}-based representations using \textit{cumulative networks} and the \textit{text-constrained} narrative units to tackle the narrative matching task, expressed in terms of F1-score. 
    Columns are organised as in previous tables.}
    \input{tables/narr_match/perf_structural_cumulative}
    \label{tab:matching_cumulative}
\end{table}

Since the cumulative network of a narrative unit is the aggregation of previous instant networks, differences between narrative units are less and less noticeable, leading to poor matching performance. The Smith--Waterman algorithm is still able to achieve mild performance in some cases ($47.18$ F1 on the \textit{Novels vs. Comics} pair), but overall matching with instant networks clearly yields more performance.

\end{document}

%% file: tables/narr_match/perf_textual.tex
\begin{tabular}{l l r r r}
    \toprule
    Sim. & Alignment & \textit{Novels} vs. \textit{Comics} & \textit{Novels} vs. \textit{TV Show} & \textit{Comics} vs. \textit{TV Show} \\
    \midrule
    \textit{tfidf} & Thresholding & 19.88 & 15.98 & 24.13 \\
     & Smith--Waterman & 48.25 & 17.21 & 30.68 \\
    \midrule
    \textit{sbert} & Thresholding & 18.01 & 8.04 & 20.28 \\
     & Smith--Waterman & \bfseries 55.24 & \bfseries 22.55 & \bfseries 35.96 \\
    \bottomrule
\end{tabular}

%% file: tables/narr_match/perf_structural_best.tex
\begin{tabular}{l r@{~}r@{~}r r@{~}r@{~}r r@{~}r@{~}r}
\toprule
Alignment & \multicolumn{3}{c}{\textit{Novels} vs. \textit{Comics}} & \multicolumn{3}{c}{\textit{Novels} vs. \textit{TV Show}} & \multicolumn{3}{c}{\textit{Comics} vs. \textit{TV Show}} \\
 & \ttfamily named & \ttfamily common & \ttfamily top-20 & \ttfamily named & \ttfamily common & \ttfamily top-20 & \ttfamily named & \ttfamily common & \ttfamily top-20 \\
\midrule
Thresholding & 29.61 & 34.74 & 13.31 & 20.72 & 23.43 & 7.34 & 43.36 & 46.95 & 33.33 \\
Smith--Waterman & 58.74 & \bfseries 62.94 & 46.81 & 28.72 & \bfseries 32.63 & 9.87 & 49.72 & \bfseries 51.40 & 46.86 \\
\bottomrule
\end{tabular}

%% file: tables/narr_match/perf_structural_blocks_best.tex
\begin{tabular}{l r@{~}r@{~}r r@{~}r@{~}r r@{~}r@{~}r}
\toprule
Alignment & \multicolumn{3}{r}{\textit{Novels} vs. \textit{Comics}} & \multicolumn{3}{r}{\textit{Novels} vs. \textit{TV Show}} & \multicolumn{3}{r}{\textit{Comics} vs. \textit{TV Show}} \\
 & \ttfamily named & \ttfamily common & \ttfamily top-20 & \ttfamily named & \ttfamily common & \ttfamily top-20 & \ttfamily named & \ttfamily common & \ttfamily top-20 \\
\midrule
Thresholding & 25.66 & 33.23 & 14.04 & 21.02 & 18.72 & 7.74 & 36.25 & 36.71 & 29.51 \\
Smith--Waterman & 71.81 & \bfseries 72.30 & 63.33 & 34.46 & \bfseries 35.09 & 30.69 & 60.42 & \bfseries 61.78 & 61.46 \\
\bottomrule
\end{tabular}

%% file: tables/narr_match/perf_combined_best.tex
\begin{tabular}{lrrr}
\toprule
Alignment & \textit{Novels} vs. \textit{Comics} & \textit{Novels} vs. \textit{TV Show} & \textit{Comics} vs. \textit{TV Show} \\
\midrule
Thresholding & 39.60 & 26.90 & 46.75 \\
Smith--Waterman & \bfseries 67.37 & \bfseries 30.95 & \bfseries 49.16 \\
\bottomrule
\end{tabular}

%% file: tables/narr_match/perf_combined_blocks_best.tex
\begin{tabular}{lrrr}
\toprule
Alignment & \textit{Novels} vs. \textit{Comics} & \textit{Novels} vs. \textit{TV Show} & \textit{Comics} vs. \textit{TV Show} \\
\midrule
Thresholding & 47.62 & 23.58 & 46.24 \\
Smith--Waterman & \bfseries 78.50 & \bfseries 39.04 & \bfseries 63.87 \\
\bottomrule
\end{tabular}

%% file: tables/narr_match/perf_structural.tex
\begin{tabular}{l l l l r r r r r r r r r}
    \toprule
    Narrative & Repr. & Measure & Alignment & \multicolumn{3}{c}{\textit{Novels} vs. \textit{Comics}} & \multicolumn{3}{c}{\textit{Novels} vs. \textit{TV Show}} & \multicolumn{3}{c}{\textit{Comics} vs. \textit{TV Show}} \\
    Units &  &  &  & \ttfamily named & \ttfamily common & \ttfamily top-20 & \ttfamily named & \ttfamily common & \ttfamily top-20 & \ttfamily named & \ttfamily common & \ttfamily top-20 \\
    \midrule
    Text-Constr. & Edges & Jaccard & Thresholding & 25.86 & 28.46 & 11.66 & 12.86 & 13.18 & 6.59 & 39.85 & 44.08 & 31.28 \\
    & &  & Smith--Waterman & 54.55 & 53.85 & 38.49 & 28.38 & \bfseries 32.63 & 1.54 & 47.78 & 46.93 & 45.71 \\
    \cmidrule{3-13}
    & & Ru\v{z}i\v{c}ka & Thresholding & 29.61 & 34.74 & 13.31 & 20.72 & 23.43 & 7.34 & 43.36 & 46.95 & 33.33 \\
    & &  & Smith--Waterman & 55.63 & 57.54 & 43.84 & 25.43 & 24.12 & 1.75 & 39.55 & 49.72 & 46.86 \\
    \cmidrule{2-13}
    & Vertices & Jaccard & Thresholding & 13.73 & 17.61 & 8.53 & 11.67 & 13.72 & 6.15 & 28.30 & 24.31 & 19.85 \\
    & &  & Smith--Waterman & 53.71 & 56.54 & 26.71 & 28.72 & 23.61 & 8.04 & 39.77 & 49.16 & 39.33 \\
    \cmidrule{3-13}
    & & Ru\v{z}i\v{c}ka & Thresholding & 15.75 & 26.57 & 6.98 & 13.38 & 18.73 & 6.81 & 35.20 & 34.74 & 19.19 \\
    & &  & Smith--Waterman & 58.74 & \bfseries 62.94 & 46.81 & 25.91 & 23.61 & 9.87 & 49.72 & \bfseries 51.40 & 38.42 \\
    \midrule
    Commens. & Edges & Jaccard & Thresholding & 24.67 & 27.95 & 11.33 & 17.97 & 16.91 & 6.71 & 37.72 & 37.86 & 34.18 \\
    & &  & Smith--Waterman & 71.52 & 71.14 & 49.35 & 33.17 & 31.58 & 22.54 & 47.62 & 47.87 & 46.99 \\
    \cmidrule{3-13}
    & & Ru\v{z}i\v{c}ka & Thresholding & 25.66 & 33.23 & 14.04 & 21.02 & 18.72 & 7.74 & 44.34 & 44.79 & 38.18 \\
    & &  & Smith--Waterman & 71.38 & 71.57 & 59.93 & 31.33 & 34.38 & 24.33 & 47.62 & 49.46 & 46.99 \\
    \cmidrule{2-13}
    & Vertices & Jaccard & Thresholding & 16.79 & 19.84 & 8.24 & 15.67 & 12.54 & 6.42 & 8.39 & 5.97 & 7.50 \\
    & &  & Smith--Waterman & 71.81 & 70.23 & 60.93 & 31.78 & 32.16 & 30.69 & \bfseries 52.91 & \bfseries 52.91 & 44.32 \\
    \cmidrule{3-13}
    & & Ru\v{z}i\v{c}ka & Thresholding & 14.80 & 23.82 & 7.77 & 15.62 & 17.63 & 6.55 & 35.35 & 26.92 & 7.14 \\
    & &  & Smith--Waterman & 70.95 & \bfseries 72.30 & 63.33 & 34.46 & \bfseries 35.09 & 25.91 & 51.06 & 51.06 & 40.88 \\
    \bottomrule
\end{tabular}

%% file: tables/narr_match/perf_combined.tex
\begin{tabular}{lllll r@{~}r@{~}r r@{~}r@{~}r r@{~}r@{~}r}
    \toprule
    Narrative & Text & Struct. & Struct. & Alignment & \multicolumn{3}{c}{\textit{Novels} vs. \textit{Comics}} & \multicolumn{3}{c}{\textit{Novels} vs. \textit{TV Show}} & \multicolumn{3}{c}{\textit{Comics} vs. \textit{TV Show}} \\
    Unit & Sim. &  Repr. & Measure &  & \ttfamily named & \ttfamily common & \ttfamily top-20 & \ttfamily named & \ttfamily common & \ttfamily top-20 & \ttfamily named & \ttfamily common & \ttfamily top-20 \\
    \midrule
    Text-Constr. & \textit{sbert} & Edges & Jaccard & Thresholding & 27.00 & 18.60 & 6.90 & 17.33 & 15.69 & 7.04 & 40.13 & 42.70 & 24.90 \\
    & &  &  & Smith--Waterman & 63.16 & 63.86 & 61.75 & 27.06 & 27.41 & 1.53 & 38.86 & 45.71 & 44.07 \\
    \cmidrule{4-14}
    & &  & Ru\v{z}i\v{c}ka & Thresholding & 11.54 & 15.95 & 13.79 & 22.58 & 24.64 & 10.98 & 45.09 & 46.75 & 25.17 \\
    & &  &  & Smith--Waterman & 57.75 & 60.35 & 58.95 & 25.16 & 26.21 & 17.80 & 39.55 & 40.68 & 38.42 \\
    \cmidrule{3-14}
    & & Vertices & Jaccard & Thresholding & 25.64 & 9.09 & 13.23 & 11.43 & 12.00 & 6.32 & 32.22 & 33.99 & 22.22 \\
    & &  &  & Smith--Waterman & 65.25 & 62.90 & 61.97 & 22.11 & 28.01 & 12.50 & 35.29 & 35.29 & 32.94 \\
    \cmidrule{4-14}
    & &  & Ru\v{z}i\v{c}ka & Thresholding & 29.32 & 29.51 & 13.78 & 15.79 & 20.07 & 6.32 & 36.67 & 46.15 & 22.73 \\
    & &  &  & Smith--Waterman & 62.46 & 65.96 & 57.45 & 28.95 & 27.40 & 21.73 & 44.07 & 44.71 & 32.94 \\
    \cmidrule{2-14}
    & \textit{tfidf} & Edges & Jaccard & Thresholding & 27.51 & 20.93 & 7.73 & 19.95 & 22.57 & 12.65 & 34.10 & 35.40 & 25.42 \\
    & &  &  & Smith--Waterman & 57.24 & 57.54 & 58.95 & 30.24 & 29.31 & 9.89 & 47.78 & 42.46 & 42.94 \\
    \cmidrule{4-14}
    & &  & Ru\v{z}i\v{c}ka & Thresholding & 21.18 & 33.67 & 14.96 & 23.73 & 26.90 & 14.02 & 33.72 & 39.14 & 24.97 \\
    & &  &  & Smith--Waterman & 60.28 & 60.28 & 58.95 & 29.10 & 27.09 & 17.02 & 46.93 & 42.46 & 42.94 \\
    \cmidrule{3-14}
    & & Vertices & Jaccard & Thresholding & 24.31 & 19.51 & 14.69 & 15.06 & 13.00 & 12.80 & 31.00 & 29.90 & 24.73 \\
    & &  &  & Smith--Waterman & 61.70 & 60.99 & 51.06 & 21.93 & 26.95 & 16.62 & 42.46 & 48.04 & 31.14 \\
    \cmidrule{4-14}
    & &  & Ru\v{z}i\v{c}ka & Thresholding & 38.74 & 39.60 & 15.17 & 20.50 & 22.60 & 12.31 & 33.59 & 39.29 & 26.89 \\
    & &  &  & Smith--Waterman & 65.26 & \bfseries 67.37 & 47.52 & 30.61 & \bfseries 30.95 & 16.44 & \bfseries 49.16 & \bfseries 49.16 & 31.14 \\
    \midrule
    Commens. & \textit{sbert} & Edges & Jaccard & Thresholding & 29.41 & 39.06 & 8.28 & 18.08 & 21.40 & 6.82 & 44.28 & 46.24 & 25.85 \\
    & & & & Smith--Waterman & 70.71 & 71.19 & 70.51 & 33.50 & 34.26 & 30.20 & 53.68 & 58.64 & \bfseries 63.87 \\
    \cmidrule{4-14}
    & &  & Ru\v{z}i\v{c}ka & Thresholding & 43.27 & 45.98 & 13.26 & 23.58 & 22.58 & 9.87 & 39.41 & 43.81 & 27.05 \\
    & & & & Smith--Waterman & 72.11 & 70.75 & 66.22 & 36.86 & 36.43 & 30.50 & 58.33 & 52.08 & 60.73 \\
    \cmidrule{3-14}
    & & Vertices & Jaccard & Thresholding & 7.84 & 11.25 & 13.14 & 17.30 & 14.96 & 6.13 & 37.35 & 33.72 & 25.23 \\
    & & & & Smith--Waterman & 69.39 & 75.93 & 68.90 & 32.36 & 35.86 & 31.78 & 59.38 & 60.42 & 61.46 \\
    \cmidrule{4-14}
    & &  & Ru\v{z}i\v{c}ka & Thresholding & 21.05 & 47.62 & 12.58 & 16.49 & 19.17 & 6.25 & 32.36 & 33.99 & 25.79 \\
    & & & & Smith--Waterman & 72.60 & 74.75 & 70.43 & 34.04 & 33.63 & 35.56 & 59.07 & 54.74 & 61.46 \\
    \cmidrule{2-14}
    & \textit{tfidf} & Edges & Jaccard & Thresholding & 22.10 & 31.96 & 13.26 & 19.17 & 15.62 & 11.71 & 31.96 & 31.58 & 24.56 \\
    & & & & Smith--Waterman & 72.97 & 74.07 & 75.84 & 32.58 & 35.79 & 33.29 & 59.69 & 58.33 & 52.08 \\
    \cmidrule{4-14}
    & &  & Ru\v{z}i\v{c}ka & Thresholding & 27.32 & 45.91 & 12.57 & 22.39 & 21.37 & 12.46 & 32.34 & 33.28 & 24.30 \\
    & & & & Smith--Waterman & \bfseries 78.50 & 74.50 & 74.92 & 35.00 & 38.30 & 33.62 & 58.03 & 58.03 & 56.54 \\
    \cmidrule{3-14}
    & & Vertices & Jaccard & Thresholding & 17.05 & 20.77 & 12.22 & 20.69 & 15.25 & 12.75 & 32.38 & 28.07 & 22.99 \\
    & & & & Smith--Waterman & 71.86 & 74.92 & 70.00 & 34.30 & 36.27 & 29.97 & 55.96 & 59.69 & 46.07 \\
    \cmidrule{4-14}
    & &  & Ru\v{z}i\v{c}ka & Thresholding & 35.24 & 46.84 & 13.33 & 16.91 & 21.08 & 11.92 & 29.58 & 31.18 & 23.37 \\
    & & & & Smith--Waterman & 75.00 & 74.75 & 72.48 & 36.18 & \bfseries 39.04 & 31.27 & 55.96 & 58.33 & 46.07 \\
    \bottomrule
\end{tabular}

%% file: tables/narr_match/tvshow-novels/perf_U2_best.tex
\begin{tabular}{p{5cm} r r r r r}
    \toprule
    Alignment & Textual Matching & \multicolumn{2}{c}{Structural Matching} & \multicolumn{2}{c}{Hybrid Matching} \\
     &  & Text-Constrained & Commensurate & Text-Constrained & Commensurate \\
    \midrule
    Thresholding & 29.01 & 37.24 & 32.56 & 37.95 & 41.18 \\
    Smith--Waterman & 36.65 & 45.00 & 62.87 & 50.78 & \bfseries 64.65 \\
    \bottomrule
\end{tabular}

%% file: tables/narr_match/perf_structural_cumulative.tex
\begin{tabular}{l l l r r r r r r r r r}
    \toprule
    Representation & Measure & Alignment & \multicolumn{3}{c}{\textit{Novels} vs. \textit{Comics}} & \multicolumn{3}{c}{\textit{Novels} vs. \textit{TV Show}} & \multicolumn{3}{c}{\textit{Comics} vs. \textit{TV Show}} \\
     &  &  & \ttfamily named & \ttfamily common & \ttfamily top-20 & \ttfamily named & \ttfamily common & \ttfamily top-20 & \ttfamily named & \ttfamily common & \ttfamily top-20 \\
    \midrule
    Edges & Jaccard & Thresholding & 4.05 & 3.94 & 3.61 & 6.36 & 5.12 & 4.89 & 19.98 & 20.20 & 19.95 \\
     & & Smith--Waterman & 16.85 & 41.40 & 10.79 & 2.00 & 22.34 & 0.00 & 19.88 & 29.71 & 21.05 \\
    \cmidrule{2-12}
     & Ru\v{z}i\v{c}ka & Thresholding & 3.80 & 3.98 & 3.62 & 5.66 & 5.17 & 4.90 & 20.21 & 20.41 & 20.22 \\
     & & Smith--Waterman & 22.14 & \bfseries 47.18 & 11.72 & 7.75 & 13.11 & 1.27 & 18.29 & 26.59 & 21.05 \\
    \cmidrule{1-12}
    Vertices & Jaccard & Thresholding & 4.10 & 3.90 & 3.53 & 4.82 & 5.18 & 4.89 & 22.24 & 20.78 & 19.79 \\
     & & Smith--Waterman & 12.69 & 40.43 & 1.01 & 1.19 & \bfseries 25.63 & 0.42 & 20.25 & 33.33 & 7.74 \\
    \cmidrule{2-12}
     & Ru\v{z}i\v{c}ka & Thresholding & 3.76 & 3.88 & 3.53 & 7.72 & 5.19 & 4.89 & 20.81 & 20.23 & 19.95 \\
     & & Smith--Waterman & 17.02 & 32.62 & 1.01 & 2.07 & 24.97 & 0.42 & 22.50 & \bfseries 35.63 & 7.74 \\
    \bottomrule
\end{tabular}